# On the Origin of Near-Infrared Extragalactic Background Light Anisotropy

Michael Zemcov[1,2], Joseph Smidt[3,4], Toshiaki Arai[5,6], James Bock[1,2], Asantha Cooray[4], Yan Gong[4], Min Gyu Kim[7], Phillip Korngut[2,1], Anson Lam[8,1], Dae Hee Lee[9], Toshio Matsumoto[5,10], Shuji Matsuura[5], Uk Won Nam[9], Gael Roudier[2], Kohji Tsumura[11] & Takehiko Wada[5]

[1] Department of Physics, Mathematics and Astronomy, California Institute of Technology, Pasadena, CA 91125, USA;
[2] Jet Propulsion Laboratory (JPL), National Aeronautics and Space Administration (NASA), Pasadena, CA 91109, USA;
[3] Theoretical Division, Los Alamos National Laboratory, Los Alamos, NM 87545, USA;
[4] Department of Physics & Astronomy, University of California, Irvine, CA 92697, USA;
[5] Department of Space Astronomy and Astrophysics, Institute of Space and Astronautical Science (ISAS), Japan Aerospace Exploration Agency (JAXA), Sagamihara, Kanagawa 252-5210, Japan;
[6] Department of Physics, Graduate School of Science, The University of Tokyo, Tokyo 113-0033, Japan;
[7] Department of Physics and Astronomy, Seoul National University, Seoul 151-742, Korea,
[8] Department of Physics and Astronomy, University of California, Los Angeles, CA 90095, USA,
[9] Korea Astronomy and Space Science Institute (KASI), Daejeon 305-348, Korea;
[10] Institute of Astronomy and Astrophysics, Academia Sinica, Taipei 10617, Taiwan, Republic of China;
[11] Frontier Research Institute for Interdisciplinary Science, Tohoku University, Sendai, Miyagi, 980-8578, Japan

**Extragalactic background light (EBL) anisotropy traces variations in the total production of photons over cosmic history, and may contain faint, extended components missed in galaxy point source surveys. Infrared EBL fluctuations have been attributed to primordial galaxies and black holes at the epoch of reionization (EOR), or alternately, intra-halo light (IHL) from stars tidally stripped from their parent galaxies at low redshift. We report new EBL anisotropy measurements from a specialized sounding rocket experiment at 1.1 and 1.6 micrometers. The observed fluctuations exceed the amplitude from known galaxy populations, are inconsistent with EOR galaxies and black holes, and are largely explained by IHL emission. The measured fluctuations are associated with an EBL intensity that is comparable to the background from known galaxies measured through number counts, and therefore a substantial contribution to the energy contained in photons in the cosmos.**

At near-infrared wavelengths, where the large zodiacal light foreground complicates absolute photometry measurements, the extragalactic background light (EBL) may be best accessed by anisotropy measurements. On large angular scales, fluctuations are produced by the clustering of galaxies, which is driven by the underlying distribution of dark matter. EBL anisotropy measurements can probe emission from epoch of reionization (EOR) galaxies (*1–3*) and direct-collapse black holes (*4*) that formed during the EOR before the universe was fully ionized by exploiting the distinctive Lyman cutoff feature in the rest-frame ultraviolet (UV), thus probing the UV luminosity density at high redshifts (*5*). However, large-scale fluctuations may also arise from the intrahalo light (IHL) created by stars stripped from their parent galaxies during tidal interactions (*6*) at redshift $z < 3$. A multi-wavelength fluctuation analysis can distinguish among these scenarios and constrain the EOR star formation rate.

A search for such background components must carefully account for fluctuations produced by known galaxy populations. Linear galaxy clustering is an important contribution to fluctuations on scales much larger than galaxies themselves. On fine scales, the variation in the number of galaxies produces predominantly Poissonian fluctuations, with an amplitude that depends on the luminosity distribution. Anisotropy measurements suppress foreground galaxy fluctuations by masking known galaxies from an external catalog.

The first detections of infrared fluctuations in excess of the contribution from known galaxies with the Spitzer Space Telescope (*7–9*) were interpreted as arising from a population of faint first-light galaxies at $z > 7$. The Hubble Space Telescope was used at shorter wavelengths (*10*) to carry out a fluctuation study in a small deep field but did not report fluctuations in excess of known galaxy populations. Measurements with the *AKARI*satellite (*11*) show excess fluctuations with a blue spectrum rapidly rising from $4.1\mu$m to $2.4\mu$m. Fluctuation measurements in a large survey field (*6*) with *Spitzer* agreed with earlier measurements (*7–9*), but were instead interpreted as arising from tidally stripped stars at $z \sim 1$ to $3$. Most recently, a partial correlation has been reported (*12*) between *Spitzer* and soft x-ray images, which Yue et al. (*4*) interpret as arising from direct-collapse black holes at $z > 12$.

We have developed and flown the specialized Cosmic Infrared Background Experiment [CIBER, (*13*)], a rocket-borne instrument specifically designed to study the spatial and spectral properties of the EBL. The imaging instrument (*14*) measures fluctuations in $\Delta\lambda/\lambda = 0.5$ bands centered at $1.1$ and $1.6\,\mu$m using two 11-cm telescopes each with a 2° by 2° field of view. Here, we report our analysis of data from two flights in 2010 and 2012.

The CIBER imager data are reduced from raw telemetered time streams, flat-field corrected based on a laboratory measurement, and masked for stars and galaxies using the Two Micron All Sky Survey (2MASS) $J-$ and $H$-band catalogs (*15*). We analyze differences between fields to reduce the effect of flat-fielding errors. The auto- and cross-power spectra of the masked, differenced images, corrected for mode coupling from the mask using a correction matrix, are shown in Figure 1. We also compute auto- and cross-spectra from a Spitzer infrared array camera (IRAC) $3.6\mu$m image that coincides with two of the five CIBER fields.

The CIBER auto- and cross-spectra show a significant excess over the predicted fluctuations from known galaxy populations (*16*) at $\ell < 5000$, where the multipole moment $\ell \simeq \pi/\theta$ and $\theta$ is the angular separation of two points on the sky. The excess is also evident in the cross-power spectra with *Spitzer*, showing that the source of the fluctuations is largely common from 1.1 to $3.6\,\mu$m. The large-scale fluctuations measured with CIBER correlate between the two flights in all combinations of bands and are independent of the detector arrays; the data pass multiple internal consistency tests (see the supplementary materials).

We rule out the following sources for producing the large-scale fluctuations (Fig. 1): (i) sunlight scattered by interplanetary dust, (ii) starlight scattered by interstellar dust (known as diffuse Galactic light), and (iii) fluctuations from faint stars. Zodiacal light fluctuations are eliminated because we observed the same fields separated by 17 months and obtained consistent results, viewed through two independent lines of sight through the interplanetary dust cloud. Constraints on zodiacal light fluctuations from *Spitzer* at $8\,\mu$m (*8*), scaled by the zodiacal spectrum to near-infrared wavelengths (*17*), lie significantly below the detected fluctuations (Fig. 1). The zodiacal spectrum is scaled by a conservative upper limit to the fluctuation amplitude measured at $7\,\mu$m from *AKARI* (*17,18*) in Fig. 2.

We estimate diffuse Galactic light fluctuations (Figs. 1 and 2) by scaling the Infrared Astronomical Satellite (IRAS) $100\,\mu$m intensity by a factor consistent with previous diffuse Galactic light measurements (*19*). We estimate the diffuse Galactic light component by calculating the CIBER $\times$ IRAS cross-spectra and fitting a single amplitude coefficient to a $C_\ell \propto \ell^{-3}$ spectrum with a fixed $1.1\mu$m / $1.6\mu$m color ratio, which is less than the detected CIBER power spectra at all spatial scales. Fluctuations from the extinction of the background light by galactic dust are also negligible.

We constrain fluctuations from unmasked stars using the UKIDSS-UDS stellar catalog (*20*), which is complete to $J = 24.9, H = 24.2$ ($5\sigma$), accounting for $> 99.9\%$ of the integrated light from stars. We compute the auto power spectrum of residual stars below our cutoff flux and find that the amplitude is negligible on large scales and follows a Poisson spectrum.

The root mean square (RMS) fluctuation amplitude $\delta\lambda I_\lambda = \langle \ell(\ell+1)C_\ell/2\pi \rangle^{1/2}$ over $500 < \ell < 2000$ has a spectral energy distribution that is approximately Rayleigh-Jeans (Fig. 2). The $1.1\,\mu$m data point lies $2\sigma$ below the best-fitting $\lambda^{-3}$ scaling, suggesting the possibility of a departure from the Rayleigh-Jeans spectrum at short wavelengths. To fit models to CIBER and *Spitzer* data, we mask *Spitzer* data to a depth $L < 16$ that is comparable to CIBER. The *Spitzer* power spectra give a non-Poissonian $\ell^{1.7}$ signal at high multipoles for the $L < 16$ flux cut (Fig. 1), which is evidence for nonlinear clustering that is not observed with a deeper flux cut (*6*) and is not predicted in linear galaxy clustering (*16*).

Having excluded explanations based on zodiacal light, diffuse Galactic light, and the clustering of known galaxies, we consider three possibilities to explain these measurements: EOR galaxies, EOR black holes, and IHL. For first-light galaxies, we use models for population II and population III stars (*21*) and combinations of the two. The direct-collapse black hole model at $z > 12$ is excluded both by amplitude and color as it has no mechanism to generate substantial $1.1$ and $1.6\,\mu$m power (*4*). For IHL, we use models based on the spectral energy distributions of old stellar populations in dark matter halos (*6*). We simultaneously fit the $1.1$, $1.6$, and $3.6\,\mu$m auto-power spectra, taking into account galaxy clustering, diffuse Galactic light, and residual flat-field errors at low $\ell$. The IHL model, summarized in Fig. 1, generally fits the data but somewhat underpredicts the amplitude at $1.1$ and $1.6\,\mu$m; neither the IHL nor the EOR models quite match the high observed $1.6/3.6\,\mu$m color ratio (see section 10 of the supplementary materials). This may indicate that additional components are reflected in the data. However, more theoretical work is required to determine whether adding non-linear galaxy clustering and non-linear

IHL production to the model can improve the fit.

We estimate the EBL intensity associated with the fluctuations by taking the measured fluctuation amplitude between $500 < \ell < 2000$, obtained by subtracting estimated contributions from low-z galaxies and diffuse Galactic light, and multiplying by a model-dependent contrast factor $\lambda I_\lambda / \delta\lambda I_\lambda$, where $\lambda I_\lambda$ is the total intensity associated with a component. For the IHL model, which has a low contrast factor, we obtain an associated EBL of $7.0^{+4.0}_{-3.5}$ and $11.4^{+5.4}_{-4.8}$ nW m$^{-2}$ sr$^{-1}$ at 1.1 and 1.6 $\mu$m, respectively. As shown in Table 1, the IHL background is of a similar magnitude to the integrated galaxy light (IGL) background derived from galaxy counts. However, we note the IHL background has a much bluer color than the IGL background. We similarly estimate the IHL background at longer wavelengths from *AKARI* and *Spitzer*. Non-linear galaxy clustering appears to contribute to the *Spitzer* fluctuations, so we quote two values that depend on the choice of flux cut, the deeper flux cut being less sensitive to the non-linear clustering contribution. The CIBER data do not appear to be as sensitive to the flux cut, perhaps due to the higher IHL to IGL ratio at these wavelengths.

The total EBL, the summation of the IHL and IGL backgrounds, is consistent with current EBL measurements. Near-infrared absolute photometric background measurements remain uncertain due to the bright zodiacal foreground, but the lowest such measurement (*22*) gives $21 \pm 15$nW m$^{-2}$ sr$^{-1}$ and $13.3 \pm 2.8$ nW m$^{-2}$ sr$^{-1}$ at 1.25 and 3.6 $\mu$m, respectively. The High Energy Stereoscopic System (HESS) measurement of the EBL using $\gamma$-ray absorption spectra (*23*), 15 $\pm 2(\mathrm{statistical}) \pm 3(\mathrm{systematic})$ nW m$^{-2}$ sr$^{-1}$at 1.4 $\mu$m, is independently consistent with the sum of the IHL and IGL backgrounds, within the uncertainties in all measurements.

An EOR interpretation of these measurements would result in a large background, as the contrast factor is larger than that of IHL, 10 to 100 for EOR models (*3,6*). The implied EOR background (see Table 1) originating from high redshifts is difficult to justify, due to the over-production of metals and x-ray background photons (*24*).

Our results indicate that a substantial fraction of the EBL at optical and near-infrared wavelengths originates from stars outside of galaxies (with boundaries as traditionally defined). This in turn adds to the cosmic energy budget and, depending on the mass characteristics and spectrum of the population responsible, could help alleviate the "photon underproduction crisis" (*25*) and the "missing baryon problem" (*26*). We see no evidence for a detected EOR background component in our data. Multiwavelength fluctuation measurements extending into the optical will help discriminate the EOR background component using the redshifted Lyman cutoff, and future spectroscopic measurements will enable tomographic measurements to determine the history of IHL production.

**Acknowledgements** Our thanks to O. Doré, J. Filippini and K. Ganga for useful conversations and comments throughout the course of this work and K. Helgason for kindly providing models of the statistics of the near-infrared CIB. The authors acknowledge the excellent support from the NASA sounding rockets program that was essential in developing, testing, qualifying, launching, and recovering our payloads. The CIBER auto- and cross-power spectra are available for public download at: http://ciber.caltech.edu/zemcovetal/ This work was supported by NASA APRA research grants NNX07AI54G, NNG05WC18G, NNX07AG43G, NNX07AJ24G, and NNX10AE12G. Initial support was provided by an award to J.B. from the Jet Propulsion Laboratory's Director's Research and Development Fund. Japanese participation in CIBER was supported by KAKENHI (20·34, 18204018, 19540250, 21340047, and 21111004) from Japan Society for the Promotion of Science (JSPS) and the Ministry of Education, Culture, Sports, Science and Technology (MEXT). Korean participation in CIBER was supported by the Pioneer Project from Korea Astronomy and Space Science Institute (KASI). M.Z. and P.K. acknowledge support from NASA Postdoctoral Program fellowships, A.C. acknowledges support from an NSF CAREER award AST-0645427 and NSF AST-1313319, and K.T. acknowledges support from the JSPS Research Fellowship for Young Scientists. This publication makes use of data products from the Two Micron All Sky Survey (2MASS), which is a joint project of the University of Massachusetts and the Infrared Processing and Analysis Center/California Institute of Technology, funded by the National Aeronautics and Space Administration and the National Science Foundation. This work made use of images and/or data products provided by the National Optical Astronomy Observatory (NOAO) Deep Wide-Field Survey (NDWFS), which is supported by the NOAO, operated by AURA, Inc., under a cooperative agreement with the National Science Foundation.

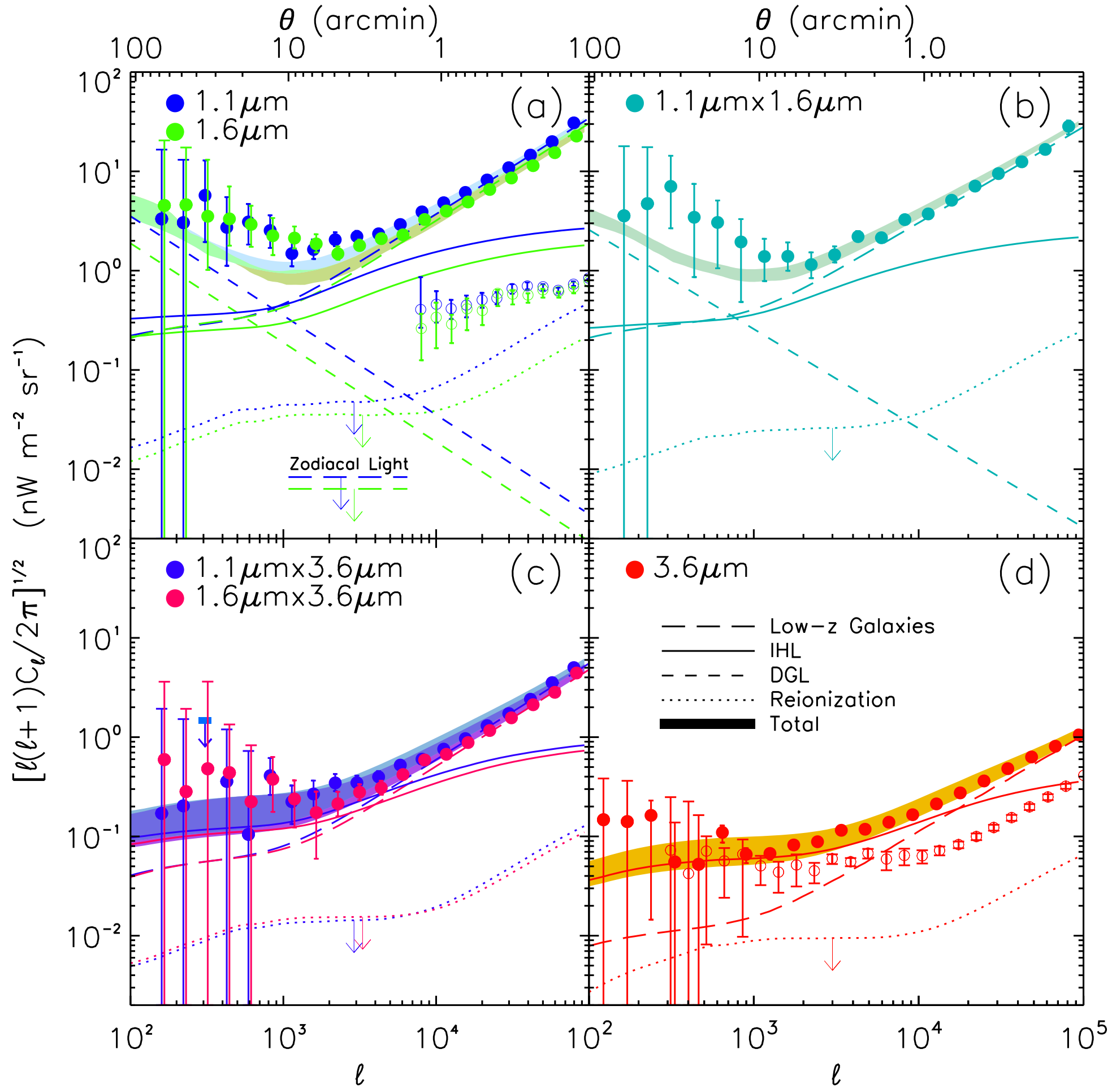

Figure 1: **CIBER and *Spitzer* auto- and cross-spectra at 1.1, 1.6 and 3.6 $\mu$m.** We show the CIBER auto-spectra for $1.1 \times 1.1\,\mu$m, and $1.6 \times 1.6\,\mu$m (panel a), the CIBER $1.1 \times 1.6\,\mu$m cross-spectrum (panel b), the CIBER–*Spitzer* $1.1 \times 3.6\,\mu$m and $1.6 \times 3.6\,\mu$m cross-spectra (panel c), and *Spitzer* $3.6 \times 3.6\,\mu$m auto-spectra (*6*) (panel d). At $1.1, 1.6$ and $3.6\,\mu$m we indicate previous measurements (open circles; (*6,10*); note the HST measurements apply a much deeper flux cut at 1.1 and 1.6 $\mu$m for masking, and the *Spitzer* flux cut is somewhat deeper than the cut we are applying at $3.6\,\mu$m). The $3.6\,\mu$m points use the same data set but are masked to a lower source flux for comparison to our spectrum which is masked to $L < 16$. The increased depth reduces some of the mid-$\ell$ power. We show constraints on astrophysical foregrounds, including unmasked stars and $z < 5$ galaxies (*16*), zodiacal light (*8*), and diffuse Galactic light. In all cases, we detect a significant excess power at $\ell < 5000$ (angular separations $\theta > 4.3'$). We model the data using components from IHL (*6*) and $z > 7$ first galaxies (*3, 5*), obtaining a 95% confidence upper limit on the EOR contribution. The fitted total indicated by the filled band includes all of the astrophysical components plus a bounded systematic error for flat-field variations. The width of the band indicates the 68 % uncertainty interval of the fit including all of the modeling uncertainties. The correlation coefficients for $1.1 \times 1.6\,\mu$m, $1.1 \times 3.6\,\mu$m, and $1.6 \times 3.6\,\mu$m are $0.76 \pm 0.10$, $0.55 \pm 0.14$ and $0.31 \pm 0.14$, respectively, with no statistically significant angular dependence in the correlations (see Section 6.2 of the Supplementary Materials for a description of this calculation).

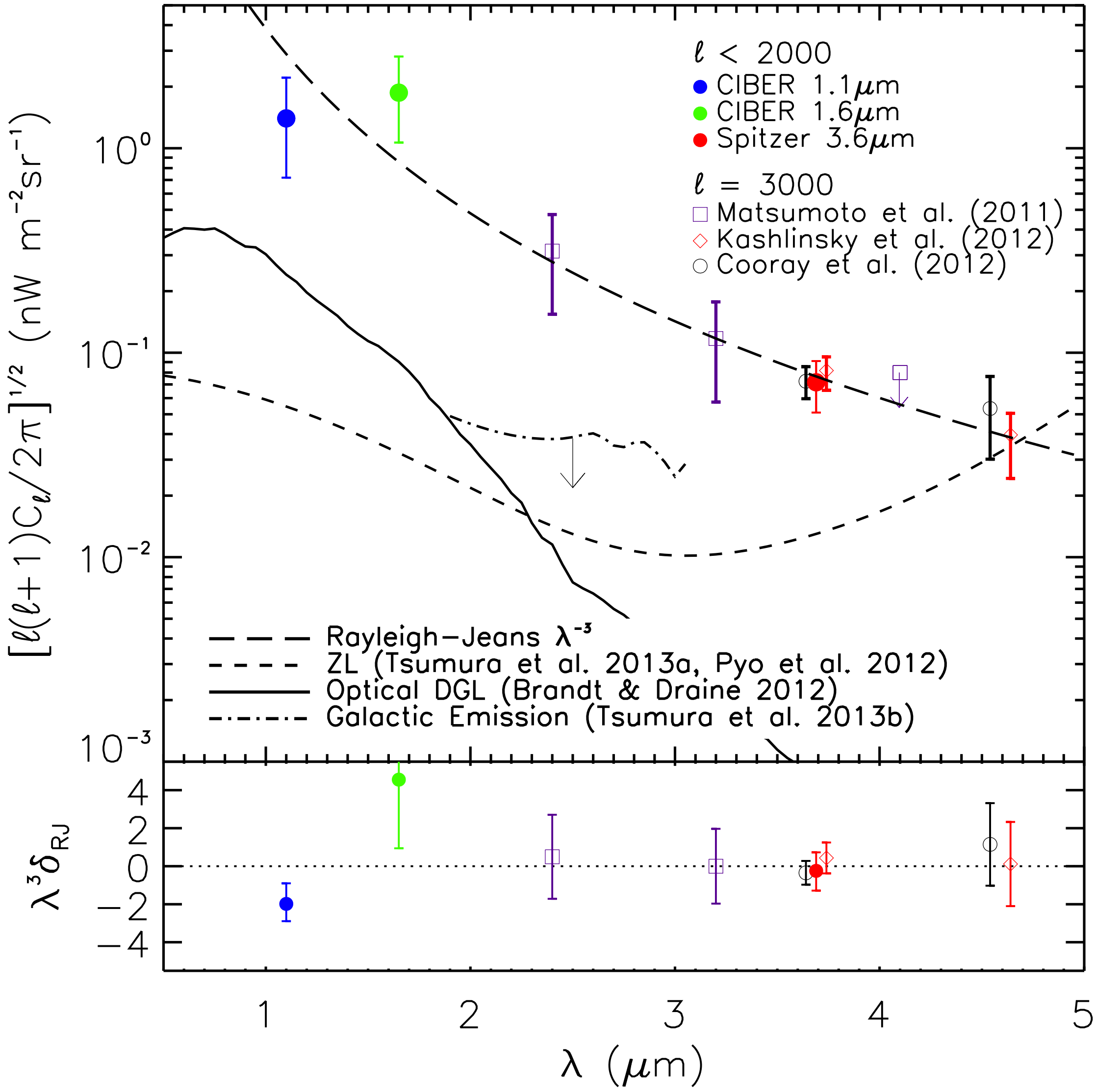

Figure 2: **The electromagnetic spectrum of the near-infrared fluctuations.** We show measurements of the fluctuation power from CIBER and *Spitzer* averaged between $500 < \ell < 2000$ (solid points). Also indicated are previous measurements from *AKARI* (*11*) and *Spitzer* (*6, 7*) at $\ell = 3000$ which use deeper masking thresholds. In all cases we subtract the contribution from the shot noise of unmasked galaxies (*16*). We indicate the best fitting Rayleigh-Jeans spectrum to the points from this analysis, estimates for diffuse Galactic light fluctuations (*19*), a conservative constraint on zodiacal light fluctuations (*17, 18*), and an upper limit on Galactic emission (*27*). The known foreground components have both smaller amplitudes and different spectra than the measurements. We show the residual from the best fitting Rayleigh-Jeans spectrum $\delta_{\rm RJ}$ in the bottom panel, scaled by $\lambda^3$ to reduce the range. The shortest wavelength measurement at $1.1\,\mu$m is $2.0\,\sigma$ below the fit, indicating a possible short-wavelength departure from a Rayleigh-Jeans spectrum.

Table 1: **Contributions to near-infrared EBL anisotropy and intensity.** At each wavelength, we list the measured fluctuation amplitude at large angular scales; the model-dependent ratio of EBL intensity to EBL anisotropy; the IGL determined by previous measurements; the ratio of the IHL and IGL intensities; and finally the inferred total background intensity from both components. We also list the background intensity that would arise assuming the measured fluctuations are entirely due to high-redshift EOR galaxies.

| $\lambda$ ($\mu$m) | Measured $\delta\lambda I_\lambda^{a}$ (nW m$^{-2}$ sr$^{-1}$) | $\frac{\lambda I_{\lambda,\mathrm{IHL}}}{\delta\lambda I_\lambda}$ | $\lambda I^{b}_{\lambda,\mathrm{IHL}}$ (nW m$^{-2}$ sr$^{-1}$) | $\lambda I^{c}_{\lambda,\mathrm{IGL}}$ (nW m$^{-2}$ sr$^{-1}$) | $\frac{\lambda I_{\lambda,\mathrm{IHL}}}{\lambda I_{\lambda,\mathrm{IGL}}}$ | $\lambda I_{\lambda,\mathrm{IHL}}$ + $\lambda I_{\lambda,\mathrm{IGL}}$ (nW m$^{-2}$ sr$^{-1}$) | $\lambda I^{d}_{\lambda,\mathrm{EOR}}$ (nW m$^{-2}$ sr$^{-1}$) |
|---|---|---|---|---|---|---|---|
| 1.1 | $1.4^{+0.8}_{-0.7}$ | 5 | $7.0^{+4.0}_{-3.5}$ | $9.7^{+3.0}_{-1.9}$ | 0.7 | $16.7^{+5.0}_{-4.0}$ | 28 |
| 1.6 | $1.9^{+0.9}_{-0.8}$ | 6 | $11.4^{+5.4}_{-4.8}$ | $9.0^{+2.6}_{-1.7}$ | 1.3 | $20.4^{+6.0}_{-5.1}$ | 38 |
| 2.4 | $0.32 \pm 0.05$* | 7 | $2.2 \pm 0.4$ | $7.8^{+2.0}_{-1.2}$ [e] | 0.3 | $10.0^{+2.0}_{-1.3}$ | 6.4 |
| 3.6 | $0.072^{+0.019}_{-0.021}$ | 9 | $0.65^{+0.17}_{-0.19}$ | $5.2 \pm 1.0$ | 0.1 | $5.9 \pm 1.0$ | 1.4 |
| 3.6[f] | $0.049^{+0.021}_{-0.007}$ | 9 | $0.44^{+0.19}_{-0.06}$ | $5.2 \pm 1.0$ | 0.1 | $5.6 \pm 1.0$ | 1.0 |
| 4.5 | $0.053 \pm 0.023$* | 7 | $0.37 \pm 0.16$ | $3.9 \pm 0.8$ | 0.1 | $4.3 \pm 0.8$ | 1.0 |

[a] RMS fluctuation amplitude computed as averages of measured data over $500 < \ell < 2000$, except for $*$ which are determined at $\ell = 3000$ using fainter mask cuts due to restricted field size (see also note [d]).

[b] The IHL background from the product of column 2 and 3.

[c] The IGL background as compiled by Ref. 28.

[d] Computed EOR background assuming EOR fluctuations with $\lambda I_\lambda / \delta\lambda I_\lambda = 20$.

[e] Determined at $K$-band corresponding to $2.2\,\mu$m.

[f] Computed using the measurements of Ref. 6 averaged over $500 < \ell < 5000$.

In these Supplemental Notes to the Report, we describe how we calculate angular power spectra and the various statistical, systematic, and astrophysical uncertainties. The design of the overall CIBER experiment (*13*) and the design of the imager instruments (*14*) are presented elsewhere.

# 1 Introduction

Accurate measurements of near-infrared extragalactic background light (EBL) anisotropy require: high fidelity on $\theta > 5'$ angular scales where clustering fluctuations dominate over shot noise, sufficient angular resolution to mask foreground galaxies, and observations of several fields to allow tests of internal consistency. The CIBER imager instrument is designed to make such accurate wide-field measurements from a sounding rocket platform, above airglow emission in the upper atmosphere.

The organization of the supplement is as follows. The design of the fluctuations survey is detailed in Section 2. We describe the reduction pipeline from raw data to calibrated sky images, masking astronomical sources in Section 3. Power spectra of the images are then calculated, and various instrumental and analysis effects are accounted for, including noise bias subtraction, beam correction, and mode coupling correction in Section 4.

The noise properties of the instrument, from which the statistical uncertainties are calculated, have complex structure in Fourier space. In Section 5 we develop a noise model which replicates the uncertainties arising from the detector read chain and photon noise. This noise model is used to predict the statistical uncertainty of the measurement, the power spectral noise bias, and to optimally weight spectra in Fourier space. The results of the power spectrum analysis are presented in Section 6.

We place limits on instrumental systematic errors arising from airglow contamination, flat-field errors, source masking, and mode coupling artifacts in Section 7. In Section 8 we perform a variety of calculations to test the internal consistency of the data between instruments and flights, and validate our models for the statistical and systematic uncertainties.

We compute astrophysical contributions to the power spectra from foreground components, including sunlight scattered from interplanetary dust (Zodiacal Light, or ZL), light from the Galactic interstellar radiation field scattered from dust particles in the Galaxy (diffuse Galactic light, or DGL), stars, and known galaxy populations in Section 9.

Finally, in Section 10 we interpret the fluctuation measurements using a model for the near-infrared EBL which includes intra-halo light (IHL) arising from stars tidally stripped from their parent galaxies as a result of structure formation and first galaxies and black holes during the epoch of reionization (EOR).

## 2 Imager Survey

The CIBER experiment has flown four times: 10:45 UTC 25 February 2009; 4:50 UTC 11 July 2010; 9:00 UTC 22 March 2012; and 3:05 UTC 6 June 2013. In this analysis we present the results from the second and third flights. The first flight had technical problems which precluded robust analysis of the imager data (*14*). We will present an analysis the fourth flight data in a future paper. CIBER's two imaging telescopes are identical and differ only in their band-defining filters. To eliminate any potential artifacts associated with the arrays, the filters were swapped between optical assemblies after the second flight. In addition, in the third flight the filter bands were shifted to reduce spectral overlap around $1.25\,\mu$m. Table 1 indicates the wavelength and detector array configuration for each flight.

The science fields listed in Table 1 were chosen for their ancillary imaging data (*14*). In particular, data from the *Spitzer* Deep and Wide Field Survey (SDWFS (*29*)) are of critical importance as they allow us to perform cross-correlation studies with the fluctuation power measured by *Spitzer* in the Boötes fields (*6*). The fields were observed in both flights to check reliability. Due to the different instrument configurations between the flights, individual arrays view the sky rotated by $90^\circ$ in each flight, reducing any array-dependent signatures.

## 3 Data Analysis and Power Spectrum Generation

### 3.1 Low-Level Data Analysis

We first reduce raw telemetered data to produce calibrated, flat-field corrected, astrometrically aligned images for each science field.

*Raw Telemetered Data to Sky Images*. The imager instruments use HAWAII-1 HgCdTe detector arrays, imaged to a $2^\circ \times 2^\circ$ region on the sky with $7''$ pixels. Data are telemetered to the ground during the flight and mapped into $1024 \times 1024$ pixel image frames, continuously sampled at $1.78$ s intervals. We fit lines to the data for each pixel to produce best estimates of the photocurrent (*30*). These slope fits are converted from raw detector digital units to $e^{-}\mathrm{s}^{-1}$ using array gain factors, a convenient intermediate choice of units used before the images are calibrated to surface brightness.

*Dark Current Correction*. The detectors have a zero-signal response termed "dark current", which includes thermally generated carriers and drifts in the array and readout electronics. We estimate a dark current image from the sum of 24 dark $50$ s integrations taken pre-flight. The average is subtracted from each science field to remove the dark current. The dark current subtraction is a small correction to the offset with a fluctuation amplitude that is always $< 2\%$ of our reported level at all band powers and is negligible compared to the measurement errors (see Figure S20).

*Flat Fielding*. Each pixel in an imager detector array has a particular responsivity to incoming photons. A measurement of the instrument's response to a spatially uniform input illumination (the "flat-field" response) is used to correct the sky images for these gain variations. We estimate the flat-field response from a laboratory measurement using a field-filling integrating sphere, a uniform radiance source with a solar spectrum (*13,14*). Alternatively, the flat-field response could be calculated from the ensemble of flight sky images. However, except for the fourth flight, there are not enough stable, independent images free from airglow contamination to avoid mixing EBL anisotropy into the gain estimate. The potential systematic uncertainties associated with using the lab flat-field measurement are investigated in Section 7.

Table S1: Imager Survey Parameters

| Field Name | R.A. (J2000)[a] | Dec. (J2000)[a] | Type | Time After Launch (s) | Altitude (km) | $\langle \lambda I_\lambda \rangle$ (nW m$^{-2}$ sr$^{-1}$) at 1.1 $\mu$m | $\langle \lambda I_\lambda \rangle$ (nW m$^{-2}$ sr$^{-1}$) at 1.6 $\mu$m | Airglow Suppression Factor |
|---|---|---|---|---|---|---|---|---|
| 2[nd] Flight, 4:50 11 July 2010 UTC, NSROC Mission # 36.265, $\lambda_{1.1\,\mu m} = 1.12\,\mu$m, $\lambda_{1.6\,\mu m} = 1.56\,\mu$m. | | | | | | | | |
| ELAIS-N1 | $16^h10^m.8$ | $54°41'.8$ | Airglow | 150 - 200 | 231 - 283 | 298 | 375 | 1.0 |
| NEP[b] | $18^h2^m.5$ | $66°26'.8$ | Science | 217 - 280 | 295 - 319 | 320 | 263 | 0.021[d] |
| Boötes A[c] | $14^h33^m.8$ | $34°43'.0$ | Science | 367 - 423 | 294 - 242 | 367 | 315 | 0.034 |
| Boötes B[c] | $14^h28^m.9$ | $33°20'.2$ | Airglow | 451 - 501 | 206 - 122 | 409 | 506 | 1.0 |
| 3[rd] Flight, 9:00 22 March 2012 UTC, NSROC Mission # 36.277, $\lambda_{1.1\,\mu m} = 1.05\,\mu$m, $\lambda_{1.6\,\mu m} = 1.79\,\mu$m[e]. | | | | | | | | |
| Lockman | $10^h46^m.1$ | $58°12'.2$ | Airglow | 126 - 172 | 206 - 264 | 530 | 650 | 1.0 |
| ELAIS-N1 | $16^h11^m.5$ | $54°37'.9$ | Science | 196 - 242 | 287 - 316 | 304 | 238 | 0.031 |
| NEP[b] | $18^h2^m.4$ | $66°19'.9$ | Science | 256 - 308 | 321 - 325[f] | 320 | 247 | 0.001 |
| Boötes B[c] | $14^h29^m.5$ | $33°22'.1$ | Science | 377 - 425 | 293 - 246 | 343 | 258 | 0.018 |
| Boötes A[c] | $14^h33^m.8$ | $34°49'.7$ | Airglow | 441 - 511 | 225 - 108 | 617 | 1173 | 1.0 |

[a] Referenced to the 1.1 $\mu$m channel field center.

[b] We differentiate between the slightly shifted NEP field centers in the second and third flights using NEP-2 and NEP-3.

[c] These fields overlap with the SDWFS *Spitzer* data.

[f] Apogee at $t = 294$ s, $h = 327$ km.

[e] Filter stacks are swapped between instrument assemblies in 2[nd] and 3[rd] flights.

[f] Through apogee at $t = 291$ s, $h = 326$ km.

*Astrometric Registration*. We astrometrically register the science images using code supplied by the ASTROMETRY.NET collaboration (*31*). The astrometry solution for the center of each of our survey fields referenced to the $1.1\,\mu$m channel is listed in Table 1, accurate to $< 1''$ per field. The two imager cameras have a slight mismatch in their field coverage due to a boresight misalignment of $\sim 5'$. Due to variance in the achieved pointing in each flight, the center of the single field common to both flights, NEP, is mismatched by $7'.1$ between the second and third flights. To differentiate between these fields in figures, we define NEP-2 as the second-flight NEP field and NEP-3 as the third-flight NEP field. The pointing mismatches lead to a small loss of sky coverage in matching comparisons.

*Point Spread Function*. The determine the effective point spread function (PSF) for each image by stacking the science field images on the positions of known 2MASS (*15*) sources, which includes the effect of pointing smear. We stack 2MASS sources to determine the core PSF. We use sources in the range $16.0 < m < 16.1$, where $m$ is the 2MASS Vega magnitude, to avoid detector non-linearity. The resulting stack is then spliced into an equivalent stack of bright sources $(7 < m < 9)$ to measure the PSF out to large radius (see Ref. 14 for details). The effect of the PSF on the power spectral analysis is presented in Section 4.2.

*Surface Brightness Calibration*. We calibrate the cameras to surface brightness by stacking sources with $13.5 < m < 15$ in 2MASS catalogs over $\delta m = 0.1$. We interpolate the three-band photometry of 2MASS to the imager bands (a $\sim 5\,\%$ modification to the flux) for each source and sum to determine the fiducial calibration. The average 2MASS source flux in each $\delta m$ bin, when compared to the photocurrent in the image stack, yields a flux conversion factor. This measurement is combined with the measured solid angle of the PSF to determine the surface brightness calibration factor (*14*). As a check, the resulting values are compared with laboratory-based measurements of the radiance responsivity (*13*), and we find these independent calibration measurements are consistent within 10% accuracy. The resulting unaligned science images for a typical observation are shown in Figure S1.

## 3.2 Image Space Mask Generation

Power from both imperfections in the detector and astronomical sources are then masked prior to the calculation of power spectra.

*Detector Mask*. We calculate a "detector mask" to remove imperfections in the detector array, comprised of a static mask that removes consistently unresponsive and hot pixels, and a dynamic mask that removes pixels affected by transient problems. The static mask consists of pixels which display consistently atypical behavior in an ensemble of laboratory images. These pixels have two clear populations: dead pixels randomly distributed over the array, and dead and noisy pixels concentrated along the edges, quadrant boundaries, and near the corners where fabrication yield is poor. Between these two populations $\sim$ 8% of pixels are lost, depending on the detector array. Dynamically-masked pixels are normally well-behaved but suffer from transients like cosmic ray hits, electronics glitches, and abnormal multiplexer behavior. These are flagged as part of the preprocessing based on the probability to exceed $\chi^2$ statistic of the line fits, as well as a later outlier pixel masking step in processed images. Typically the dynamic mask removes $\sim 0.5\,\%$ of pixels from analysis, and the population of these pixels changes in each integration.

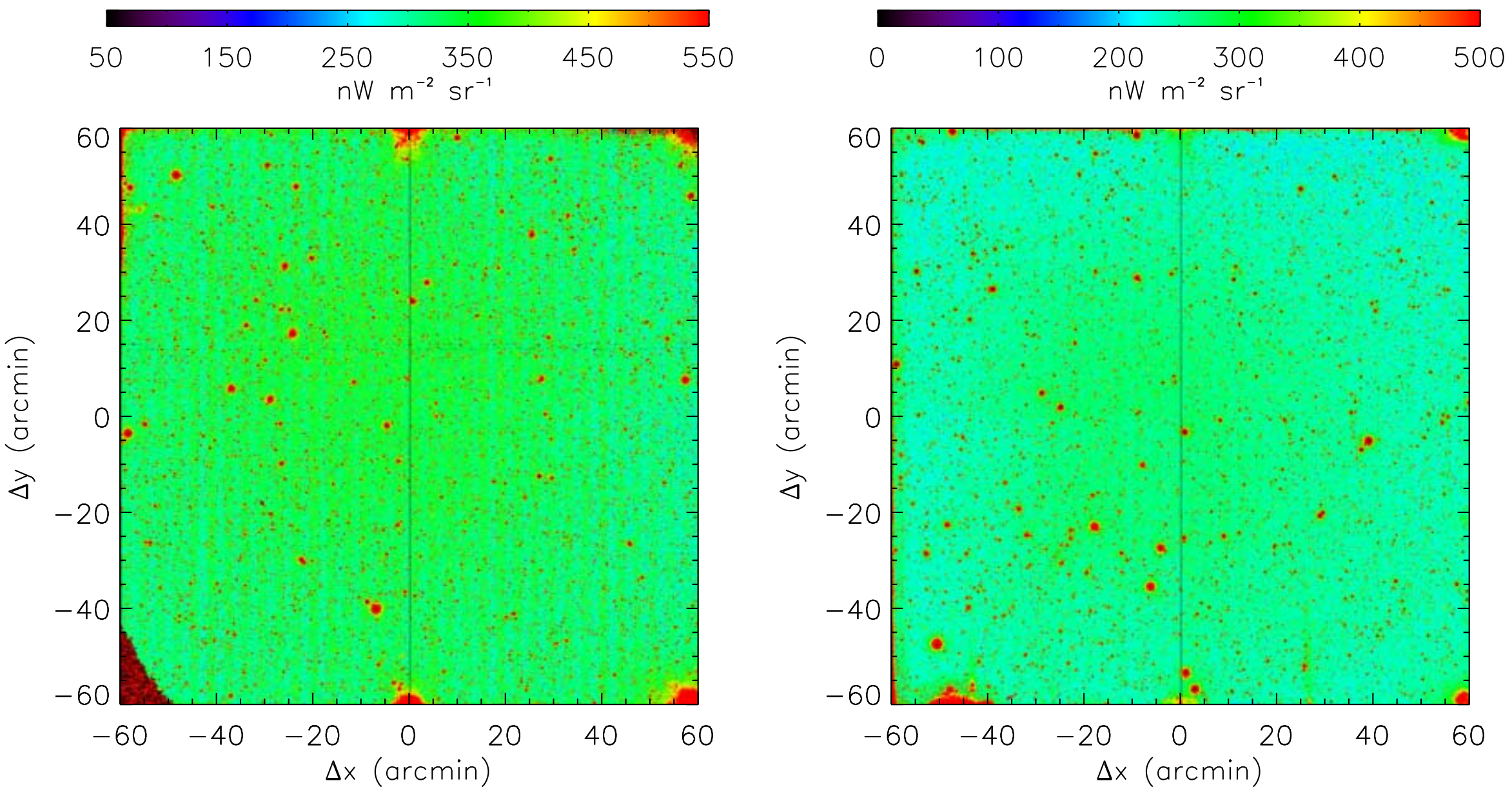

Figure S1: **Imager flight images.** The 1.1 $\mu$m image (left hand panel) and the 1.6 $\mu$m image (right hand panel) for NEP in the third flight are shown. No masking is applied to these images, which are the output of the pipeline presented in Section 3.1. These images are in array coordinates and are not yet astrometrically aligned.

*Astronomical Mask*. We mask stars and galaxies from an external catalog to reduce Poisson fluctuations so that we can detect the clustering amplitude of fluctuations at $\ell < 5000$. In addition to a flux cut, we must reduce the residual signal from the extended PSF outside of the chosen mask radius. Specifically, our masking requirements are as follows:

1. At the full depth of the masking catalog, the residual fluctuation power due to the extended PSF outside the mask radius ("mask halos") must be less than the statistical sensitivity of the instrument.

2. Poisson fluctuations from galaxies fainter than the mask cutoff be a low enough to allow a detection of clustered fluctuations.

3. The images retain as many unmasked pixels as possible after requirements (1) and (2) are met.

Given our measured fluctuation amplitude, through simulations and existing clustering models (*16*), we determined that masking local galaxies to $\sim 17^{\rm th}$ Vega magnitude at $1.6\,\mu$m (*14*) will meet requirements 1 and 2. Down to such a depth, we can also mask our images to retain about 40% to 50% of the pixels in each of the images; as we discuss later, our fluctuation measurements make use of difference images of two fields. Down to $\sim 17^{\rm th}$ Vega magnitude at $1.6\,\mu$m such images retain around 30% of the pixels for fluctuation measurements.

We optimize the astronomical mask by simulating sky images from the NEWFIRM (*32*) survey catalog to $J < 21$ completeness, convolved with the imager PSF. To determine the mask, we generate a trial source mask with radius $r$ as a function of the source magnitude $m$ for each source in the catalog:

$$r(m) = \alpha_m m + \beta_m, \tag{1}$$

where the coefficients $\alpha_m$ and $\beta_m$ are dictated by our requirement to reduce fluctuations from mask halos to an acceptable level while leaving sufficient pixels for analysis. To determine the value of $\alpha_m$ and $\beta_m$ for each imager band, we perform a simulation where sources with $m < m'$ are included, the trial mask is applied, and the power spectrum of the resulting masked image is calculated. We run a range of $\alpha_m$ and $\beta_m$ values to optimize for the requirements 1–3 above. The best masking parameters are found to be $\alpha_m = -6.25''$ mag$^{-1}$ and $\beta_m = 110''$. The PSF and the density of stars are sufficiently similar between fields to use this prescription for all fields and bands.

The deeper NEWFIRM data are only available for the Boötes field, so we use the extended 2MASS $J$ and $H$-band catalogs (*15*) to build the flight masks using the algorithm derived from the deep NEWFIRM data. The extended 2MASS catalog is 75% complete at $J = 17.5$, which translates to 17.5 and 17.0 in CIBER's $1.1\,\mu$m and $1.6\,\mu$m bands, respectively, and all fields are masked to this depth. Figure S2 shows example science field images after masking has been applied. Typical single-field masks remove $\sim 30\,\%$ of the pixels due to astrophysical sources, and $\sim 40\,\%$ of pixels overall. In field differences (discussed later, see Figures S4 and S5), this fraction increases to $50-70\,\%$.

We validate our choice of cutoff flux by varying the cutoff depth using NEWFIRM, as shown in Figure S3. We convolve the NEWFIRM catalog in a $2^\circ \times 2^\circ$ field with the CIBER PSF, where the deepest possibile cutoff flux is given by the NEWFIRM depth of $J = 20$. We then produce synthesized images with sources brighter than a chosen flux, and apply the optimized masking procedure. For the case where all the sources are brighter than the $J > 17.5$ cutoff flux, the only source of power is from mask halos. At the depth of the 2MASS catalog, the power from mask halos is equal to the statistical sensitivity.

When we include sources fainter than the $J = 17.5$ cutoff flux, the power increases, since they are not masked. The fluctuation level calculated for $J < 20$ sources from NEWFIRM data agrees with the calculated residual from Ref. *16* for this flux range. The NEWFIRM catalog resolves $\sim 60\,\%$ of the integrated galactic light (IGL). We extrapolate the total power from unmasked sources for the entire IGL by multiplying the full depth NEWFIRM simulation by $(0.6)^{-2}$. Though there is residual power from point sources at the chosen $J = 17.5$ cutoff flux, it is significantly less than the power we measure for $\ell < 5000$ (see Section 6).

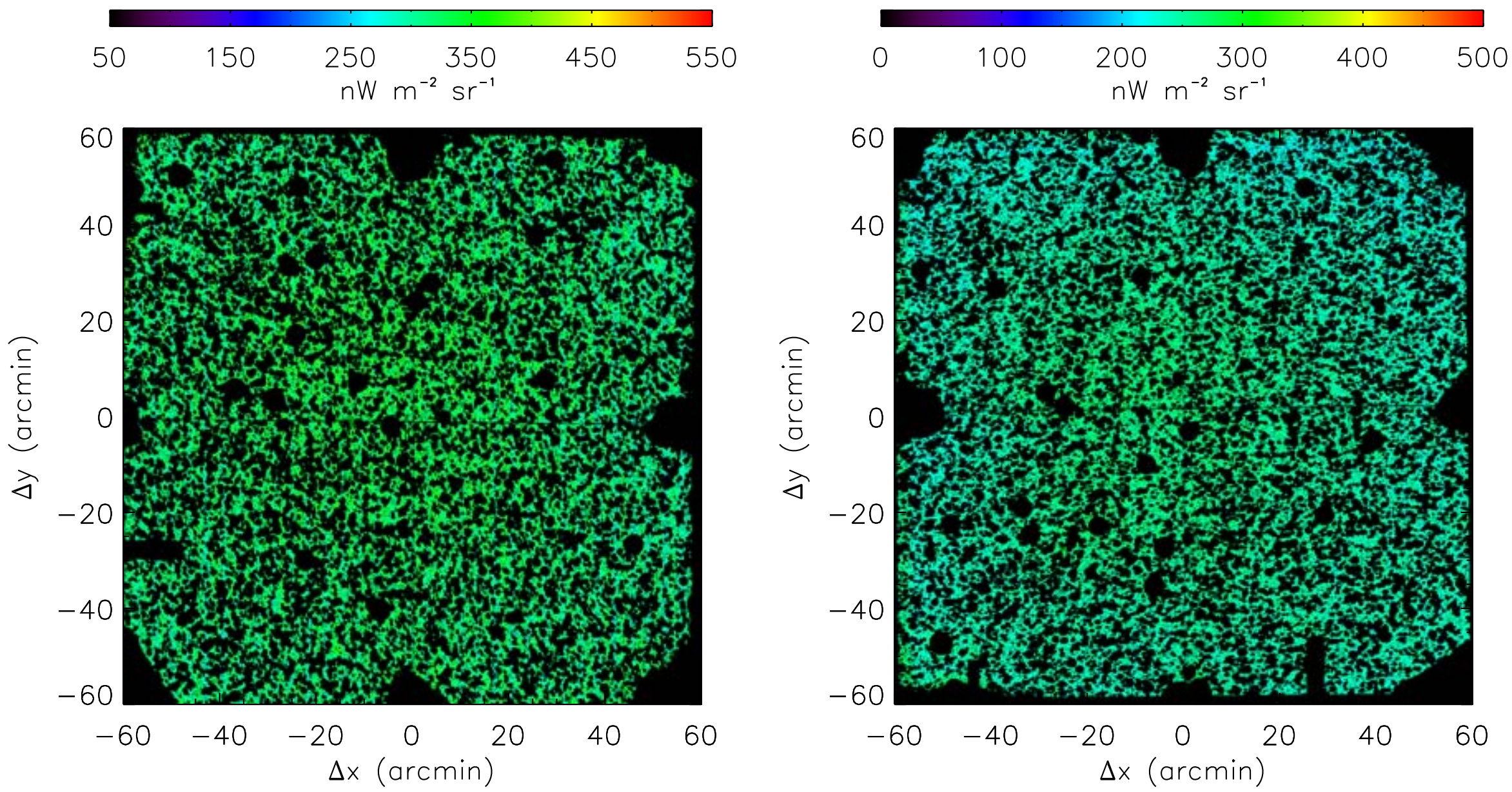

Figure S2: **Masked flight images.** We show the 1.1 $\mu$m (left hand panel) and 1.6 $\mu$m (right hand panel) images for NEP-3, masked with the algorithm derived in Section 3.2 using the 2MASS catalog. These images can be compared directly to the unmasked images shown in Figure S1.

The single-field images are potentially contaminated by flat-field error, which is common between fields (see Section 7). For this analysis, we difference two science field images to reduce this error. Examples of field-differenced images are shown in Figures S4 and S5. Because the scale of these images is large, in Figures S6 and S7 we show details of the masked images in $12' \times 12'$ regions, and in Figures S8 and S9 we show the fully masked images convolved with a Gaussian kernel with FWHM$= 7.2'$ (corresponding to $\ell = 3000$). This kernel is chosen as it is a generic functional form which is both significantly larger than the typical masking kernel, and to pick out large angular scales in the image.

# 4 Power Spectrum Estimation

We calculate spatial auto- and cross-power spectra using a version of the MASTER formalism (*33*) in which the true sky $\widetilde{C_\ell}$ is related to the measured sky power $\langle C_\ell \rangle$ by

$$\widetilde{C_\ell} = \frac{\sum_{\ell'} M^{-1}_{\ell\ell'}(\langle C_{\ell'} \rangle - N_{\ell'})}{B^2_{\ell'}}, \qquad (2)$$

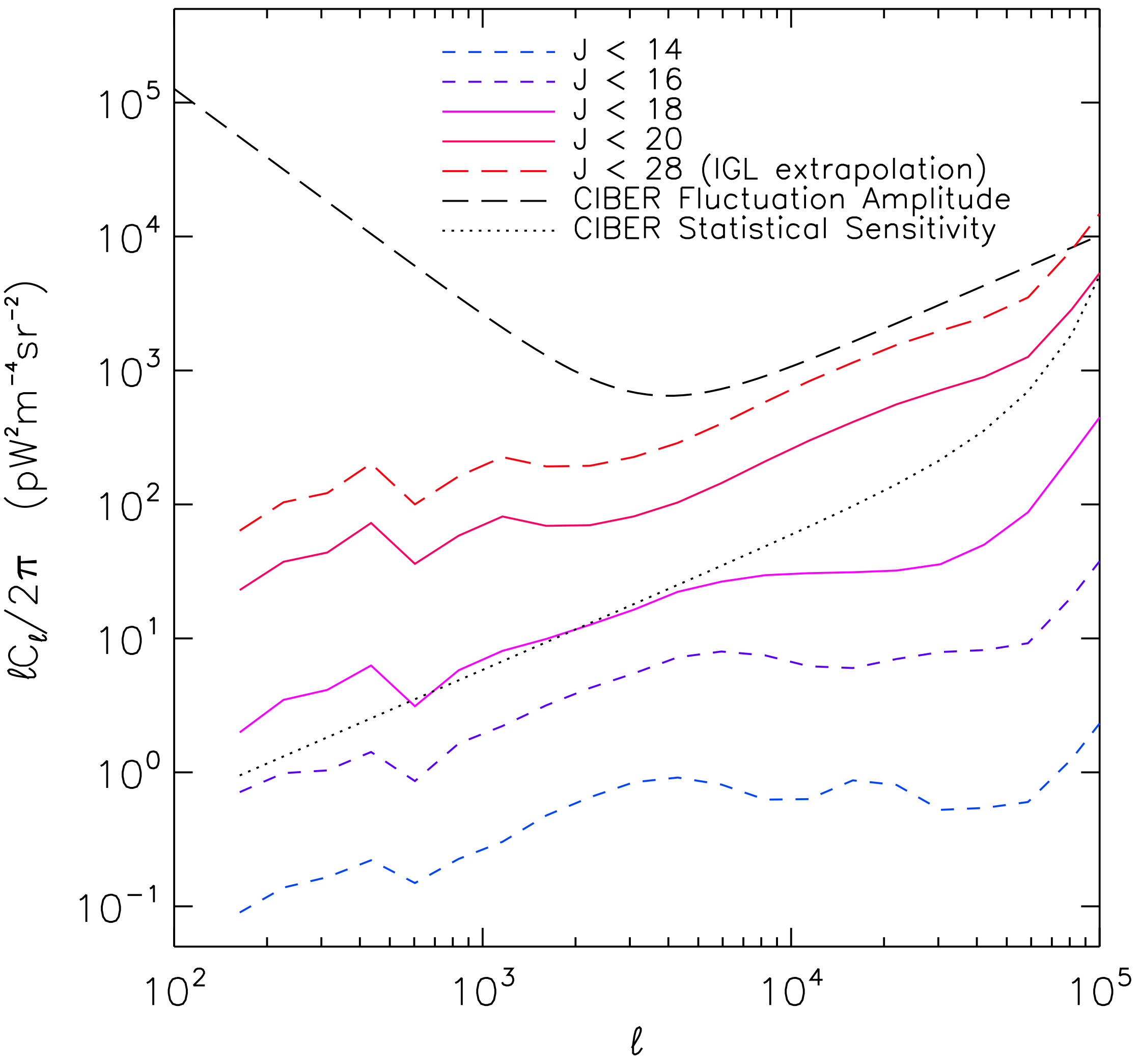

Figure S3: **Mask validation.** In this calculation, we generate simulated images from the NEWFIRM catalog including all sources to $J' < J$, convolved with the full PSF. These images are then masked, and their power spectra are calculated. For simulations including only sources brighter than the $J < 17.5$ cutoff flux (dashed lines), the residuals are from mask halos, which are below the statistical noise of the measurement (dotted line), estimated from the theoretical sensitivity of the instrument assuming uncorrelated white noise. Simulations with sources below the cutoff flux give power from galaxy clustering fluctuations that exceed the statistical sensitivity. To estimate clustering fluctuations from all galaxies, we extrapolate the $J = 20$ result by the fraction of the IGL which remains above this magnitude. The total residual galaxy fluctuations are less than the measured fluctuation power at low $\ell$. Note the $\ell C_\ell$ notation on the y-axis.

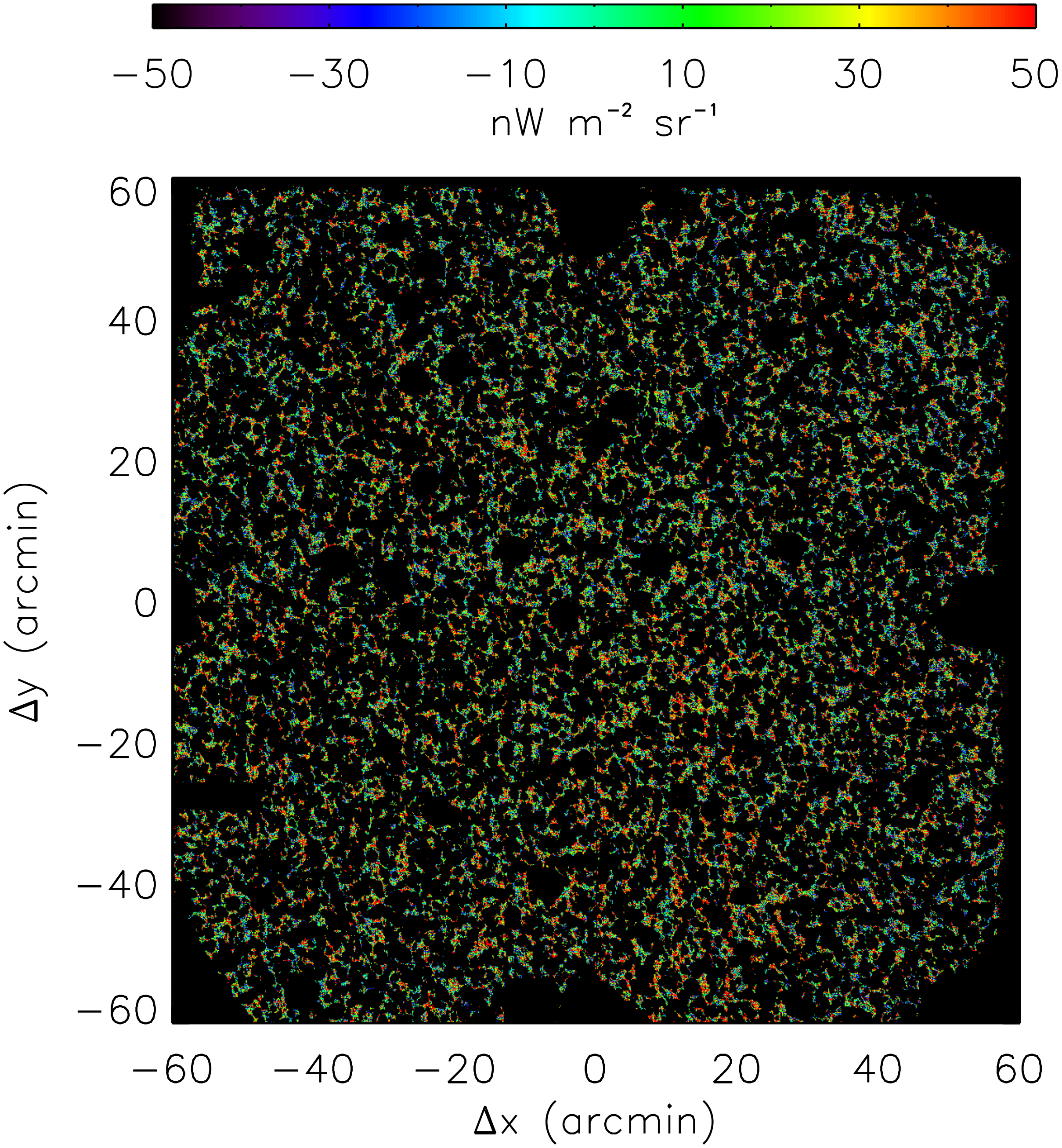

Figure S4: **Field differenced third flight** 1.1 $\mu$**m image.** The 1.1 $\mu$m image for the field combination (NEP − ELAIS-N1) in flight third flight is shown. The color scale is significantly smaller than in previous figures to highlight the faint structures in the images. In this combination, 67 % of the pixels are masked.

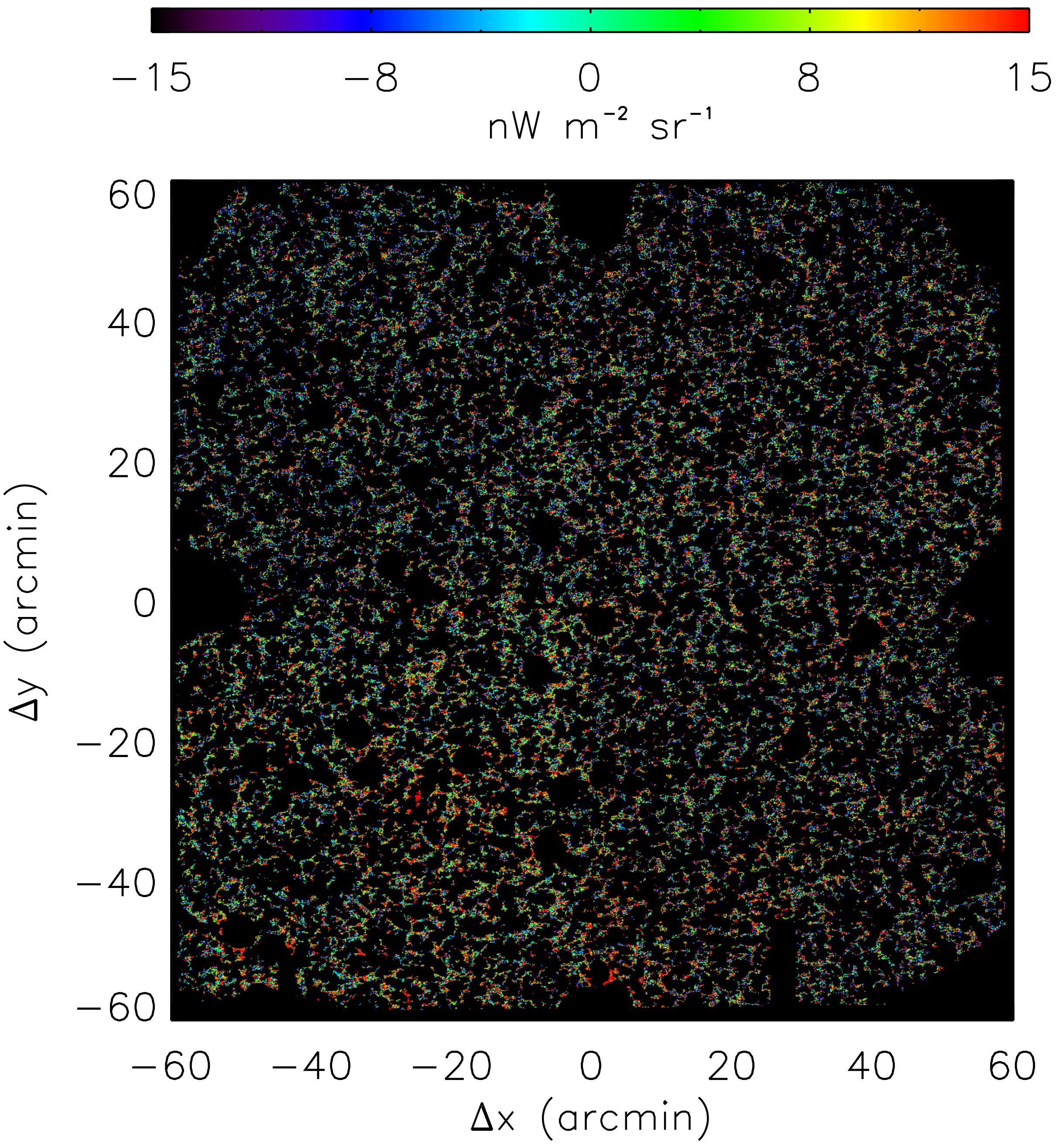

Figure S5: **Field differenced flight** $1.6\,\mu$**m image.** The $1.6\,\mu$m image for the field combination (NEP - ELAIS-N1) in flight third flight is shown, in which $65\,\%$ of the pixels are masked. As it is referenced to the instrument, this image is rotated by $90^\circ$ compared to Figure S4.

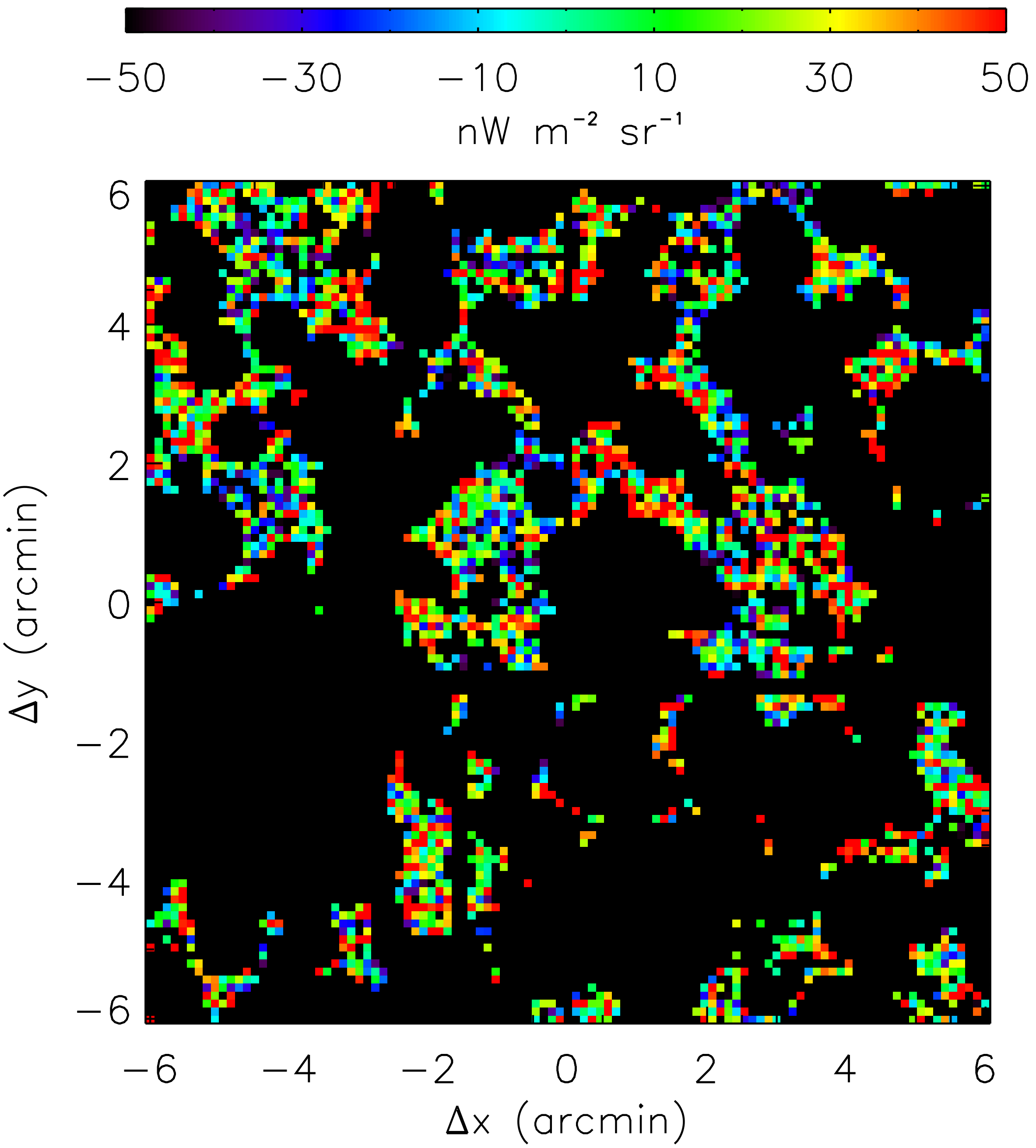

Figure S6: **Zoom of field differenced third flight** $1.1\,\mu$**m image.** The $1.1\,\mu$m image for the field combination (NEP − ELAIS-N1) in flight third flight is shown for a region $12' \times 12'$ to show the detailed properties of the mask in a small region.

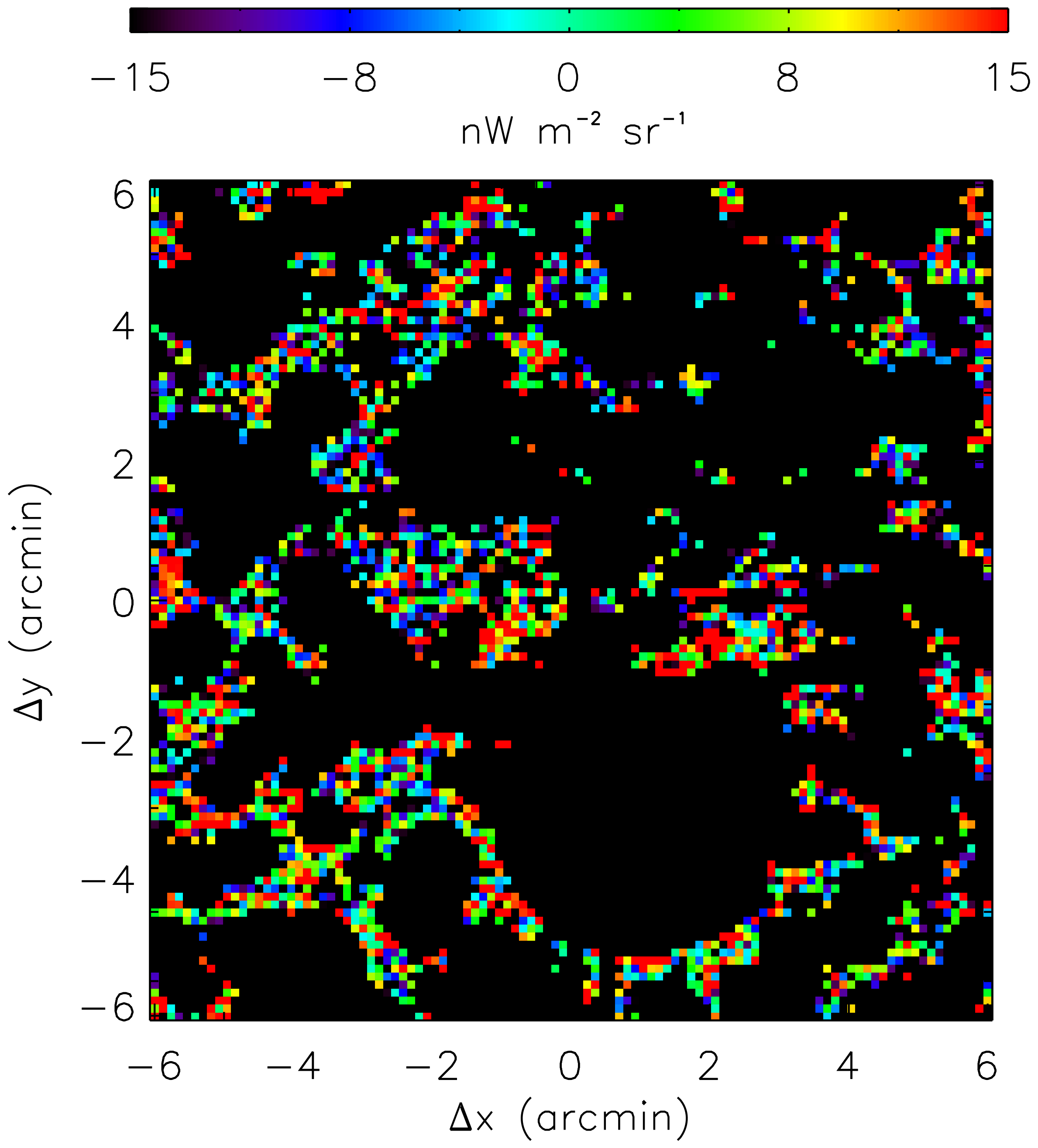

Figure S7: **Zoom of field differenced flight** $1.6\,\mu$**m image.** The $1.6\,\mu$m image for the field combination (NEP - ELAIS-N1) in flight third flight is shown, again in a $12' \times 12'$ region. As it is referenced to the instrument, this image is rotated by $90^\circ$ compared to Figure S6.

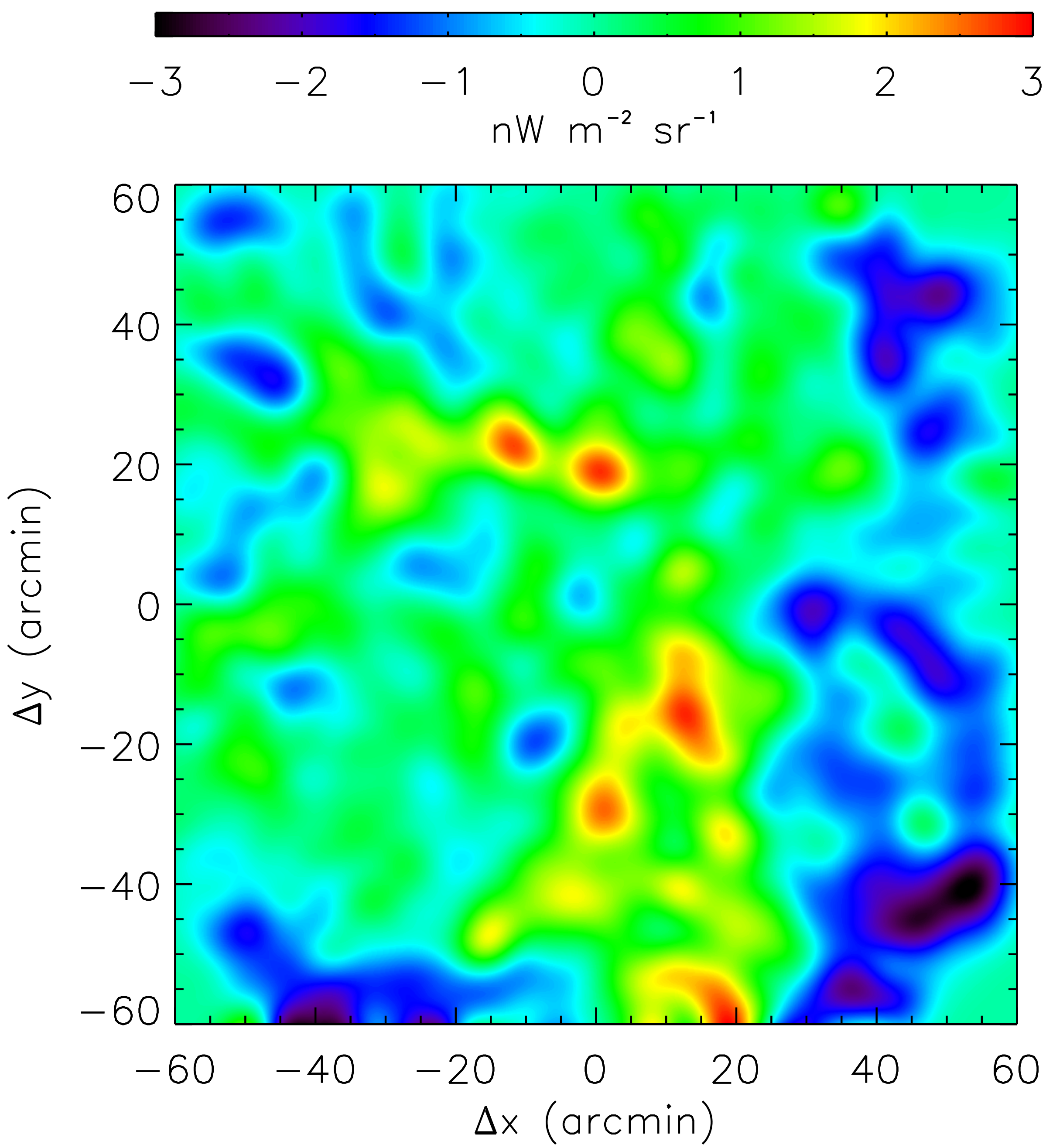

Figure S8: **Large angular scale structure in a field differenced third flight** $1.1\ \mu$**m image.** The $1.1\ \mu$m image for the field combination (NEP − ELAIS-N1) in flight third flight are shown. The image has been smoothed with a Gaussian function with FWHM$= 7.2'$ (corresponding to $\ell = 3000$) to highlight large scale structure in the image.

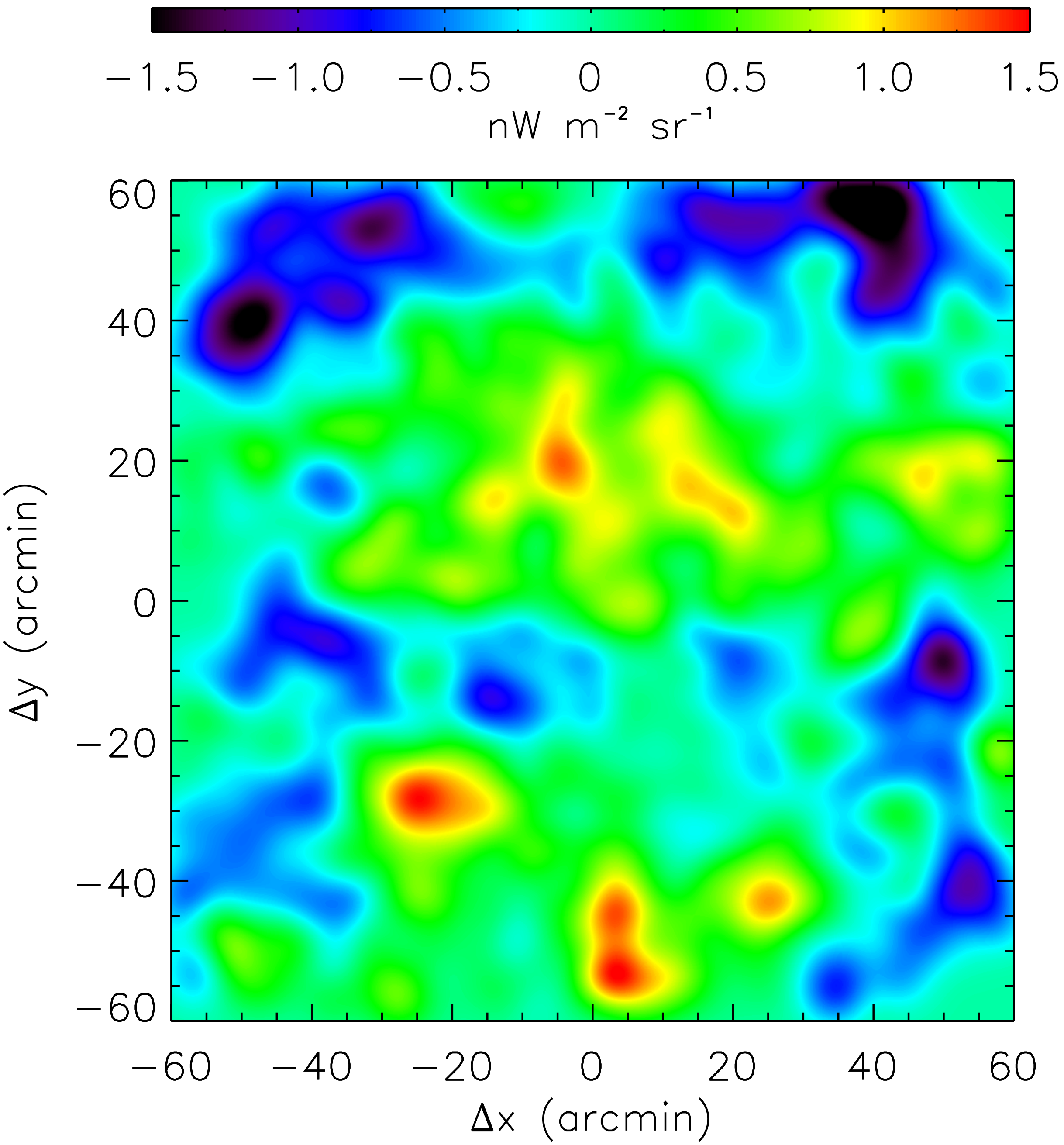

Figure S9: **Large angular scale structure in a field differenced flight** $1.6\,\mu$**m image.** The $1.6\,\mu$m image for the field combination (NEP - ELAIS-N1) in flight third flight is shown, smoothed with the same Gaussian FWHM= $7.2'$ kernel as used in Figure S8. As it is referenced to the instrument, this image is rotated by $90°$ compared to Figure S8.

where $M_{\ell\ell'}$ is the mode-mode coupling matrix, $N_\ell$ is the noise bias, and $B_\ell$ is the beam transfer function. Algorithmically, the elements required to solve this Equation are:

1. The raw power spectrum $\langle C_\ell \rangle$, derived from a sky image,
2. The noise bias $N_\ell$, computed from a noise model,
3. The beam transfer function $B_\ell$, calculated from the measured PSF, and
4. $M_{\ell\ell'}$, derived from simulations of pure tones propagated through the mask.

The raw power spectrum $\langle C_\ell \rangle$ is calculated from input images $\mathcal{M}_\mathrm{A}$ and $\mathcal{M}_\mathrm{B}$ via a two-step process. First, the Fourier transform of each map is computed via:

$$\widetilde{\mathcal{M}}(\ell_x, \ell_y) = \frac{1}{n_x n_y} \sum_{x=0}^{n_x-1} \sum_{y=0}^{n_y-1} \mathcal{M}(x,y) e^{-2\pi i(\ell_x x/n_x + \ell_y y/n_y)}, \tag{3}$$

where $n_x$ and $n_y$ are the numer of discrete points in the two dimensions of the map. We form $\langle C_\ell \rangle$ in spatial wave-mode bin $\ell$ by cross correlating $\mathcal{M}_\mathrm{A}$ and $\mathcal{M}_\mathrm{B}$, by computing the weighted mean of squared Fourier modes $\widetilde{\mathcal{M}}_\mathrm{A} \widetilde{\mathcal{M}}_\mathrm{B}^*$ and averaging over the wave-mode bin:

$$\langle C_{\ell_i} \rangle = \frac{\sum_{\ell_1}^{\ell_2} w(\ell_x, \ell_y) \widetilde{\mathcal{M}}_\mathrm{A}(\ell_x, \ell_y) \widetilde{\mathcal{M}}_\mathrm{B}^*(\ell_x, \ell_y)}{\sum_{\ell_1}^{\ell_2} w(\ell_x, \ell_y)}, \tag{4}$$

where the summand runs from $\ell_1 \leq \sqrt{\ell_x^2 + \ell_y^2} < \ell_2$, $\ell = (\ell_1 + \ell_2)/2$, $\Delta\ell = \ell_2 - \ell_1$, and $w(\ell_x, \ell_y)$ is a weighting function (see Section 5.3). This formalism computes arbitrary cross-spectra; auto-spectra are given by the case $\mathcal{M}_\mathrm{A} = \mathcal{M}_\mathrm{B}$.

## 4.1 $\mathrm{M}_{\ell\ell'}$ Correction

A mask applied to a sky map couples the underlying Fourier modes, mixing power between band-averaged spectral bins (*33*) (or "band powers"). To correct for mode mixing, we compute a mode-coupling matrix $M_{\ell\ell'}$ which describes how each individual Fourier mode is affected by the spatial mask. This is an exact computation under the hypothesis that band powers are averages in Fourier space. Following Cooray et al. (*6*), we first generate an image from a pure mode centered on band power $\ell$:

$$C_{\ell'} = \begin{cases} 1 & (\ell' = \ell) \\ 0 & (\ell' \neq \ell). \end{cases}$$

This pure-tone image is then masked with the real space source mask. We then calculate the power spectrum, which encodes how the mask mixes power in mode $\ell$ into each other mode, using the same $\ell$ binning to calculate $M_{\ell\ell'}$ as the power spectrum. This new power spectrum becomes a row in the $M_{\ell\ell'}$ matrix. We repeat this process for all $\ell$ modes in the power spectra to calculate the transformation from the unmasked power spectrum to the masked power spectrum. Because the phase information in a particular $\ell$ mode realization may interact with the phase information in the spatial mask in a non-trivial way, we perform multiple realizations of the input phase and average over the resulting ensemble for each mask. The departure from the input power spectrum is calculated in the same way as for the statistical accuracy on a Gaussian field (*34*), and has amplitude $\delta C_{\ell,M}/C_{\ell,M} = \sqrt{2/\ell\Delta\ell N}$; the mean of $N$ realizations approaches the true $C_\ell$ as $1/\sqrt{N}$. Simulations show that in 100 trials the rms deviation of the corrected band powers from their fiducial spectrum is $\delta C_\ell/C_\ell \lesssim 0.02$ for $\ell < 5000$, and decreases rapidly at higher band powers. This average error from the correction is significantly smaller than the statistical error in the measurement.

The inverse of $M_{\ell\ell'}$ transforms from the masked to the unmasked power spectrum. Importantly, this $M_{\ell\ell'}$ calculation method depends only on the input mask function and $\ell$ binning of the output spectrum. Because it is not a function of the input power, $M_{\ell\ell'}$ is not a function of the Fourier weighting. Some example $M_{\ell\ell'}{}^{-1}$ matrices are shown in Figure S10.

## 4.2 $B_\ell$ Correction

The beam correction $B_\ell$ is constructed using PSF measurements for each field separately to account for beam-smearing from pointing jitter and drift. For a single-field power spectrum, this correction is simply the power spectrum of the PSF in that field (*35*). The two dimensional $B_{\ell_x,\ell_y}$ spectra are azimuthally averaged to yield the effective beam corrections, $B_\ell$. By extension, the $B_\ell$ for difference fields is given by the sum of the two individual mapping functions in real space, which corresponds to the unit area-normalized sum of the two PSF functions (*36*). For cross-spectra, the $B_\ell$ correction is the cross-spectra of the individual field PSFs. For difference cross-spectra, it is the PSF sum and cross-spectral combination which matches the manipulations applied to $\langle C_\ell \rangle$. Finally, it is necessary to tie $B_\ell$ to the appropriate normalization in the power spectrum, by computing the window function for a delta function in the native CIBER pixels and dividing the $B_\ell$ by that normalization factor. We show an example set of typical CIBER $B_\ell$ functions in Figure S11. The uncertainty on $B_\ell$ is estimated by fitting a parametric function to the annularized beam profiles in real space, and then determining the allowed variation in the resulting $B_\ell$ as the fitting function parameters are varied by their uncertainties. To account for the covariance between parameters, we simulate an ensemble of realizations consistent with the fit uncertainties and chose the $68\,\%$ percentile as the quoted uncertainty $\delta B_\ell$. The $\delta B_\ell$ (see Figure S11) are negligible compared to the statistical and systematic uncertainties in the $\ell$ modes of interest, $\ell < 10^4$.

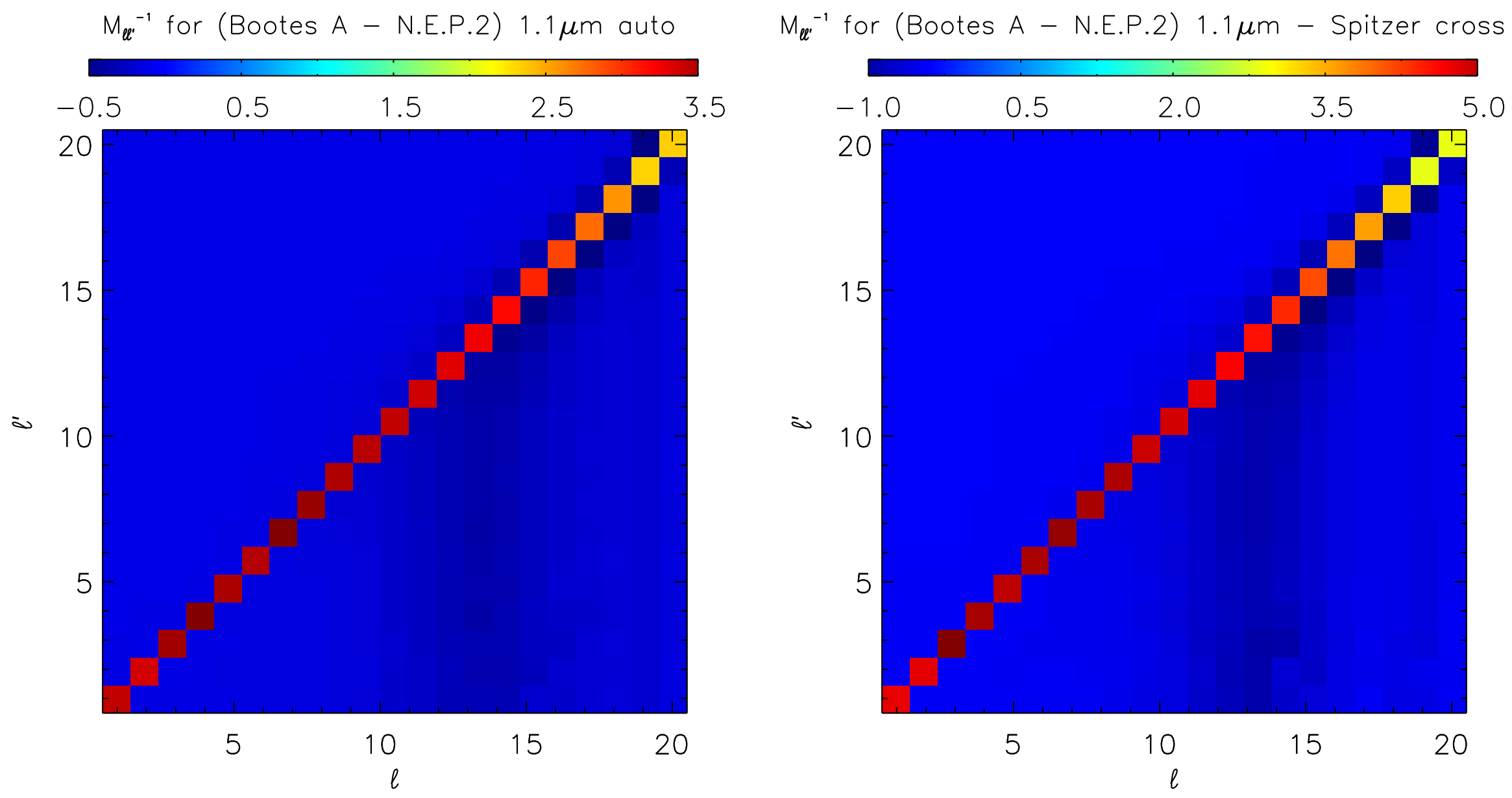

Figure S10: $\mathrm{M}_{\ell\ell'}^{-1}$ **matrices for sample fields.** We show two example $M_{\ell\ell'}{}^{-1}$ matrices, one for the auto spectrum of the $1.1\,\mu$m channel Boötes A $-$ NEP-2 (left panel) and the cross power spectrum of $1.1\,\mu$m Boötes A $-$ NEP-2 with *Spitzer* (right panel). For a given $\ell_i$, the $M_{\ell\ell'}$ operation computes the sum of the product of the uncorrected spectrum with the $i^{\mathrm{th}}$ column of this matrix to rectify the bandpower. The strongest effect of the matrix is an overall multiplicative factor on the diagonal, which corrects for loss of power due to the mask. The next largest effect is an anti-correlation from mid-$\ell$ to low-$\ell$, which is at least an order of magnitude smaller than the diagonal elements.

# 5 Noise Modeling

The analysis requires a detailed model for the noise properties of the instrument. Due to the short duration of the flights, it is difficult to accurately determine the noise properties from flight data. Instead, we construct a noise model from a large set of data from ground testing and check its veracity in flight data differences. The noise model is used for three purposes:

1. To calculate $N_\ell$, the noise bias term appearing in auto-power spectra (Eq. 2),
2. To generate noise-only simulations which are used to calculate uncertainties on the output spectra, and
3. To compute a Fourier weight function which takes into account variations in the noise properties of the instrument in Fourier space.

The noise model itself comprises two components: (1) read noise in the detector and readout, and (2) shot noise from the random variation in photons at the detector. We construct the noise

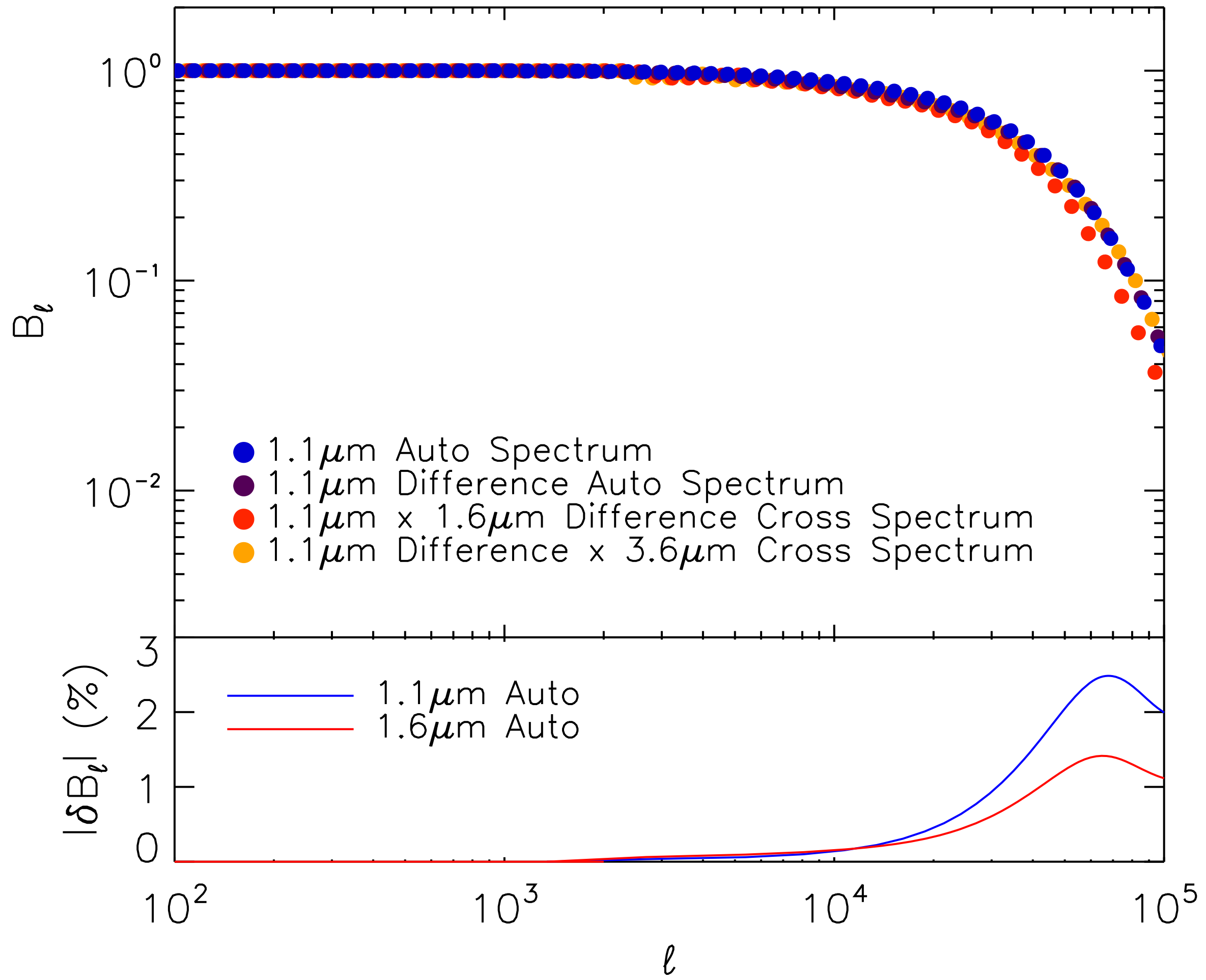

Figure S11: $B_\ell$ **functions and errors for typical CIBER cross-spectra.** The top panel shows typical $B_\ell$ functions used to correct power spectra for the effect of beam apodization using the power spectral formalism of Section 4.2. In auto-spectra, the beam is slightly more compact than in the instrument cross-spectra, since some cross-combinations include broadening in both beams. The bottom panel shows the fractional error in two example $B_\ell$, which have only small deviations at $\ell < 10^4$.

model in Fourier space.

## 5.1 Noise Model Construction

We generate the read noise model from data taken through the telemetry system while the rocket is on the launcher during pre-launch tests. These data are taken in a dark environment in which the read noise of the detectors and electronics dominates over the shot noise from photon and dark current. We take care to use data which are as similar to flight data as possible, ensuring the detector temperatures are stable and avoiding transients from the calibration lamps.

The read noise model is constructed in several steps. First, we fit slopes to each of the dark data sets using the same number of frames as each flight integration. The noise model is therefore tailored for each field, since there is a $\sim 30\,\%$ variation between the shortest and longest integrations. The dark current in these input data sets is small, $\lesssim 0.5\,e^{-}\mathrm{s}^{-1}$, so that the read noise dominates over dark shot noise by a factor of $\gtrsim 3$.

Next, we difference two successive integrations pair-wise to produce a flight-equivalent field-difference image, and then compute the two-dimensional power spectrum. This process is repeated for as many pair-wise image differences available in the dark data set, and the mean and standard deviations of the resulting ensemble of real dark realizations are calculated. Figure S12 shows the average and the standard deviation of the Fourier spectra of pair-wise frame differences used to build the noise model for both arrays. The average is smooth at large multipole numbers. However, variations are evident at low multipoles, localized in Fourier space. In particular, the power spectrum shows enhanced noise at low spatial frequencies along the read direction that is largely independent of the cross-read spatial frequency, symptomatic of correlated noise in the readout.

We use the statistics of the measured 2D Fourier structure to generate noise image realizations. For each Fourier mode $(\ell_x, \ell_y)$ we randomly generate a noise amplitude using the mean and variance of the ground-based data. The probability density which describes the distribution of power spectra in each $(\ell_x, \ell_y)$ mode is not Gaussian, but rather $\chi^2$ distributed. To make an image realization, we generate a $\chi^2$ distribution with two degrees of freedom with mean and variance equal to those of the read noise power spectra. The resulting realization is Fourier transformed into real space by taking the square root, multiplying by a unit-norm Gaussian random field, and inverse Fourier transforming the result.

The photon noise is straightforward to model, given by (*30*)

$$\sigma_\gamma^2 = \frac{6F(N^2+1)}{5T_{\mathrm{int}}(N^2-1)}, \tag{5}$$

where $F$ is the photocurrent in a pixel, $N$ are the number of frames in the fit, and $T_{\mathrm{int}}$ is the integration time. In our model, we simply take the flight photocurrent and apply Equation 5 to derive the photon noise. This noise estimate is then summed with the read noise model image to yield the total noise.

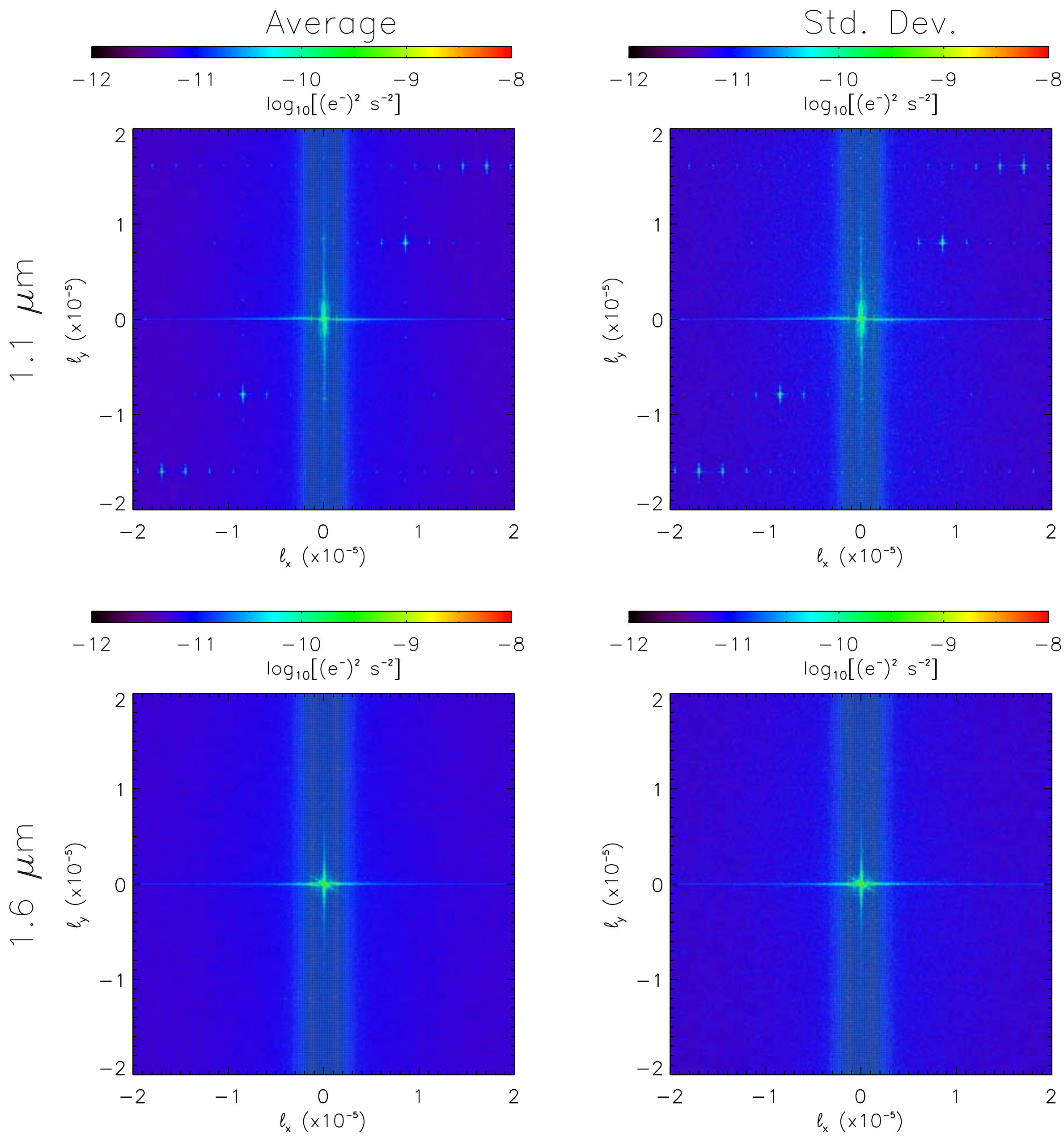

Figure S12: **Input data sets used in the read noise model construction.** The rows are for the $1.1\,\mu$m (upper) and $1.6\,\mu$m (lower) channels in the second flight for Boötes A, respectively. The columns show the average power spectrum of the input data sets (left) and the standard deviation of the input data sets (right). The structure of the noise is non-trivial in Fourier space and the noise model must adequately capture this behavior. The formulation of the noise model we use reproduces the observed mean and variance per $(\ell_x, \ell_y)$ mode for each combination of instrument, field and flight to accurately model the properties of the instrument.

An attractive feature of this model is its ability to capture the behavior of individual Fourier modes in $(\ell_x, \ell_y)$. At bandpower values $\ell < 500$ where the measured per-mode variance is large and only four $(\ell_x, \ell_y)$ bins contribute, the noise behavior is not well described by naive variance. The noise model takes the observed variance in each mode in the ensemble data set, and generates consistent realizations. This variance behavior is then propagated into all auto- and cross-combinations, where noisy modes mix into the final spectra.

## 5.2 Noise Model Validation

We compare the noise model with flight data, using a difference combination that reflects the noise properties of the instrument but contains no astrophysical signal, to verify that the model accurately represents the properties of the instrument in flight. We form half-field integrations by fitting lines to the first and second halves of an integration on a given science field, and then difference these to cancel the astrophysical signal. In order to match the noise properties of such field-halves, we generate equivalent noise model realizations and difference them in the same way as the flight data. Since the residuals from differenced point sources contribute noise power if left unmasked, we mask the half-field differences with the flight source mask and $M_{\ell\ell'}$ correct the resulting power estimate.

Using these modeled half-field images, we validate the flight noise model by:

1. Generating $M_{\rm dark}$ noise model first-half images, and an equal number of second-half integration images,

2. Differencing these to yield $M_{\rm dark}$ half-integration difference images, and forming their one-dimensional power spectra, and

3. Computing the mean and the standard deviation of the power spectra, and comparing these to the equivalent flight data difference.

We find that the modeled noise properties are statistically indistinguishable from flight noise based on the field-halves (see Section 7 and Figure S19).

## 5.3 Fourier Weight Construction

In figure S12 we show that the noise properties of the instrument as mapped into Fourier space are not white, but rather have structure associated with correlations as the detector pixels are read, caused by correlated noise and pickup in the common readout amplification chain. The correlated noise manifests as a broad hump of large variance in the scan direction (the large power $\ell_x$ modes along $\ell_y = 0$), but also as a strong low-wavenumber noise along the $\ell_y$ axis. The fact that the noise power is not azimuthally symmetric in Fourier space leaves room for optimizing the estimation of the sky power spectrum (*37*). In this analysis, these noisy modes are down weighted by applying a Fourier weight function $w(\ell_x, \ell_y)$.

To derive $w(\ell_x, \ell_y)$, we simply use the noise model appropriate for a given field combination to compute the standard deviation of the Fourier modes over many realizations. The inverse of this standard deviation is then used as the weight function in Equation 4. Because the weighted mean $\langle C_\ell \rangle$ is an unbiased estimator for the true mean, applying this weight function to power spectra does not affect the $M_{\ell\ell'}$ correction. Both the $\langle C_\ell \rangle$ and $N_\ell$ terms in Equation 2 are weighted with $w(\ell_x, \ell_y)$, although the estimator is designed to minimize $N_\ell$.

## 5.4 Power Spectrum Noise Estimates

The statistical error on an auto-power bandpower in a power spectrum is given by:

$$\delta C_{\ell_i}^2 = \frac{2}{n_{\ell_i}} (C_\ell^{\mathrm{A}} + N_\ell^{\mathrm{A}})^2, \tag{6}$$

where $\delta C_{\ell_i}$ is the cross-bandpower uncertainty in bin $\ell_i$, $C_\ell^{\mathrm{A}}$ is the power spectrum of quantity A, and $N_\ell^{\mathrm{A}}$ is the noise power spectrum. The term $n_\ell = f_{\mathrm{sky}}(2\ell+1)\Delta\ell$, where $\Delta\ell$ is the $\ell$ bin width and $f_{\mathrm{sky}}$ is the fraction of the sky observed, accounts for the number of independent Fourier modes contributing to a particular band power. For the cross-power, the equation becomes:

$$\delta C_{\ell_i}^2 = \frac{1}{n_{\ell_i}} \left[ (C_\ell^{\mathrm{A}} + N_\ell^{\mathrm{A}})(C_\ell^{\mathrm{B}} + N_\ell^{\mathrm{B}}) + (C_\ell^{\mathrm{A}\times\mathrm{B}})^2 \right], \tag{7}$$

where $C_\ell^{\mathrm{B}}$ and $N_\ell^{\mathrm{B}}$ are the auto-power and auto-noise power spectra of quantity B, and $C_\ell^{\mathrm{A}\times\mathrm{B}}$ is the cross-power spectrum (see e.g. (*38*)).

We compute the products in Equations 6 and 7 as cross-powers (squared) in two-dimensional space, allowing statistical noise correlations to constructively and destructively interfere, which is a more accurate representation of the system than the one-dimensional calculation using azimuthally symmetrized band powers. In the auto-power case, the band power uncertainties are given by:

$$\delta C_{\ell_i}^2 = \frac{2}{n_{\ell_i}} \left[ \frac{\sum_{\ell_1}^{\ell_2} w(\ell_x, \ell_y) \widetilde{\mathcal{M}_{\mathrm{A}}}(\ell_x, \ell_y) \widetilde{\mathcal{M}_{\mathrm{A}}}^*(\ell_x, \ell_y)}{\sum_{\ell_1}^{\ell_2} w(\ell_x, \ell_y)} + \frac{\sum_{\ell_1}^{\ell_2} w(\ell_x, \ell_y) \widetilde{\mathcal{N}_{\mathrm{A}}}(\ell_x, \ell_y) \widetilde{\mathcal{M}_{\mathrm{A}}}^*(\ell_x, \ell_y)}{\sum_{\ell_1}^{\ell_2} w(\ell_x, \ell_y)} + \right.$$
$$\left. \frac{\sum_{\ell_1}^{\ell_2} w(\ell_x, \ell_y) \widetilde{\mathcal{M}_{\mathrm{A}}}(\ell_x, \ell_y) \widetilde{\mathcal{N}_{\mathrm{A}}}^*(\ell_x, \ell_y)}{\sum_{\ell_1}^{\ell_2} w(\ell_x, \ell_y)} + \frac{\sum_{\ell_1}^{\ell_2} w(\ell_x, \ell_y) \widetilde{\mathcal{N}_{\mathrm{A}}}(\ell_x, \ell_y) \widetilde{\mathcal{N}_{\mathrm{A}}}^*(\ell_x, \ell_y)}{\sum_{\ell_1}^{\ell_2} w(\ell_x, \ell_y)} \right], \tag{8}$$

where the variables are defined as in Equation 4 and $\widetilde{\mathcal{N}}$ is the noise realization equivalent of $\widetilde{\mathcal{M}}$. The same Fourier weights are used both in the data and noise estimates for a given data combination. The analogous relation for the cross-power uncertainties is used in that case.

We use the noise model to derive the statistical uncertainties on the output power spectra in the following way. For each real space image combination $\mathcal{M}_{\mathrm{A}}$ and $\mathcal{M}_{\mathrm{B}}$, we compute $n_{\mathrm{sim}}$ noise realizations of those images. These noise model images are masked as in the real data, and their two-dimensional power spectra are computed. To compute the noise bias term, the resulting power spectrum realizations are averaged over $n_{\rm sim}$ realizations. We compute Equation 8 for each realization, and take the standard deviation over $n_{\rm sim} = 100$ realizations to determine $\delta C_\ell$. This procedure naturally captures the underlying variance of the system, and any interactions of the noise model with spatial structure from the sky.

There is an additional complication in estimating the sample variance terms going as $(C_\ell^{\rm A} C_\ell^{\rm B})$ or $C_\ell^{\rm A\times B}$ appearing in Equations 6 and 7. We assume that the measured band power values are drawn from a statistical process with an underlying mean following a smooth curve as a function of $\ell$. Sample variance has the effect of varying the measured bandpowers about that curve in proportion to the inverse of the number of contributing Fourier bins. For large $\ell$ modes this variance is small due to the large number of modes contributing, but at small $\ell$ where the number of modes is small, the sample variance is large. It is therefore incorrect to use the measured band power to estimate the size of the sample variance on a given $C_\ell$, since the measurement already includes the effect of sample variance (similar to the "cosmic bias" problem (*39*)). To correct for this, we perform simulations of our analysis pipeline on input spectra and measure the statistics of the data processing on random draws as a function of the input power. We perform this simulation for each true measurement case (*i.e.* including the particular mask, Fourier weight, etc.), computing the underlying probability distribution sample variance of each band power as a function of its measured value. These are estimates of the true sample variance contribution to the power spectra.

We perform an additional correction on the sample variance due to the Fourier weighting. The term $n_\ell$ appearing in accounts for the number of Fourier modes which are summed to calculate each band power average. With non-uniform Fourier weighting, this mode term must be modified to reflect the effective number of modes contributing to each band power, calculated by computing the sum of the Fourier weights. This correction typically increases the sample variance by 1.3 at the lowest bandpowers, and has little effect at high $\ell$.

To verify the CIBER data analysis pipeline and simulation suite is working properly, we developed a simulator which models the various astronomical foregrounds based on empirical measurements, as well as the low-$\ell$ fluctuation power. Noise realizations are added to this simulation, and the simulated data are passed through the data analysis pipeline. We find that the data analysis successfully reproduces the input power spectrum with no bias, as shown in Figure S13, independent of the shape of the input power.

# 6 Power Spectrum Results

Using the computational machinery presented in the preceding sections, we calculate the auto- and cross-power spectra of the CIBER flight data, and the cross-spectra of CIBER and *Spitzer* measurements of the NDWFS field. Due to flat field errors, all auto- and cross-spectra are based on imager field differences (see Section 7).

Figure S14 shows the auto-power spectra of the $1.1\,\mu$m and $1.6\,\mu$m channels in four field difference combinations: Boötes A $-$ NEP for flight 2, and Boötes B $-$ NEP, Boötes B $-$ ELAIS-N1, and NEP $-$ ELAIS-N1 for flight 3. Three combinations of third flight data show the consistency of the spectra. The cross-power between CIBER $1.1\,\mu$m and $1.6\,\mu$m in each field combination indicates that the fluctuation power is almost entirely correlated between the bands. In Figure S14 we also plot curves for galaxy and star fluctuations, as well as the sum of all instrumental systematics derived in Section 7.

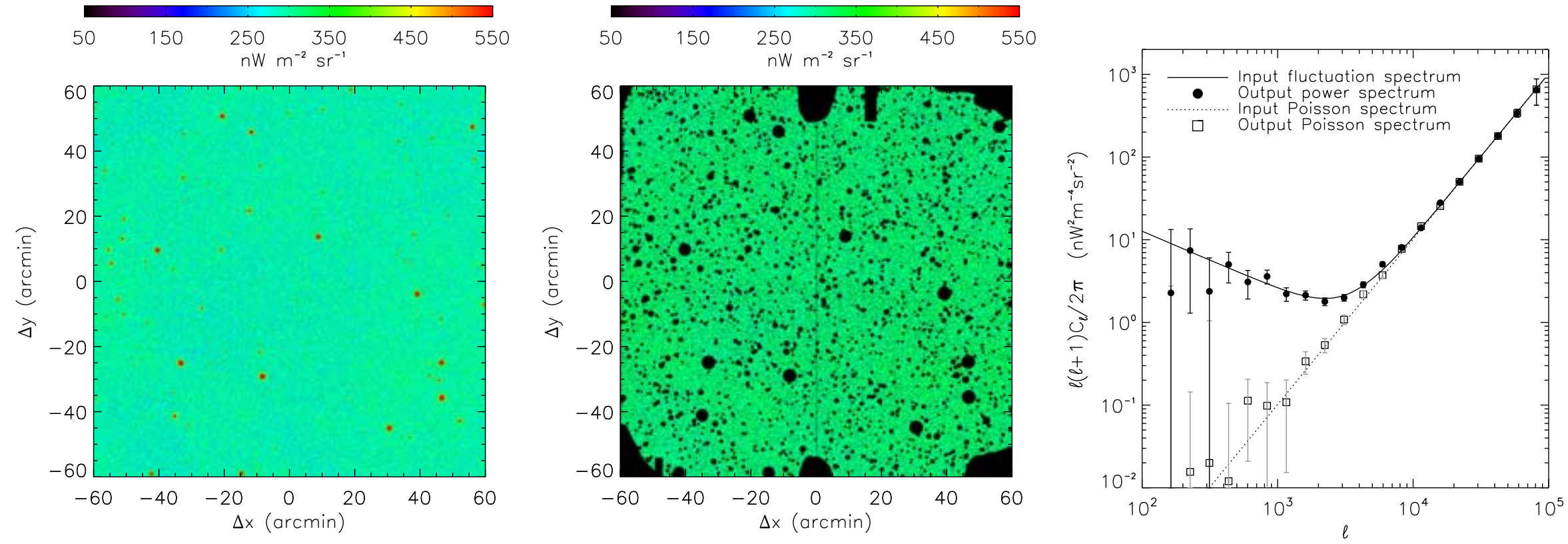

Figure S13: **Simulation verifying the pipeline estimate of input power spectra.** In this figure we show a simulated data image generated from empirical models for the astrophysical emission and noise fluctuations (left panel), the image masked using the same algorithm as for flight data (central panel), and a comparison of the power spectra processed through the pipeline to the fiducial input (right panel). The solid curve corresponds to a power spectrum similar to that measured by CIBER. The image panels show this fluctuation power spectrum summed with empirical models for stars and galaxies, plus a noise model realization. After passing these images through the data analysis pipeline (circles in right panel), we reproduce the input power spectrum with no bias in a single realization. The plotted error bars show the total uncertainty of the measurement from the combination of statistical noise and sample variance. As another test case, we propagate a simulation *without* a low-$\ell$ fluctuation component (i.e. the sum of Poisson distributed stars and galaxies, and instrument noise) through the pipeline (dotted line, right panel). Again, the band powers measured after processing match the input spectrum.

The power spectra in Figure S14 exhibit a signal level of $\sim 1.5\,$nW m$^{-2}$ sr$^{-1}$ in both CIBER bands at $\ell \lesssim 5000$, after subtracting fluctuations from residual galaxies and stars (see Section 9). The fluctuations are clearly evident in all field-difference combinations in both flights, though the errors vary depending on the combination. Also note that field-differenced cross-spectra combinations discussed in Section 7 and shown in Figure S22 are consistent with the auto-spectra.

We may combine spectra to reduce statistical uncertainty and sample variance. The field combinations shown in Figure S14 are not all independent, and share the NEP field in common between flights and field combinations within a flight. Because the NEP field has an $88\%$ overlap in area between flights, we exclude the third flight combinations using NEP from the average. We calculate the mean of the second flight Boötes A $-$ NEP and third flight Boötes B $-$ ELAIS-N1 measurements, which are fully independent with little common systematic uncertainty, as the result presented in Figure 1 of the main paper, and Figure S16 below.

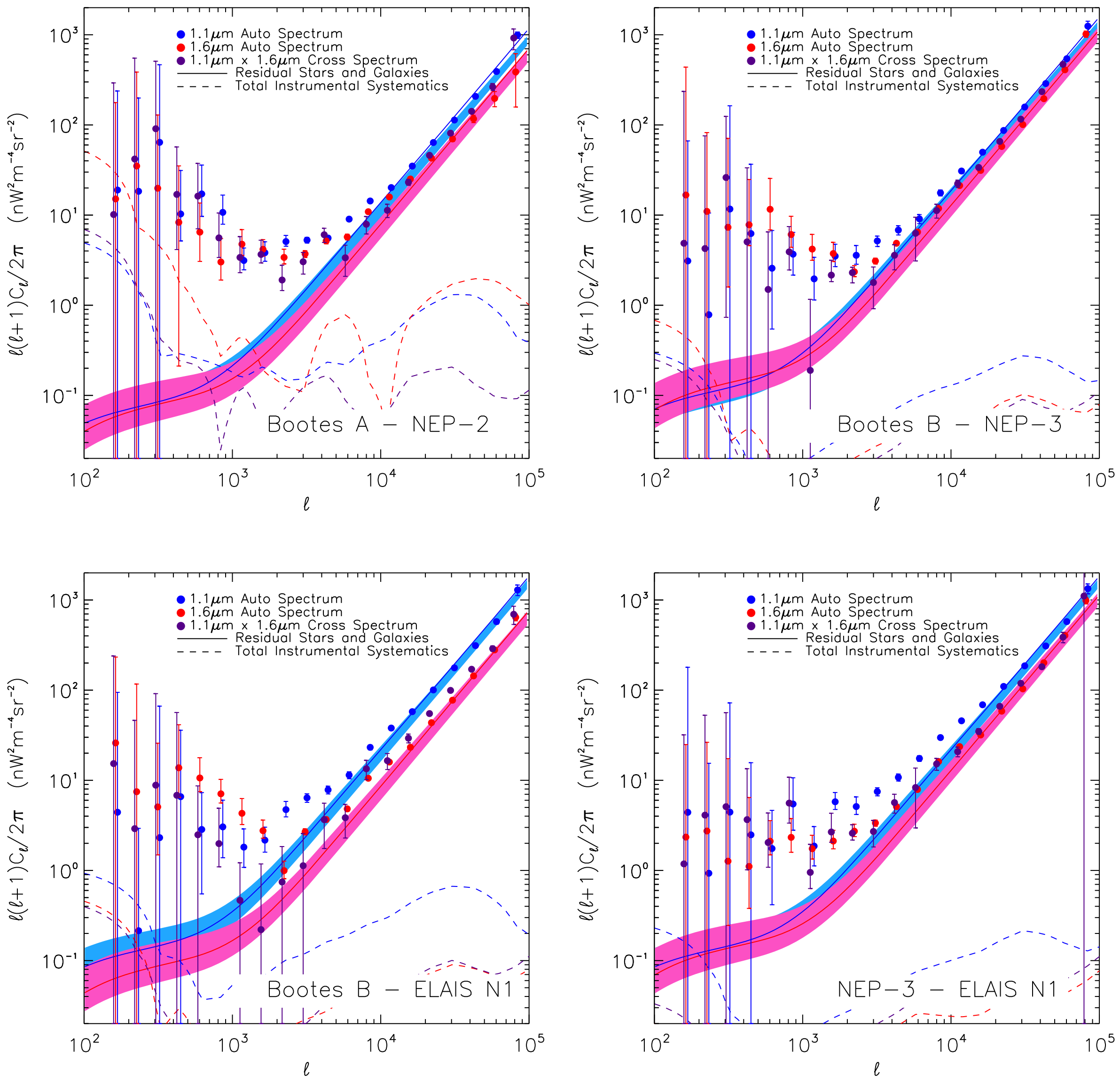

Figure S14: **The CIBER measured power spectra.** The $1.1\,\mu$m$\times1.1\,\mu$m and $1.6\,\mu$m$\times1.6\,\mu$m auto-power spectra, and the $1.1\,\mu$m$\times1.6\,\mu$m cross- power spectrum (blue, red, and purple points, respectively) for each independent field difference in the CIBER second and third flights are shown. We show a model for the contribution from galaxies below our masking threshold (*16*) as the colored bands, and the best-estimate of galaxy and star fluctuations as the solid curves. Finally, the dashed lines show the sum of the estimated instrumental systematic uncertainties from flat-field error, residual airglow, and mask halos following the color convention for each band combination.

To calculate the combination, we determine those band powers for which cosmic variance dominates over statistical error. For band powers where statistical uncertainties dominate, we combine the data based on statistical weights. For band powers where cosmic variance dominates, we use uniform weighting appropriate to sample variance. The uncertainties are computed by first computing the combined statistical uncertainties of each band power, and then combining with a new estimate for the cosmic variance computed from the values of the new combined band powers using the formalism presented in Section 5.4.

## 6.1 *Spitzer* Cross-Spectra

The fluctuation power measured by CIBER is similar to previous measurements with *Spitzer* (*8*,*6*). We cross-correlate SDWFS maps with CIBER images to probe for a correlated signal. Astrophysically, the amplitude of the correlation depends on both the commonality of the fluctuating sources.

We use the $3.6\,\mu$m full co-addition of the SDWFS mosaicked images. All of the ancillary data analysis products, including the source mask, beam correction function, mosaicking transfer function and $M_{\ell\ell'}$ correction are taken from Ref. *6*. The only difference between the original presentation and the points we plot here is that we estimate the *Spitzer* noise using a data-based noise model constructed by generating gaussian realizations consistent with *Spitzer* field differences. We first create two-dimensional Fourier images by subtracting (epoch1-epoch2) and (epoch3-epoch4). A template two-dimensional Fourier image is produced by taking the mean of these two Fourier planes, and noise model realizations are drawn from it analogously to the CIBER noise model. This method more naturally captures the low $\ell$ discrepancy in the *Spitzer* epoch difference auto power spectrum than in the Ref. *6* analysis since it matches both the expected Gaussian noise at high $\ell$ and the auto-correlations of the epoch differences at low $\ell$. The predicted uncertainties in the new noise model are very similar to those of Ref. *6*, though with $\sim 10\%$ larger noise for $\ell > 2000$.

We show the cross-power spectra of the CIBER (Boötes A $-$ NEP) and (Boötes B $-$ ELAIS-N1) difference images with the SDWFS $3.6\,\mu$m image for both $1.1\,\mu$m $\times$ $3.6\,\mu$m and $1.6\,\mu$m $\times$ $3.6\,\mu$m in Figure S15. Because the CIBER fields are differenced to reduce the effect of the flat-field error, there is additional variance introduced by the non-Boötes fields. In all of the band powers where the SDWFS data have $> 1\sigma$ detections, we detect significant cross-correlation between CIBER and *Spitzer*. The positive cross-correlation strongly indicates that the source of the fluctuation power is common. These measurements are averaged in the same way as described in the previous section for the CIBER-only combination to produce the Figure 1 in the main paper.

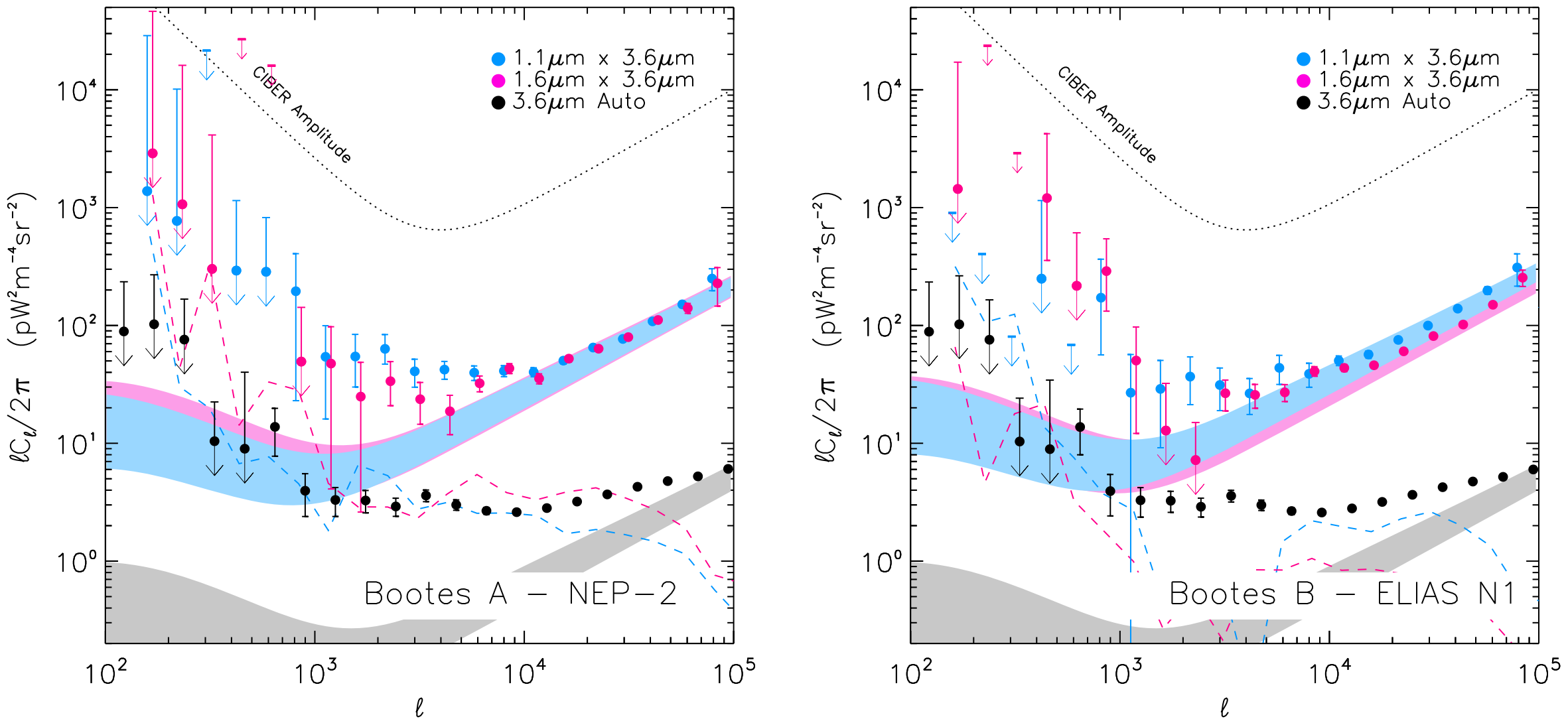

Figure S15: **The CIBER – *Spitzer* cross-spectra.** The *Spitzer* 3.6 $\mu$m auto-spectrum and the CIBER – *Spitzer* 3.6 $\mu$m cross-spectra in both Boötes A – NEP and Boötes B – ELAIS-N1. Error bars which pass through zero are plotted as upper limits, and band powers with negative expectation values are shown as barred $2\sigma$ upper limits. The level of the CIBER auto-spectrum fluctuations is indicated by the dotted line. The *Spitzer* auto-spectrum is computed for the entire SDWFS field, and extends to lower $\ell$ modes than the cross-spectra available with CIBER. The solid colored regions show foreground galaxy models (*16*) for the $1.1\ \mu$m$\times 3.6\ \mu$m, $1.6\ \mu$m$\times 3.6\ \mu$m, and $3.6\ \mu$m$\times 3.6\ \mu$m combinations. Finally, we indicate the total CIBER instrumental systematic error estimates as dashed lines for each cross-spectrum. Note that *Spitzer* only detects fluctuations for $\ell \geq 800$, and the clustering power departs from galaxy models at $\ell \sim 10^5$ due to the greater point source masking depth in *Spitzer*. The fluctuations clearly depart from galaxy models at $\ell \sim 10^4$ in the cross-correlation, at higher $\ell$ than in the auto-spectrum, also due to the greater *Spitzer* masking depth. Note the $\ell C_\ell/2\pi$ convention in these plots.

## 6.2 Correlation Coefficients

In addition to the cross power spectra, the cross correlation coefficient $r_\ell$ can be computed as:

$$r_\ell = \frac{C_\ell^{\mathrm{A}\times\mathrm{B}}}{\sqrt{C_\ell^{\mathrm{A}\times\mathrm{A}} C_\ell^{\mathrm{B}\times\mathrm{B}}}}, \tag{9}$$

where all combinations of $1.1, 1.6$ and $3.6\,\mu$m can be computed. These cross correlations have uncertainties given by:

$$\delta r_\ell = \sqrt{r_\ell^2 \left[\delta C_\ell^{\mathrm{A}\times\mathrm{B}}/C_\ell^{\mathrm{A}\times\mathrm{B}} - \delta C_\ell^{\mathrm{A}\times\mathrm{A}}/2C_\ell^{\mathrm{A}\times\mathrm{A}} - \delta C_\ell^{\mathrm{B}\times\mathrm{B}}/2C_\ell^{\mathrm{B}\times\mathrm{B}}\right]^2}, \tag{10}$$

where the $\delta C_\ell$ include sample variance terms. To capture the behavior of the large angle component we compute the mean $r$ over all points at $\ell < 2000$, and use only points with $\delta r < 1$ so that the approximation $\delta r \approx \sigma_r$ is applicable. The mean over $\ell$ of the correlation coefficients $\langle r \rangle$ for $1.1 \times 1.6\,\mu$m, $1.1 \times 3.6\,\mu$m, and $1.6 \times 3.6\,\mu$m are $0.76 \pm 0.10$, $0.55 \pm 0.14$ and $0.31 \pm 0.14$, respectively, and we do not detect a significant change in $r$ with $\ell$. The systematic uncertainties are calculated by computing the difference in the computed values of $\langle r \rangle$ when the full systematic uncertainty shown in Figure S17 is subtracted and are $\delta r_{\mathrm{sys}} < 0.05$ in all three cases. The equivalent correlation values for the Helgason model (*16*) are $0.98$, $0.97$, and $0.93$, respectively, evidence that the large angle correlations we measure are different than those expected for galaxies. More precise determinations of these quantities requires a larger data set with greater statistical information.

# 7 Instrumental Systematic Uncertainties

We estimate the maximum possible systematic error from residual airglow emission, residual flat field uncertainties, a variety of errors due to incomplete masking, and uncertainties arising from the mode-coupling correction.

## 7.1 Airglow

We estimate the contamination from residual airglow by extrapolating the time-dependent emission observed at the beginning and end of each flight to our science fields. We then compute the auto-power spectrum of the resulting airglow image based on measurements of the spatial morphology observed early and late in the flight, typically corresponding to altitudes below $250\,$km (see Table 1). We scale this spectrum to estimate the worst-case contamination from airglow fluctuations.

We estimate the spatial structure of the airglow emission from line fits to raw data. Instead of a simple linear model, we model the time series for a given pixel as:

$$Q_{ij}(t) = A_{0,ij} + A_{1,ij} t + A_{2,ij} \Theta(t), \tag{11}$$

where the charge $Q$ in pixel $i, j$ at time $t$ is modeled as a linear combination of an offset $A_0$, a constant current $A_1$, and a term $A_2$ times an airglow template $\Theta(t)$. The airglow template $\Theta(t)$ is derived using a parametric polynomial fit to the average time behavior of the data, using a Markov Chain Monte Carlo (MCMC) calculation to maximize the likelihood of the polynomial coefficients under the constraint that the airglow behavior is smooth over a flight. These images are obtained for the ascending and descending airglow fields listed in Table 1 where airglow is bright. The airglow images do not quite match the array flat field matrix, possibly due to the differing electromagnetic spectra of airglow and ZL emission (*40*).

We scale the airglow image $A_2$ to the science fields by estimating the time dependence of airglow emission throughout the flight. We fit an exponential model to the mean airglow time stream over the airglow field, and extrapolate to the science fields. The ratio of the amplitude of the exponential term in the science field to the amplitude in the airglow field gives the airglow suppression factor in that field; the resulting suppression factors are listed in Table 1. Finally, we estimate the airglow contamination by computing the auto-spectrum of the image:

$$A_{\text{airglow}} = S_{\text{airglow}} \cdot A_2, \tag{12}$$

where $A_{\text{airglow}}$ is the estimate for the image of the airglow emission in the science field of interest. These estimates are masked with the source mask for the airglow field to remove potential contamination from bright sources. The auto-power spectrum of $A_{\text{airglow}}$ yields the airglow contamination appropriate for a single field auto-spectrum, while the difference of two such estimates is used for difference-field spectra. The airglow contamination estimated for each combination of science difference fields is negligible, as shown in Figure S16.

We estimate the airglow contamination in the CIBER – *Spitzer* cross-spectra by performing the same analysis as above, but using the interpolated *Spitzer* image as one of the members of the cross-spectrum. The resulting correlation is consistent with noise, leading us to conclude that randomly correlated airglow contamination is significantly below the noise floor in the CIBER – *Spitzer* cross-spectra.

## 7.2 Flat Field Errors

We bound the amplitude of the flat-field errors based on the difference between the laboratory-based flat-field estimate $F_{\text{lab}}$ and the flight-based flat-field estimate $F_{\text{flight}}$. We construct $F_{\text{flight}}$ for each science field by taking the combination of the remaining $T_{\text{int}} > 30$s non-airglow fields. Each of these is masked as discussed in Section 3.2, and scaled to unity median. To compute $F_{\text{flight}}$, we calculate the mean of the unmasked pixels in the set of field measurements. Pixels which have no unmasked measurements in the union of fields are removed from analysis. This image is then taken to be the best estimate for the flight flat-field response. We note $F_{\text{flight}}$ overestimates flat-field errors due to astrophysical fluctuations in the flight images, due to the paucity of independent fields.

To construct the flat-field systematic error estimate $\delta I$, we first compute the difference between consistently normalized $F_{\text{lab}}$ and $F_{\text{flight}}$, scaled to the surface brightness of the science field of interest $\delta S$,

$$\delta I = \delta S(F_{\rm lab} - F_{\rm flight}). \tag{13}$$

As our analysis is based on field differences, $\delta S$ is taken to be the difference in the brightness of the two science fields. We then compute the power spectra of these quantities in the usual way. These estimates indicate the worst-case contamination from the flat field error; Figure S16 shows auto- and cross-spectra systematic uncertainty appropriate to the data combinations. Note that the lab flat-field measurements are noisy for the second flight, which significantly increases the error estimate for that flight. The noise arises from the relatively high background in the second flight calibration measurements, which adds photon noise and reduces the useful linear-response integration time. The error estimate is dominated by the lowest modes in the residual flat-field image, leading to a steep power spectrum.

Since we do not have a *Spitzer* flat-field error estimate, we calculate the flat-field uncertainty in the CIBER – *Spitzer* cross-spectra by using the *Spitzer* SDWFS image itself. This estimate is equivalent to asserting that the entire *Spitzer* power spectrum arises from flat-field errors, which is clearly an overestimate, but indicates the effect of the CIBER flat-field errors on the cross-spectra are acceptably small, as shown in Figure S15.

## 7.3 Masking Residuals

The source masking algorithm removes extended power from sources as a function of their flux. We estimate the contamination power from mask halos by simulating images of the sky with sources equal to or brighter than the nominal masking flux cutoff and computing the power spectrum of the masked simulation. For this calculation we use the NEWFIRM catalog at $J$-band and extrapolate to $1.1\,\mu$m and $1.6\,\mu$m using simple color scalings. As the NEWFIRM catalogs preserve spatial correlations between sources, this estimate naturally encodes the large scale fluctuations from galaxy clustering. We convolve these simulated images with the imager PSF, which is characterized down to a level of $10^{-5}$ from a combination of faint and bright sources (*14*). We mask the images using the real CIBER source mask for each field, and compute the $M_{\ell\ell'}$ corrected power spectra. This calculation yields the residual source structure estimates shown in Figure S16, which are similar to the $J = 18$ curve in Figure S3.

## 7.4 M$_{\ell\ell'}$ Errors

We simulate the statistical effect of the $M_{\ell\ell'}$ correction on a variety of input power spectral shapes to characterize the uncertainties arising from $M_{\ell\ell'}$ errors. We find no net bias is induced for any input spectrum $C_\ell \propto \ell^n$ with $-3 < n < 3$. We also simulate spectra with shapes similar to those of our measurements, which have a significant amount of low-$\ell$ power, and again find no bias. We find that for $n_{\rm sim} \gtrsim 100$ the additional variance from the corrections is significantly smaller than the statistical errors, so we do not explicitly correct for $M_{\ell\ell'}$ errors.

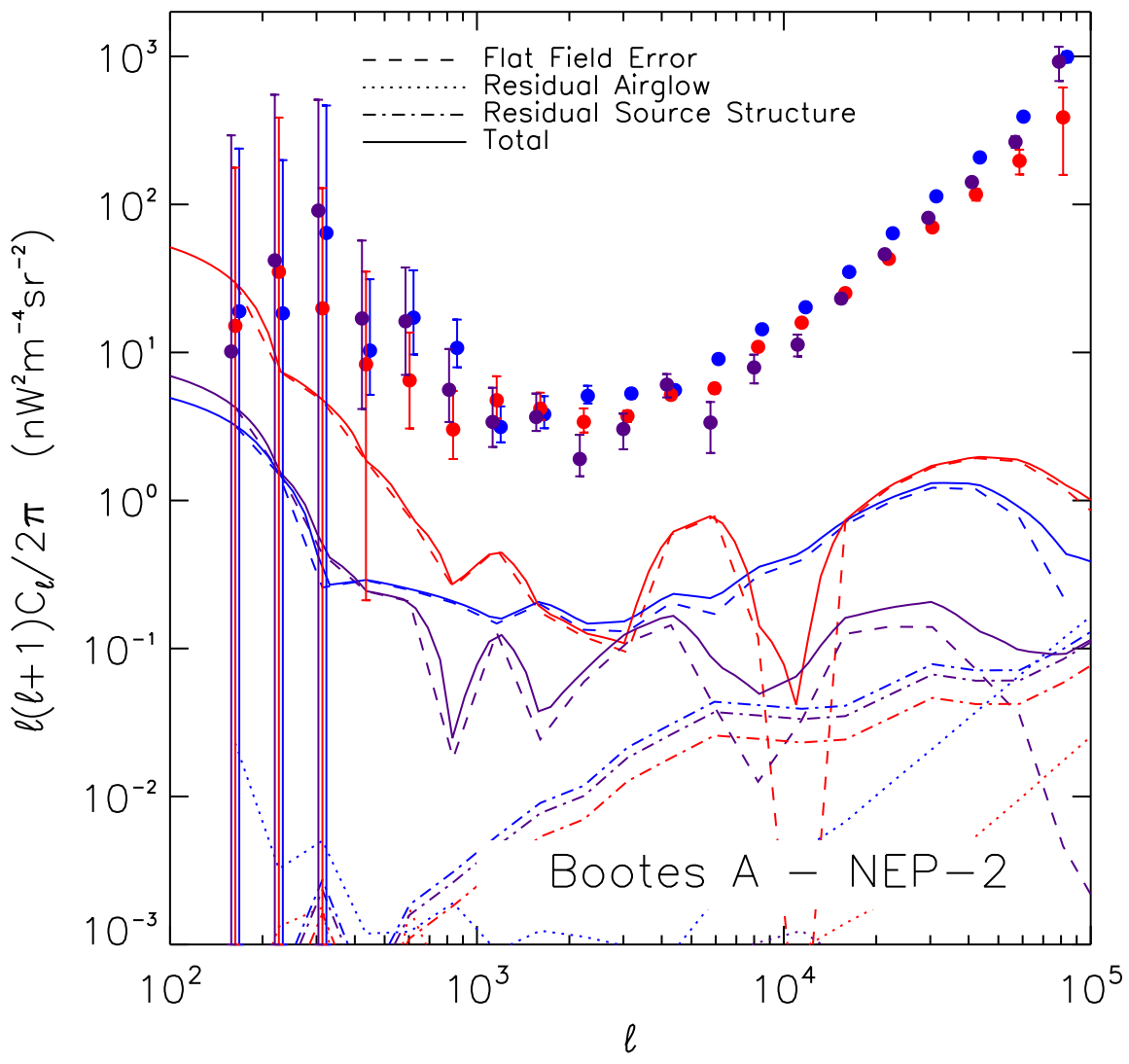

Figure S16: **Instrumental systematic uncertainties in the CIBER fluctuations measurement.** The CIBER auto- and cross-spectra for Boötes A − NEP in the second flight (left) and Boötes B − ELAIS-N1 in the third flight (left) are shown for $1.1\,\mu$m$\times 1.1\,\mu$m (blue), $1.6\,\mu$m$\times 1.6\,\mu$m (red), and $1.1\,\mu$m$\times 1.6\,\mu$m (purple). In each case, the estimated systematic uncertainty associated with flat field error (dashed), residual airglow (triple dot dash), and residual halos from masked sources (dot dash) are indicated. The total instrumental systematic uncertainty is shown as a solid line for each band and field combination.

## 7.5 Systematics in Combined Spectra

We combine the systematic uncertainties in the final averaged power spectra by computing the mean systematic uncertainty between the two data sets using a flat weight. Because the statistical noise between the flights is similar in Boötes A − NEP and Boötes B − ELAIS-N1, choosing a weighting scheme based on the statistical uncertainties yields very similar results. The systematic errors for the combined spectra are shown in Figure S17.

# 8 Internal Consistency Tests

We apply a series of internal consistency tests to either rule out or constrain sources of systematic error using data over two flights. These tests do not quantify errors in the analysis, but show the data are consistent with our assessment of statistical and systematic error. We focus particularly on the two largest potential sources of systematic error, errors in the noise model and errors in the flat-field response of the instrument. A summary of the internal consistency tests is presented in Table S2.

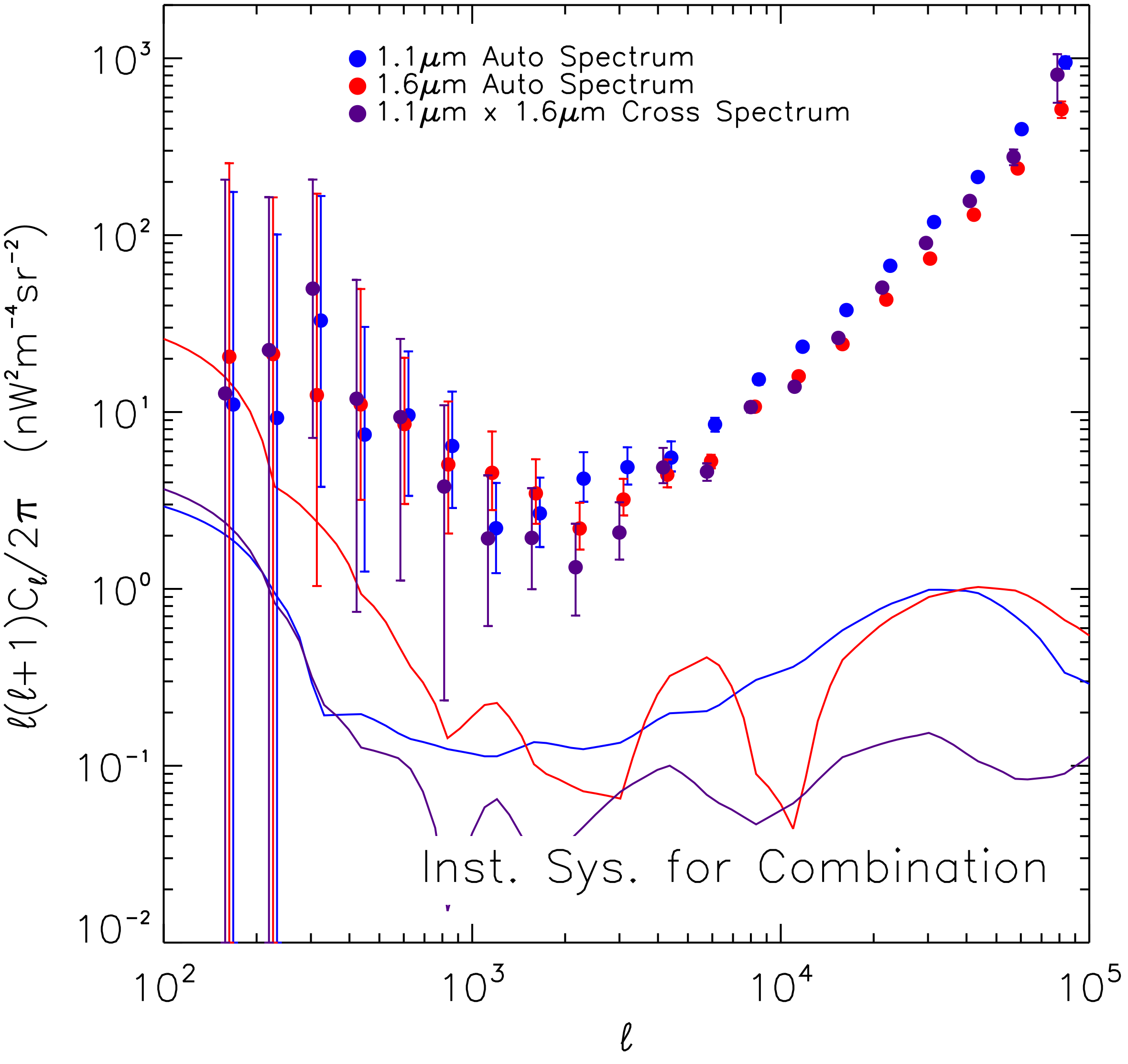

Figure S17: **Total instrumental systematic uncertainty in the CIBER fluctuations measurement.** The CIBER auto- and cross-spectra for the mean of Boötes A - NEP-2 and Boötes B − ELAIS-N1 are shown for $1.1\,\mu$m$\times$$1.1\,\mu$m (blue), $1.6\,\mu$m$\times$$1.6\,\mu$m (red), and $1.1\,\mu$m$\times$$1.6\,\mu$m (purple). The total instrumental systematic uncertainty is shown as a solid line for each band combination. This figure can be compared directly to Figure 1 of the main paper.

Table S2: Internal consistency checks.

| Consistency Check | Data Combination | Figure |
|---|---|---|
| Mask Flux Threshold | Mask to progressively fainter flux limits | S18 |
| Stability over time | CIBER F2 x F3 cross-spectra, CIBER x *Spitzer* | S22, S15 |
| Consistency between wavelengths | CIBER 1.1 $\mu$m x 1.6 $\mu$m cross-spectra, CIBER x *Spitzer* | S14, S22, S15 |
| Noise model validation | CIBER temporal halves - vs. noise model equivalent | S19 |
| Null power in the absence of astronomical signal | Cross-spectra of uncorrelated CIBER fields | S20 |
| Flat field error estimate | Null field-differenced cross-spectra | S20 |
| | Single-field cross-spectra vs. error estimates | S21 |

## 8.1 Flux Threshold Cuts

We compute the auto-spectra as a function of the masking flux threshold to test if the mask depth affects the low-$\ell$ amplitude. The resulting power spectra for a sample of field combinations is shown in Figure S18, in which the threshold magnitude is changed from $J = 7$ to the limiting magnitude $J = 17.5$ in the $1.1$ and $1.6\,\mu$m bands. The low $\ell$ structure converges to a constant level for the faint flux thresholds in both bands, while the shot noise component is progressively reduced, which indicates our results are insensitive to masking below $\sim 15^{\rm th}$ magnitude.

## 8.2 Noise Model

To check the noise model, we form half-field difference images using the first and second half of each flight integration that should be statistically consistent with zero signal. These difference combinations have five times higher variance from readout noise than single flight images, due to the shorter integration times and the field differencing. In Figure S19 we compare flight data, formed by computing the power spectrum of the flight half-field differences $N_\ell^{\rm data}$, with the equivalent noise model realizations $N_\ell^{\rm model}$, according to the prescription in Section 5. We show the sum of the set of individual half-field differences in the flight data to improve the statistical accuracy of the comparison. The probability to exceed $\chi^2$ are $0.17$ and $0.84$ for the two different detector readout chains, indicating that the noise model accurately represents the statistical properties of the flight data.

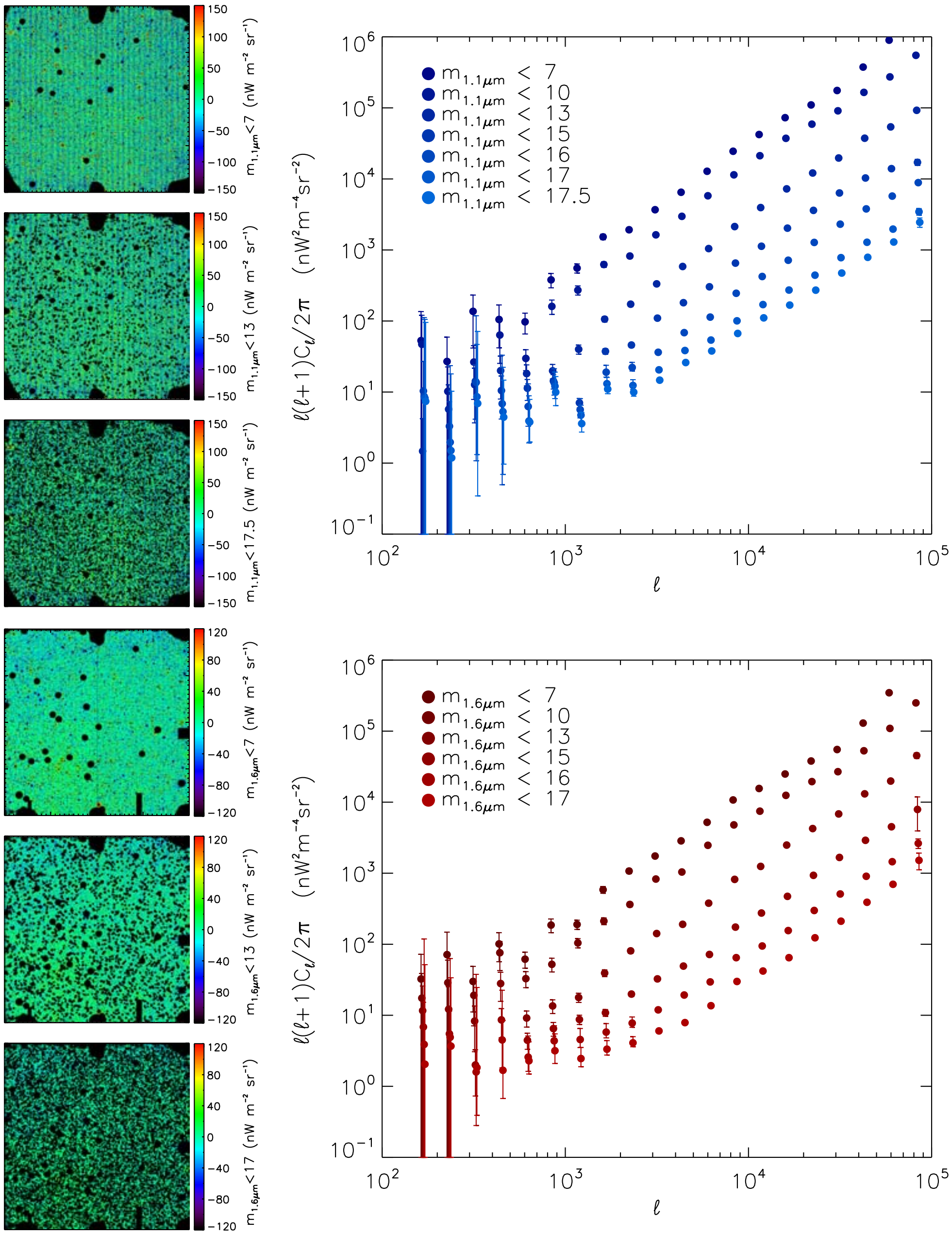


Figure S18: **Effect of mask depth on auto-power spectra.** The large panels show auto-spectra of the NEP-3 − ELAIS-N1 field difference at 1.1 (upper) and 1.6 (lower) $\mu$m as a function of masking flux threshold for $7 < m < m_{\rm lim}$, where $m_{\rm lim}$ is the masking flux. Images corresponding to sample mask thresholds are shown in the left hand column for both bands, corresponding to $m \geq 7$ (upper) $m \geq 13$ (middle) and $m \geq m_{\rm lim}$ (lower). As the limiting threshold magnitude is increased, the shot noise component of the power spectra decreases as expected. The low-$\ell$ amplitude first rapidly decreases with the mask threshold, and then converges to an approximately constant power, as expected for a constant astrophysical fluctuation component.

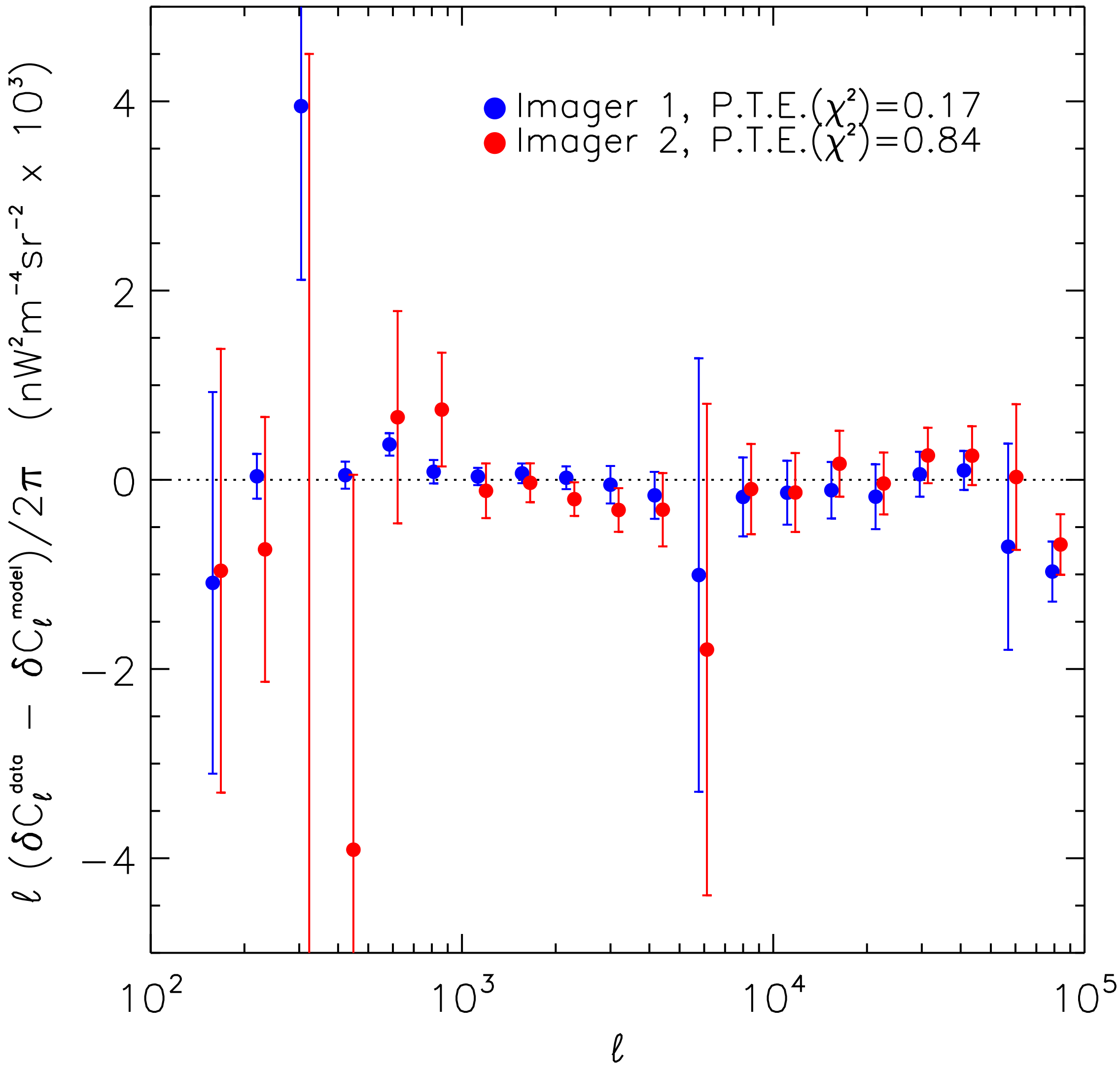


Figure S19: **Validation of flight noise model.** The difference between the flight noise and the equivalent noise model realizations as parametrized by the quantity $\delta C_\ell^{\rm data} - \delta C_\ell^{\rm model}$ where $\delta C_\ell$ is estimated as Equation 8 over $\ell$ bins is shown. The points give the noise model weighted variance of this difference in the five science fields from both flights for a given readout chain. The plotted uncertainties are the total standard deviation of the noise model assuming the mean goes as $1/\sum 1/\sigma_N^2$, summing over the science field combinations. The PTE of these residuals (listed in the legend) indicate that the noise model is a good description of the flight data.

## 8.3 Flat Field Errors

Between flight 2 and flight 3 we swapped the spectral bands associated with each detector array to mitigate any potential error associated with a fixed pattern specific to an array. The auto-spectra shown in Figure S14 were taken with different arrays in each band, indicating that array-specific gain errors do not appreciably change the power spectra.

It is possible that an overall correlated error could be introduced into both channels by non-uniformity of the flat-field source, an integrating sphere which was common in all the measurements. To test this possibility, we construct cross-spectra on field-difference combinations between flights designed to produce zero astrophysical fluctuation signal. These difference cross combinations are calculated in array coordinates, so flat field errors over the instrumental field of view are maximized. As shown in Figure S20, the cross-spectra are consistent with zero to within the systematic error bound.

In addition to these cross-spectra, we can validate the flat-field systematic uncertainty estimate using cross-spectra of single fields. These have uncorrelated astrophysical signal, but are potentially correlated in their flat field structure, an effect that is maximized in single fields. These single-field cross-spectra, shown in Figure S21, have significant deviations from zero statistically on large angular scales, but are within the limits enclosed by our flat-field uncertainty estimate.

## 8.4 Reproducibility of EBL Fluctuations

We calculate cross-spectra across flights and spectral bands to demonstrate overall consistency. Detector noise in a cross-spectrum is uncorrelated and does not introduce a bias, and the consistency of the plotted cross-spectra with the equivalent auto-spectra indicate that any errors in the noise model are not introducing a significant systematic error. In Figure S22 we show cross-spectra of Boötes A $-$ NEP-2 and Boötes B $-$ NEP-3 in all combinations of flights and bands. These cross-spectra have a $\sim 50\,\%$ increase in variance because of the differencing with non-common fields (Boötes A and Boötes B) and so are somewhat noisier than their single-flight counterparts shown in Figure S14. However, they exhibit fluctuations consistent with NEP being $100\,\%$ correlated between flights, strong evidence that the power is constant on the sky. The fluctuation signal is constant over time, and thus unlikely to be due to ZL from structure in the interplanetary dust cloud. Furthermore the cross-spectra in Figure S22 are made in all possible detector array combinations across flights and bands. These cross-combinations also include positional shifts in array coordinates, generally reducing the correlation of flat-field errors significantly compared with those in the auto-spectra. No significant deviation between these cross-spectra and the auto-spectra is apparent.

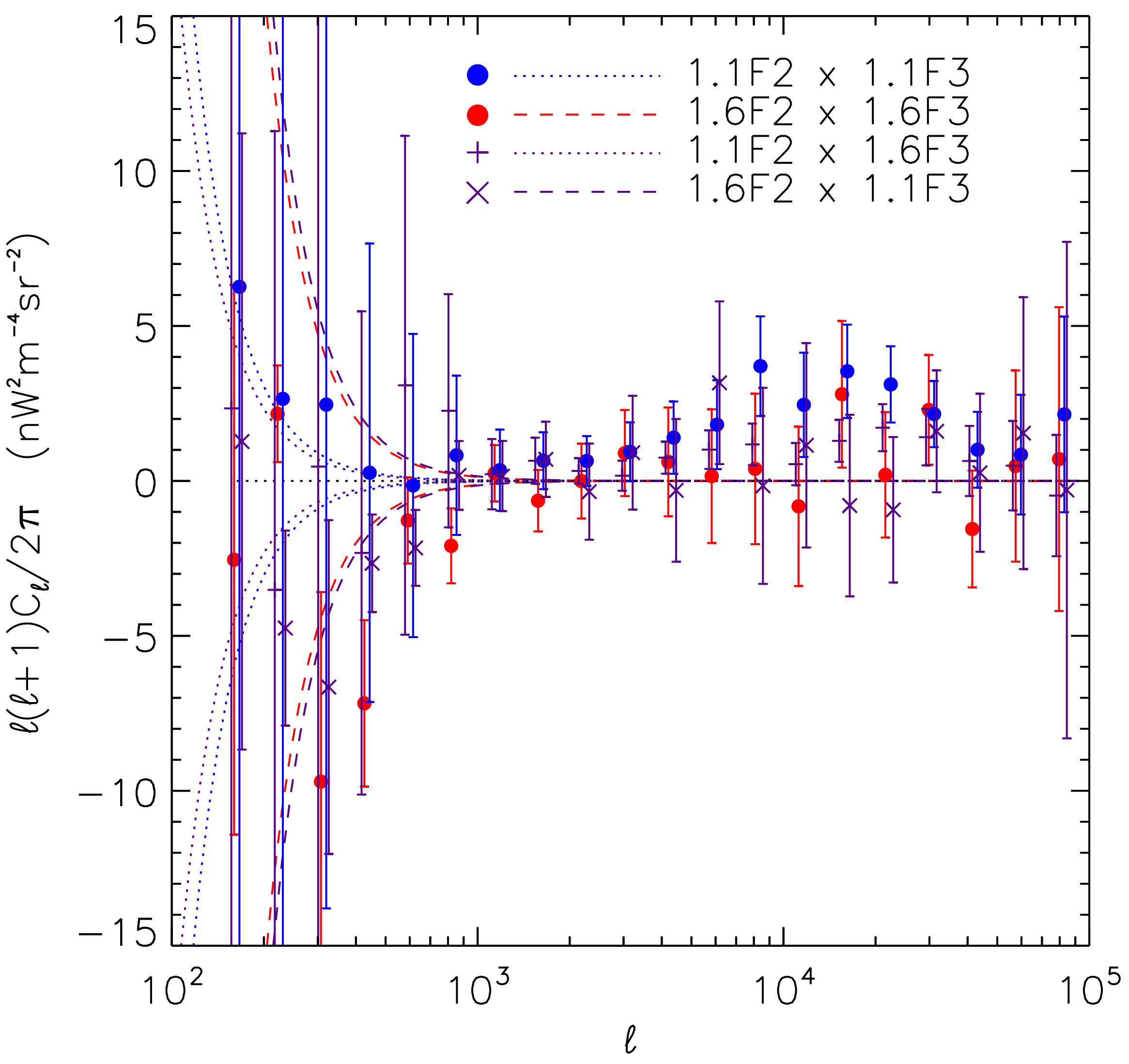


Figure S20: **Flat-field error cross-correlation test.** The cross-spectra between Boötes A − NEP in the second flight and Boötes B − ELAIS-N1 in the third flight are shown. The plotted statistical errors are derived from the noise model and include sample variance. The flat field systematic uncertainties in each case are plotted as dashed or dotted lines. At low $\ell$ no bandpowers significantly exceed the flat field error, evidence that the systematic error bound is conservative. The probability to exceed $\chi^2$ for the null signal hypothesis in these cross-spectra are $\{0.40, 0.06, 0.40, 0.92\}$ (in the order appearing in the legend), indicating that these nominally uncorrelated images are consistent with zero. There is no evidence for additional systematics beyond those already identified in these data.

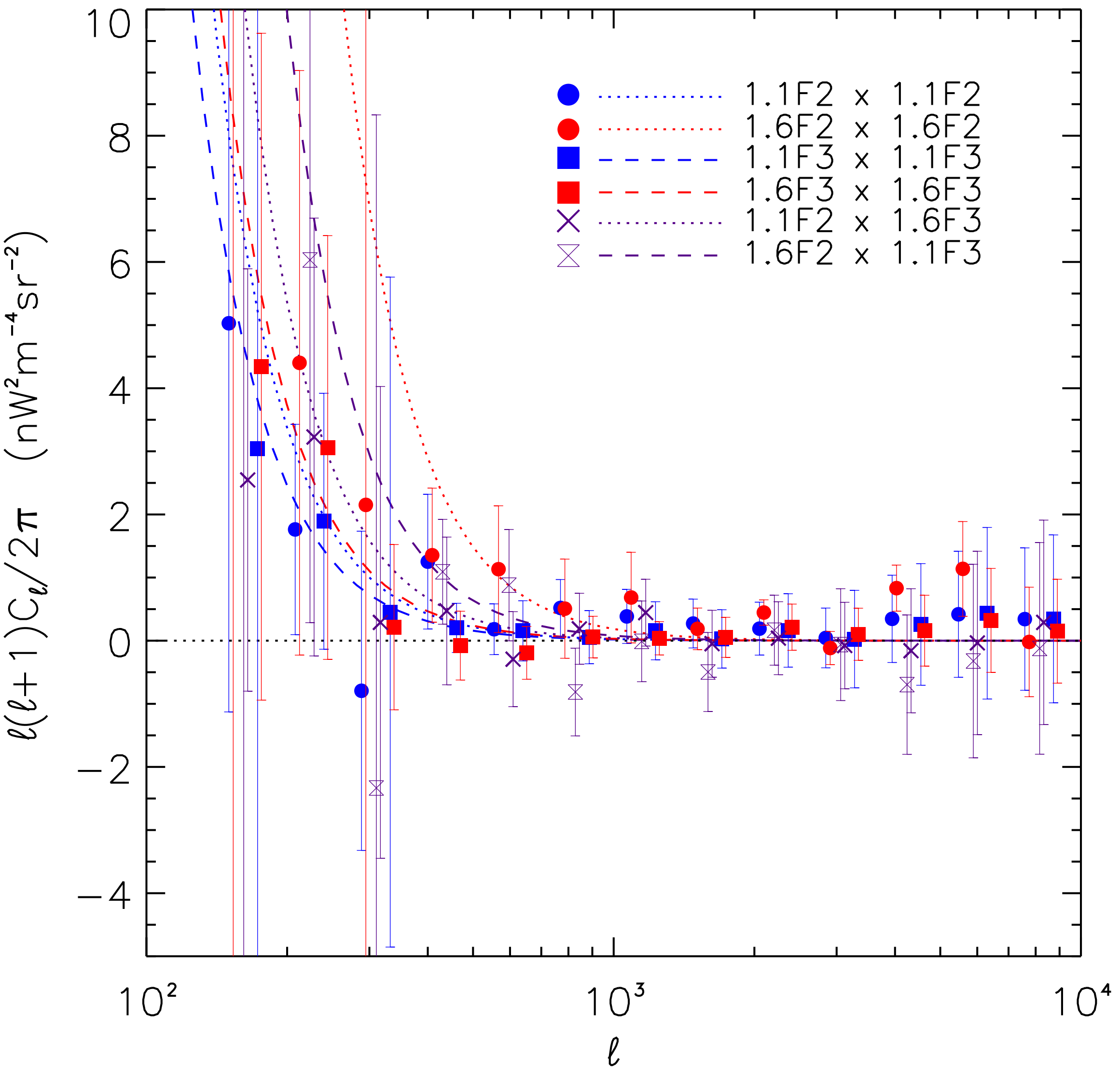


Figure S21: **Single-field cross-spectra.** We compute the cross-spectra of individual flight images which share no astrophysical power, but may have common flat-field error. The spectra are computed in array coordinates to maximize the flat-field error. The lines show the flat-field systematic uncertainty estimate generated for each instance as an upper limit. The error bars include both statistical and sample variance. None of the auto-spectra exceed the flat-field error estimate beyond $1\sigma$ in any band power, evidence that our estimate is reasonable. The fields used in these power spectra are Boötes A crossed with NEP-2 Boötes B crossed with ELAIS-N1in the third flight, and Boötes A crossed with ELAIS-N1 in between flights.

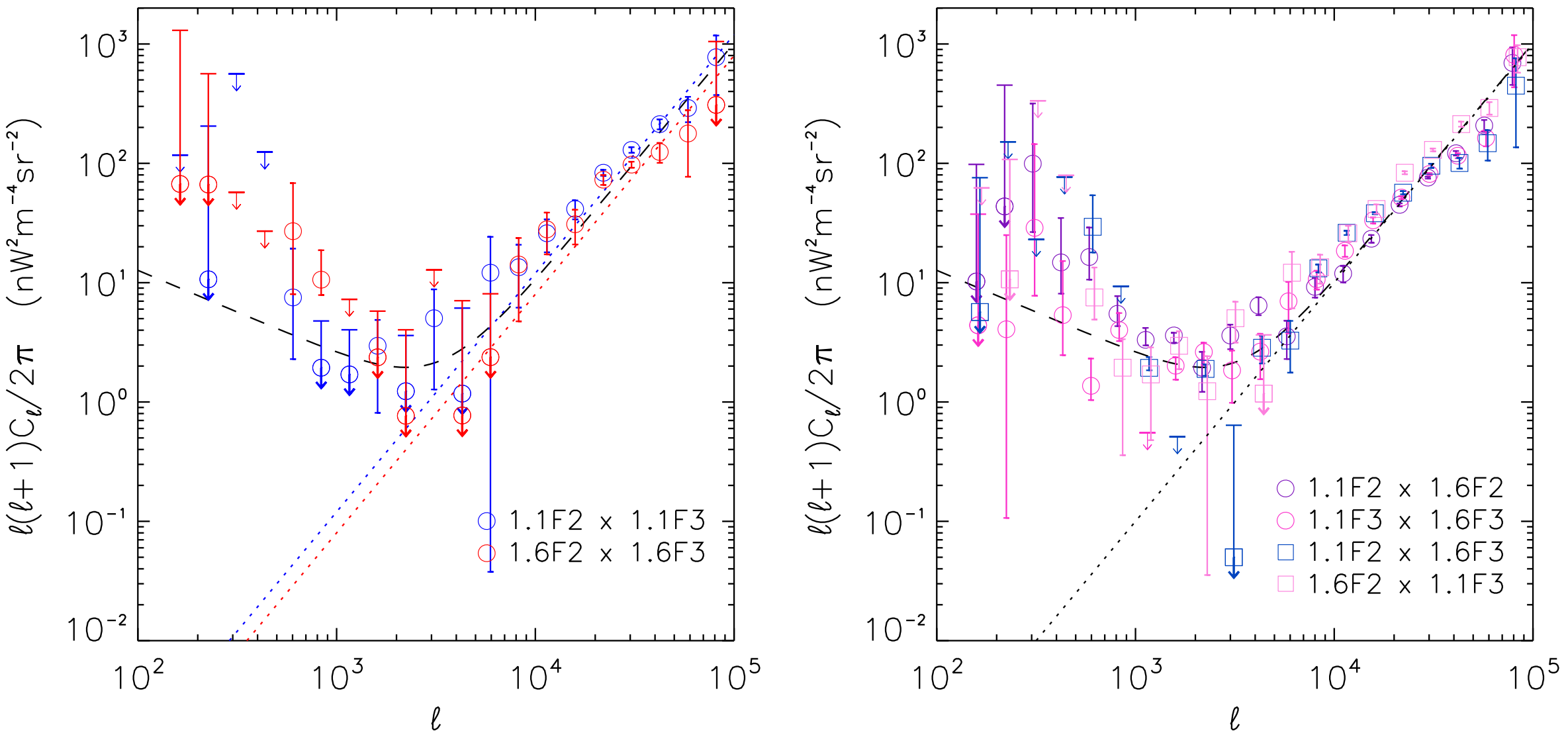


Figure S22: **Difference cross-spectra between instruments and flights.** The cross-power spectra for every combination of Boötes A − NEP-2 and Boötes B − NEP-3. The left plot shows the two combinations of 1.1 $\mu$m×1.1 $\mu$m and 1.6 $\mu$m×1.6 $\mu$m between flights, and the right plot all combinations of 1.1 $\mu$m×1.6 $\mu$m between flights. Lower uncertainties that pass through zero are shown with a standard $1\sigma$ upper uncertainty, but as limits on the lower error bar. These cross-spectral combinations suffer additional variance arising from the field differencing used in this analysis, but are consistent with fluctuations detected in auto-spectra (Figure 1 of the main paper, represented here by long-dashed lines), most importantly at the low-$\ell$ modes. The dotted lines show a simple shot-noise spectrum scaled to high $\ell$ to aid in discriminating the deviation at mid- and low-$\ell$ scales.

# 9 Astrophysical Foregrounds

There are several possible astrophysical foregrounds which may contribute near-infrared fluctuations. In order of distance from the Earth, these are ZL, DGL, and fluctuations from faint stars and galaxies below our detection limit. We place limits on these sources of emission and show that the measured fluctuations are not explained by any combination of these sources.

## 9.1 Zodiacal Light

The correlation of CIBER data between flights shown in Figures S15 and S22 indicates that the fluctuation power does not change over a 17 month time scale. Over this time interval, the line of sight passes through a different path through the interplanetary dust cloud and the ZL brightness differs by $\sim 10\,\%$ (*41*).

Several upper limits to ZL fluctuation power have been published at mid-infrared wave-

lengths, including measurements at $25\,\mu$m from ISO (*42*), $8\,\mu$m from *Spitzer* (*7,8*), and $7-25\,\mu$m from AKARI (*18*). In addition, DIRBE (*41*), AKARI (*17*), and CIBER (*40*) have all measured the near- to mid-infrared electromagnetic spectrum of the ZL which allows us to extrapolate the mid-infrared ZL fluctuation measurements to the near-infrared. Though extrapolating over such a large wavelength interval is not ideal, the decorrelation between fluctuations measured at near- and mid-infrared wavelengths has been estimated as being at most a few percent (*11*). These upper limits are shown in Figure S24, which are far below the detected fluctuation power.

## 9.2 Diffuse Galactic Light

At high galactic latitudes, DGL has a $C_\ell \propto \ell^{-3}$ spatial power spectrum following the structure of the interstellar dust (ISD) emission (*43*). Since DGL is due to scattering of the interstellar radiation field (ISRF) from dust in the galaxy, its surface brightness is simply the product of the ISRF times the scattering cross-section, which is proportional to the dust grain size and density, integrated along the line of sight. Thus the same clouds responsible for cirrus emission in the far-infrared also produce DGL, assuming that the ISRF is uniform and the temperature of the dust is constant.

We estimate the potential contamination from DGL by scaling far-infrared emission to near-infrared wavelengths using an empirical factor, and cross-correlating the resulting image with the CIBER images in each field and band. We use $100\,\mu$m IRAS dust maps (*44*) interpolated to the imager field coverage and differenced in the same way as for our science fields. The IRAS beam size is much larger than the imager beam, so conclusions can only be drawn for $\ell < 2000$ where the IRAS beam dilution is small. The maps are scaled from $100\,\mu$m intensity in MJy sr$^{-1}$ to near-infrared intensity in nW m$^{-2}$ sr$^{-1}$ using the empirical correlation determined from an analysis of the CIBER low resolution spectrometer (LRS; paper in preparation), which yields scalings of 15 nW m$^{-2}$ sr$^{-1}$ (MJy sr$^{-1}$)$^{-1}$ and 8 nW m$^{-2}$ sr$^{-1}$ (MJy sr$^{-1}$)$^{-1}$ at $1.1\,\mu$m and $1.6\,\mu$m, respectively. Our near- to far-infrared conversion factor is consistent with the upper range of conversion factors in previous measurements (*19, 45–48*) which produces a conservative estimate for the DGL foreground. We assume a constant scaling, appropriate for optically thin dust in these high Galactic latitude fields.

We compute both the auto-spectrum of the $100\,\mu$m map scaled to near-infrared wavelengths, and the cross-spectra between the scaled map and CIBER. The curves in Figure S23 show the amplitude of the scaled $100\,\mu$m map, the cross-spectra, and a fit of the model $C_\ell \propto \ell^{-3}$ to the cross-power spectra for the bandpowers $\ell < 3500$. In the field combinations including NEP for the fit including all band powers, we find a correlation between the imager data and DGL, while in the fainter Boötes B $-$ ELAIS-N1 combination no correlation is detected. In Boötes A $-$ NEP-2 the cross-correlation is about half the level predicted by the scaled IRAS auto-spectrum in power. We calculate a simultaneous joint fit to the DGL cross-spectra assuming the DGL color falls between 1.1 and $1.6\,\mu$m in the ratio of $8/15$, and has an $\ell^{-3}$ spatial dependence. This fit is plotted in Figure S24 and Figure 1 of the Main Paper.

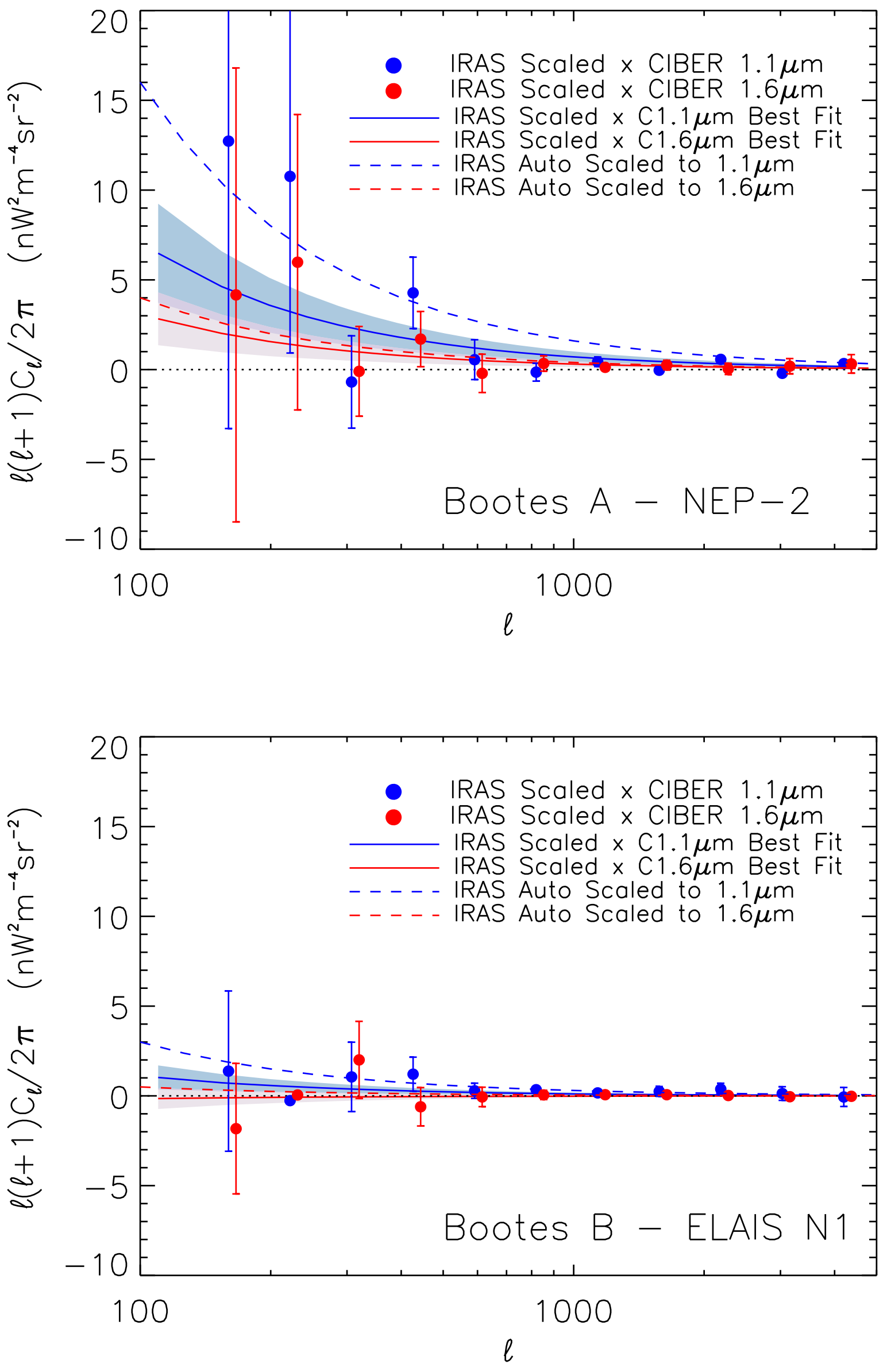


Figure S23: **DGL scaling limits and the CIBER – DGL cross-correlations.** The auto-spectra of 100 $\mu$m IRAS far-infrared maps scaled to the near-infrared (dashed lines), the cross-correlation of the CIBER images with the scaled 100 $\mu$m maps (points), and fits of the $C_\ell \propto \ell^{-3}$ DGL fluctuation power spectra to the CIBER – DGL cross-power spectra for $\ell < 3500$ (shaded intervals denoting $\pm 1\sigma$ about best fit) are shown. Panels show Boötes A - NEP-2 (left) and Boötes B $-$ ELAIS-N1 in the third flight (right). We marginally detect a cross-correlation in the NEP combinations, where DGL is brightest, but not in the Boötes B $-$ ELAIS-N1 combination, where DGL is faint.

To estimate whether optical extinction from the dust in the galaxy could source fluctuations, we compute extinction maps at $1.1$ and $1.6\,\mu$m using $100\,\mu$m-derived extinction images (*44*), assuming $R_V = 3.1$, appropriate to the diffuse interstellar medium. We multiply a uniform EBL of $\lambda I_\lambda = 15\,$nW m$^{-2}$ sr$^{-1}$ by the extinction map, and the power spectra of these extincted images are computed. We find that large angular scale power generated by extinction is negligible.

### 9.3 Faint Stars

Fluctuations from stars fainter than the masking cutoff flux can be estimated by populating a simulated image with $J > 17.5$ stars from the UKIDSS-UDS DR9 stellar catalog (*20*) which covers $0°.8 \times 0°.8$, and computing its power spectrum. The completeness of the UDS survey is $24$ mag at $J$-band, a depth at which virtually all of the integrated star light (ISL) is captured by the UDS catalog (*49*). The resulting $J$-band power spectrum, extrapolated to the CIBER bands using a faint star ISL template electromagnetic spectrum, is shown Figure S24. The contribution from ISL is almost entirely Poissonian, and negligible for $\ell < 10^4$.

### 9.4 Residual Galaxies

To estimate the contribution from faint galaxies, we employ a recent empirical model (*16*) in which luminosity functions from the UV to near-infrared are used to generate a distribution of galaxies up to $z = 5$ which includes a halo model to describe clustering. We use the galaxy fluctuation model for each combination of the CIBER and *Spitzer* power spectra, calculated for galaxies below the masking flux cutoff. Both minimal and maximal surface brightness models are calculated, which defines the range of galaxy fluctuation power consistent with the model. These model predictions, shown in Figure S24, are below the measured galaxy count fluctuations for $\ell < 5000$.

## 10 Astrophysical Interpretation

To interpret the CIBER measurements we make use of existing models for near-infrared intensity fluctuations, including (i) high-$z$ galaxies present during reionization with PopII stars, PopIII stars, or some combination of the two (*3*,*5*); (ii) IHL (*6*); and (iii) DCBHs (*4*) which have been invoked to explain the cross-correlation between $3.6\,\mu$m fluctuations and the X-ray background (*12*).

For the high-$z$ galaxy model, we follow the prescription of Ref. *5* using the spectral energy distributions (SEDs) of Ref. *3*. Our fiducial model generates near-infrared anisotropies at $z > 7$, but it is also possible to consider models where this minimum redshift is varied between 6 and 12. We have developed an analytical model to estimate the EOR contribution to the EBL mean intensity and anisotropy that depends on the star formation efficiency $f_*$ and the escape

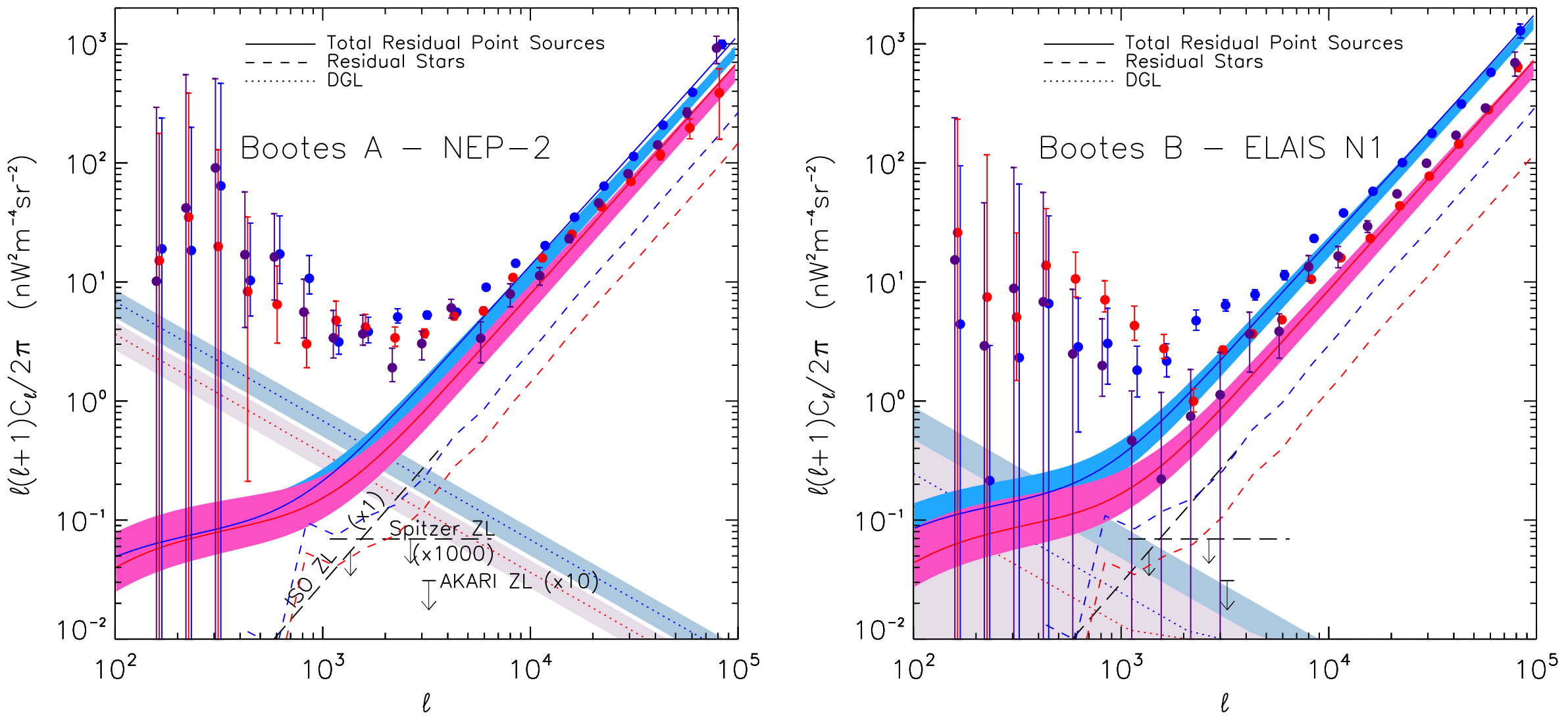


Figure S24: **Estimated fluctuations in astrophysical foregrounds.** The CIBER auto- and cross-spectra for Boötes A – NEP-2 (left) and Boötes B – ELAIS-N1 (right) are shown for $1.1\,\mu\mathrm{m}\times 1.1\,\mu\mathrm{m}$ (blue), $1.6\,\mu\mathrm{m}\times 1.6\,\mu\mathrm{m}$ (red), and $1.1\,\mu\mathrm{m}\times 1.6\,\mu\mathrm{m}$ (purple). The colored regions show the fluctuations of galaxies below the CIBER masking flux threshold (*16*). The estimated foreground contributions from residual stars below the CIBER masking flux threshold are indicated by dashed lines, and the total foreground from residual point sources (the sum of unmasked galaxies and stars) is indicated by a solid line for each band and field combination. DGL fluctuations are estimated by fitting the CIBER – $100\,\mu$m cross-spectra to a simple $\ell^{-3}$ model shown by the shaded regions denoting $1\sigma$ about the best fit (dotted lines). Finally, upper limits to ZL fluctuations are indicated for ISO (*42*), *Spitzer* (*8*), and AKARI (*18*) (dashed lines).

fraction of ionizing photons $f_{\mathrm{esc}}$. We fix these to standard values and capture all modeling uncertainties using a single fitter parameter, a normalization of the signal $A_{\mathrm{high-z}}$ which captures the component which appears as Lyman dropout between the CIBER bands.. A measurement of this amplitude leads us to a measurement of the SFRD during reionization. The CIBER 1.1 and $1.6\,\mu$m bands probe the Lyman cutoff at $z < 9.5$ and $z < 16$, respectively, hence the redshift ranges for $C_{\ell,\mathrm{high-z}}^{\lambda\lambda'}$ are different for different cross-power spectra, i.e. $7 < z < 9.5$ for $C_{\ell,\mathrm{high-z}}^{1.1\times 1.1}$, $7 < z < 16$ for $C_{\ell,\mathrm{high-z}}^{1.6\times 1.6}$, $7 < z < 30$ for $C_{\ell,\mathrm{high-z}}^{3.6\times 3.6}$, $7 < z < 9.5$ for $C_{\ell,\mathrm{high-z}}^{1.1\times 1.6}$ and $C_{\ell,\mathrm{high-z}}^{1.1\times 3.6}$, and $7 < z < 16$ for $C_{\ell,\mathrm{high-z}}^{1.6\times 3.6}$.

For the IHL component, we use a halo model with diffuse stars distributed over the halo following the dark matter profile, following Ref. *6*. We assume an IHL fraction relative to the total luminosity of the halo and describe the mass dependence of the IHL fraction as a power-

law with:

$$f_{\rm IHL}(M) = A_{\rm IHL}\left(\frac{M}{M_0}\right)^{\alpha}. \tag{14}$$

Here we normalize to $M_0 = 10^{12}\,M_\odot$, and $A_{\rm IHL}$ and $\alpha$ are free parameters that are fitted to the data to account for the amplitude and mass dependence of IHL fraction. The IHL luminosity is taken to be a fraction of the total luminosity:

$$L^{\nu}_{\rm IHL}(M,z) = f_{\rm IHL}(M)L^{\nu}_{\rm tot}(M,z), \tag{15}$$

where $L^{\nu}_{\rm tot}(M,z)$ is the maximum halo luminosity with mass $M$ at $z$. We assume $L^{\nu}_{\rm tot}(M,z)$ to take the form:

$$L^{\nu}_{\rm tot}(M,z) = L_0(M,z=0)(1+z)^{\beta}F_{\nu}(\nu(1+z)), \tag{16}$$

where $L_0(M,z=0)$ is the total halo luminosity at $z=0$. Note that $L^{\nu}_{\rm tot}(M,z)$ as defined in Ref. 50 is the maximum possible halo luminosity, which is not realized in time-averaged star formation in galaxies. Furthermore galactic star formation is a function of halo mass, and is less efficient in low-mass halos. Following Ref. *6*, we use a fitting function for total luminosity in halos at $2.2\,\mu$m from Ref. 51 with:

$$L_0(M,z=0) = 5.64\times10^{12}h_{70}^{-2}\left(\frac{M}{2.7\times10^{14}h_{70}^{-1}M_\odot}\right)^{0.72}L_\odot. \tag{17}$$

In general galaxies trace this total luminosity in high mass halos, but depart rapidly away from this relation in low-mass halos due to a decrease in efficiency of galaxy formation. Despite not hosting a central galaxy, dark matter halos at small masses can host IHL since galaxy merges and tidal stripping can populate stars in a fraction of the small mass halos even though the stars originate elsewhere. Because the above relation is only available at $z=0$ and $\lambda=2.2\,\mu$m, we assume it evolves as $(1+z)^{\beta}$ and make use of the SED $F_\nu$ to transfer $L_0$ to the other frequencies. Note that galaxy luminosity evolution is such that for galaxies, $\beta\sim3$.

We assume the IHL is described by stellar populations with a fixed age for the formation of stars (*52*). The SED in Figure S25 is calculated assuming stellar populations with ages between 0.2 and 8 Gyr. For model fits related to the power spectra, quoted in the main paper, we fix the SED to the 3 Gyr stellar population case. In addition to these parameters we also set the minimum and maximum halo mass sourcing IHL fluctuations to be $10^{10}$ and $10^{13}$ $M_\odot$, respectively. Our results are insensitive to assumptions about the maximum mass, but do depend on the minimum mass. Because of this, we match the minimum mass to the minimum galaxy halo mass scale in standard halo models to allow comparison with our results. In all the models presented here we assume a flat ΛCDM model with $\Omega_{\rm M}=0.27$, $\Omega_{\rm b}=0.046$, $\sigma_8=0.81$, $n_{\rm s}=0.96$ and $h=0.71$ (*53*).

In addition to the IHL component, we also allow for a residual galaxy clustering signal for $z<5$ sources (*16*) modeled by $f_{\rm low-z}$, the fraction of residual low-$z$ galaxies relative to the Helgason model. The total amplitude of the angular power spectrum then is given by:

$$C^{\lambda\lambda'}_{\ell,\rm tot} = C^{\lambda\lambda'}_{\ell,\rm IHL} + C^{\lambda\lambda'}_{\ell,\rm low-z} + C^{\lambda\lambda'}_{\ell,\rm high-z}. \tag{18}$$

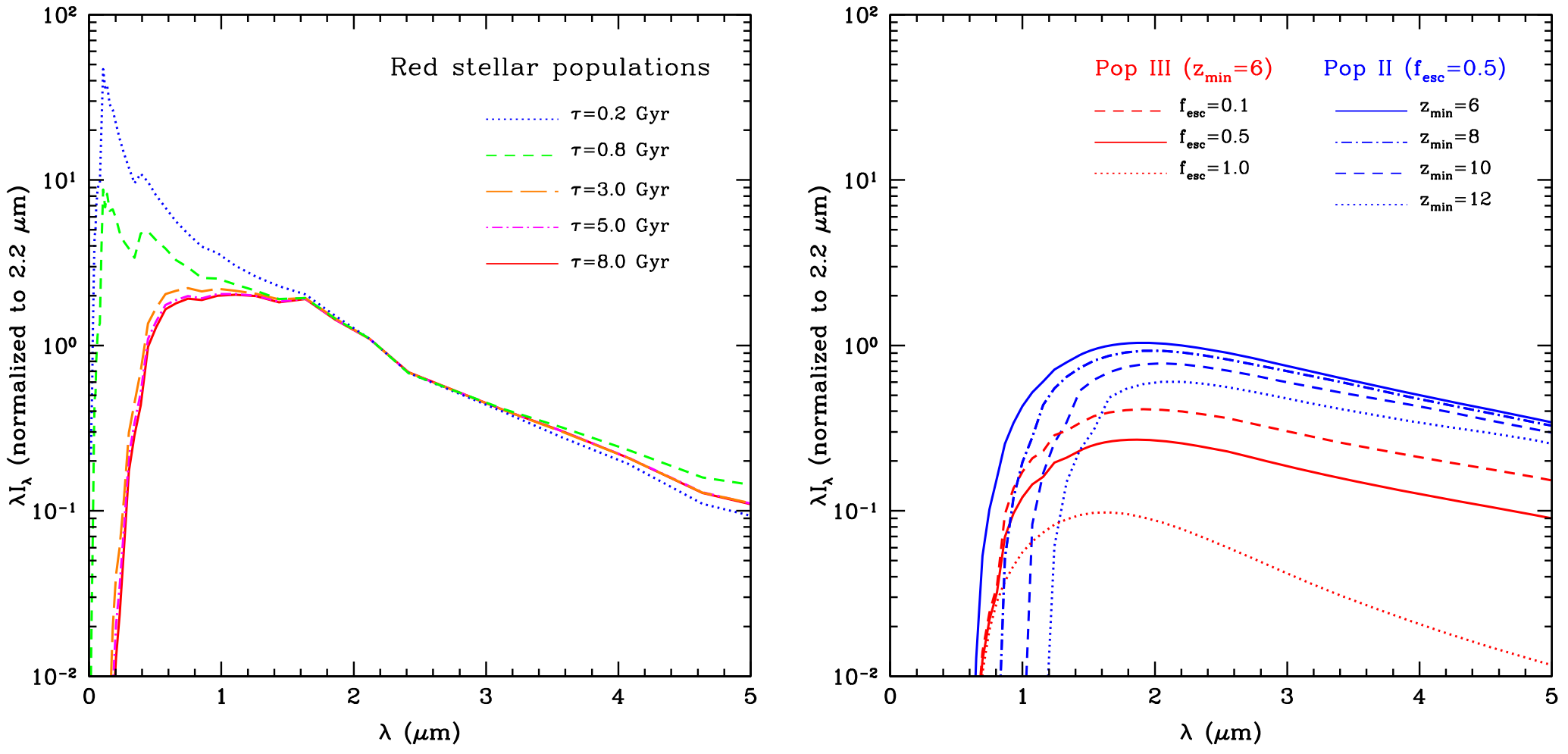


Figure S25: **The spectral energy distribution of IHL stars and high-redshift galaxies.** *Left:* We show here the rest-frame SED templates used to describe IHL with all SED models normalized at $2.2\,\mu$m. They are from models related to old stellar populations with ages at 0.2, 0.8, 3, 5 and 8 Gyr from the literature. These SED models have a broad peak around $1\,\mu$m. Thus the fluctuations measured at 1.1 and $1.6\,\mu$m directly probe this peak at $z < 1$, but at $3.6\,\mu$m, fluctuations arise from a somewhat wider range of redshifts out to $z \sim 2$. In the future, a precise assessment of the correlation between 1.1 and $3.6\,\mu$m could constrain IHL evolution. The SEDs have a ratio of $\sim 7$ from $1.6\,\mu$m to $3.6\,\mu$m. *Right:* Here we show the SED templates of PopII and PopIII stars (*3*). These stars are expected in $z > 6$ galaxies during the epoch of reionization. For the PopII case, we assume a constant escape fraction with $f_{\rm esc} = 0.5$ and show SEDs as a function of the redshift. For the case with PopIII stars, we fix the minimum redshift and show three cases with $f_{\rm esc} = 0.1, 0.5$ and 1.0. The case with $f_{\rm esc} = 1.0$ results in no nebular emission lines and shows simply a PopIII stellar spectrum. This case has the largest color ratio between 1.6 and 3.6 $\mu$m with a value of $\sim 3$. Due to nebular and free-free emission, all other cases have a 1.6 to 3.6 $\mu$m color ratio that is $\lesssim 3$.

We calculate the theoretical auto- and cross-spectra for 1.1, 1.6 and $3.6\,\mu$m and employ a MCMC method to fit the model to the data. The Metropolis-Hastings algorithm is used to determine the probability of accepting a new chain point. The $\chi^2$ distribution is adopted to estimate the likelihood function $\mathcal{L} \propto \exp(-\chi^2/2)$, where we have:

$$\chi^2 = \sum_{i=1}^{N_d} \frac{(C_{\ell,\mathrm{tot}}^{\lambda\lambda',\mathrm{obs}} - C_{\ell,\mathrm{tot}}^{\lambda\lambda',\mathrm{th}})^2}{\sigma_\ell^2}. \tag{19}$$

Here $N_d$ is the number of all data points of six auto- and cross-spectra for 1.1, 1.6 and $3.6\,\mu$m, $C_{\ell,\mathrm{tot}}^{\lambda\lambda',\mathrm{obs}}$ are the observed angular power spectra, and $\sigma_\ell$ are the uncertainties in each $C_{\ell,\mathrm{tot}}^{\lambda\lambda',\mathrm{obs}}$. As the 1.1 and $1.6\,\mu$m $C_{\ell,\mathrm{tot}}^{\lambda\lambda',\mathrm{obs}}$ are dominated by DGL and flat-field systematic error at $\ell < 500$, we also include these components in the model. The MCMC fits a single scaling to either component, where the maximum allowable value is given by the curves shown in Figure 1 of the main paper.

We assume a uniform prior probability distribution for the free parameters in our model. The model parameters and their ranges are as follows: $\log_{10}(A_{\mathrm{IHL}}) \in (-6, 0)$, $\beta \in (-5, 5)$, $\log_{10}(A_{\mathrm{high-z}}) \in (-4, 7)$, and $f_{\mathrm{low-z}} \in (0, 1)$. The $A_{\mathrm{IHL}}$ is the amplitude of the IHL fraction in Eq. (14), and we fix $\alpha = 0.1$ in the MCMC fitting according to the result obtained in Ref. *6*. Since our power spectra measurements are matched in terms of the relative depth of the mask in each of the three bands, including in *Spitzer* with the use of a shallow mask, we find that a uniform model with just one normalization parameter for IHL can describe the data.

We generate sixteen parallel MCMC chains and $200,000$ chain points are sampled after convergence. After the burn-in process and thinning the chains, we merge the chains together and collect about $10^4$ chain points to generate the probability distributions of the free parameters in our model. More details of our MCMC method can be found in Ref. 54. In Figure 1 of the main paper we show the best-fitting models of the low-$z$, IHL, and high-$z$ galaxy contributions. With the current data we are only able to place a 95% confidence upper limit on the high-redshift amplitude$A_{\mathrm{high-z}}$, which corresponds to a star-formation rate density (SFRD) of about 1 $\mathrm{M}_\odot$ yr$^{-1}$ Mpc$^{-3}$ at $z > 7$ at 95% confidence. The required star-formation rate to reionize the Universe is roughly an order of magnitude lower than this value.

To generate the DCBHs model, we use the calculations reported in Ref. *4*. The DCBH model, developed to explain the $3.6\,\mu$m auto-power fluctuations and the apparent cross-correlation between the X-ray background and the $3.6\,\mu$m fluctuations, requires a minimum redshift of about 12. Due to the Lyman cutoff, DCBHs do not contribute to intensity fluctuations at $1.1\,\mu$m and only have a small contribution at $1.6\,\mu$m. Therefore, as a single explanation for fluctuations, we find that DCBHs cannot fit the observed color ratios. A scenario in which DCBHs generate fluctuations at $3.6\,\mu$m, but the 1.1 and $1.6\,\mu$m fluctuations are sourced by some other population, could produce the appropriate colors. However, such a scenario would not produce a measurable cross-correlation between $3.6\,\mu$m fluctuations and CIBER fluctuations unless the unknown 1.1 and $1.6\,\mu$m component is at high-redshift, in which case it would also have a Lyman cutoff. Given the measured positive cross-correlation between $3.6\,\mu$m and both CIBER bands, we can safely state that DCBH models, even ones invoking a second source population, are unlikely.

Figure S25 shows the spectral energy distributions of IHL, PopII, and PopIII stars. Because each model component has different behavior in color space, we show a variety of models using the SEDs of IHL, high-$z$ galaxies, and DCBHs. These SEDs are used to calculate the color-color diagram shown in Figure S26. Fluctuations arising from $z > 7$ galaxies and DCBHs have neither the amplitude nor the color characteristics required to match the measurements. The DCBH model colors do not match the data because they cannot produce sufficient short wavelength power. The color of the EOR model does not match the data because SEDs for PopII and PopIII stars do not have a $\lambda^{-3}$ spectrum between 1.6 and 3.6 $\mu$m.

Of the three hypotheses we test here, the IHL model most closely matches the amplitude of the power spectrum (as shown in Figure 1 of the main paper), but does not match the color characteristics of the measurements. Based on this, the general IHL hypothesis seems to be a plausible explanation of the large scale fluctuation power, though the exact prescription detailed here does have shortcomings, most prominently a significantly redder spectrum than the measurements indicate. However the IHL model does not include the non-linear effects, triggered star formation and non-linear clustering, associated with tidal interactions. Furthermore non-linear IGL clustering is not included in our low-z galaxy model, although there is clear evidence for such a contribution in the *Spitzer* data.

With these caveats noted, the best fitting IHL model is sourced at $z < 0.5$ and is responsible for a large fraction of the large angular scale fluctuations from 1.1 to 3.6 $\mu$m. In comparison, IGL intensity peaks at $z \sim 1.5$ to 2 during the peak epoch of galaxy formation. More specifically, while the IGL luminosity evolves as $(1+z)^3$, we find that the IHL luminosity, in our best-fit model, evolves as $(1+z)^{-1.0^{+0.5}_{-0.1}}$ with an IHL fraction of 1%. We find that the total backgrounds given by our IHL model are $4.6 \pm 0.8$ and $3.5 \pm 0.7$, nW m$^{-2}$ sr$^{-1}$ at 1.1 and 1.6 $\mu$m respectively (Figure S27). Note these levels are somewhat smaller than the values presented in Table 1 of the main paper, resulting from the IHL model fits falling somewhat lower than the measurements at low-$\ell$ in Figure 1 of the main paper. This inferred IHL is 30-50% of the IGL due to differences in their redshift evolution and, to a lesser extent, the halo mass dependence. As a point of comparison, existing observational and numerical studies argue that the IHL fraction could be as large as $\sim 30\,\%$ for Milky Way-sized galaxies (*55*).

There is a significant difference between the redshift evolution of the galaxy luminosity when compared to the total IHL luminosity. In this model, the IHL signal is primarily from $z < 0.5$, while galaxy intensity peaks around $z \sim 1.5$ to 2. The difference in the square of the luminosity distance, necessary to convert from luminosity to an intensity on the sky, accounts for a factor of 15 to 20 increase in the IHL intensity relative to galaxy intensity, assuming a fixed fraction of 1% at all mass scales. The halo mass dependence accounts for the second difference, at a given redshift, since smaller, more abundant, halos are more likely to contain IHL but are not hosting central galaxies as galaxy formation is less efficient in those halos. However, to avoid over-populating all small halos with IHL stars we have introduced a minimum mass cut-off for the calculation at a halo mass scale of $10^{10}$ M$_{\odot}$, comparable to and perhaps slightly lower than the minimum halo mass scale of galaxies in the halo models of clustering. There is some degeneracy between the IHL amplitude and the redshift dependence and the effect of that is to move the IHL signal to much lower redshifts $(z < 0.5)$ than the case of involving *Spitzer* fluctuations alone (*6*).

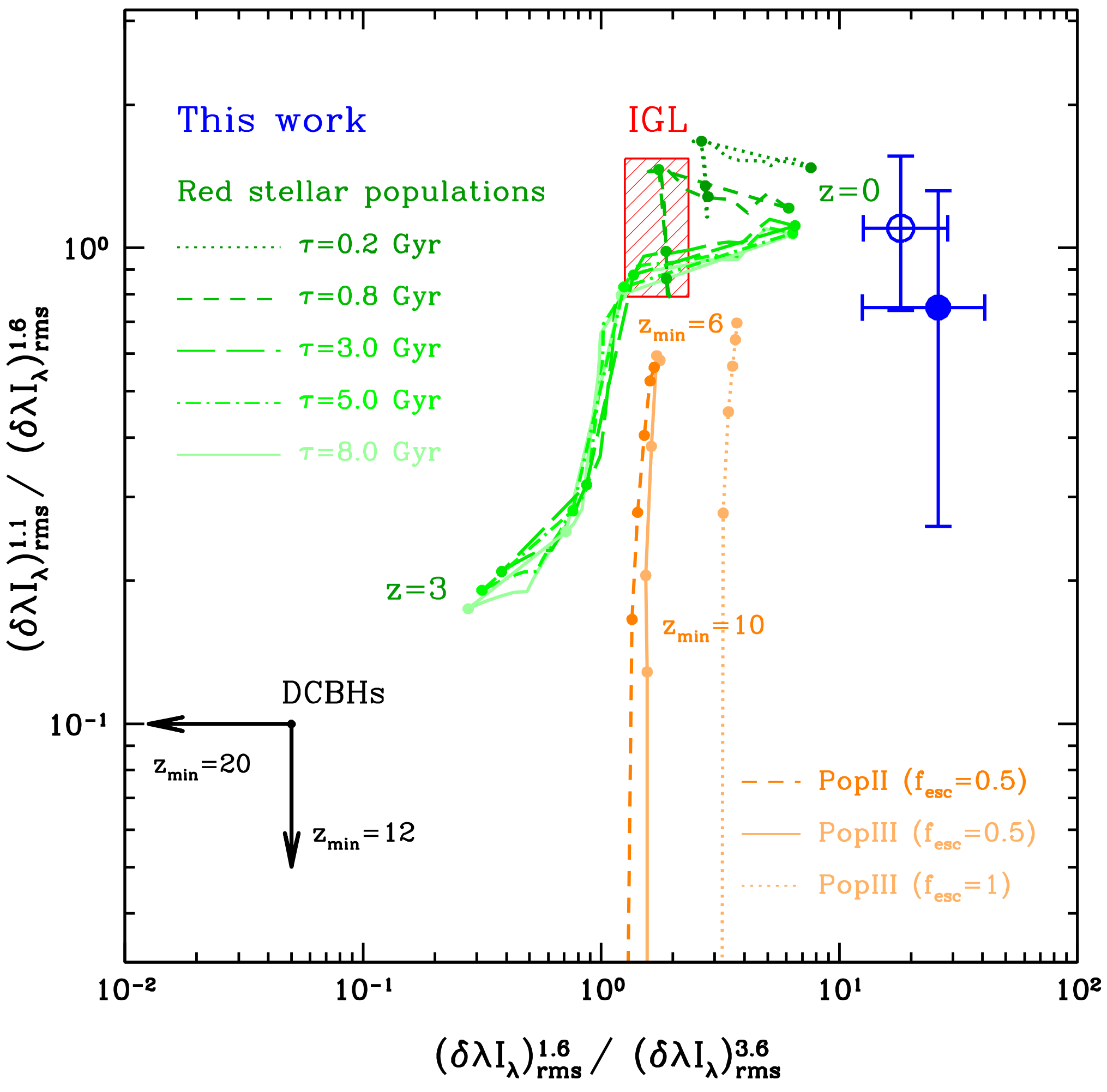

Figure S26: **Color of the rms fluctuation amplitude at 1.6 $\mu$m to 3.6 $\mu$m and 1.1 $\mu$m to 1.6 $\mu$m at $\ell = 3000$.** To account for the effects of masking, we plot the CIBER and *Spitzer* rms fluctuation ratios in a narrow $500 < \ell < 2000$ bin (filled blue circle) and a broader $500 < \ell < 5000$ bin (open blue circle), after subtracting the contribution of residual galaxies. The narrower $\ell$ range point has larger uncertainties but is a more reliable estimate of the constant-$\delta I$ component of the large angular scale power in *Spitzer*. The indicated $1\sigma$ errors are derived from the combination of statistical and foreground galaxy model uncertainties. The measured fluctuations color is much bluer than the color of the IGL background, estimated from measurements of deep galaxy counts (*28*), and indicated by the red hatched region. The green family of lines shows the color ratios of IHL as a function of redshift from $z = 0$ and $z = 3$ in $\delta z = 1$ steps for red stellar populations with stars aged 0.2, 0.8, 3, 5, and 8 Gyr. The orange lines show the expected color of EOR galaxies containing PopII and PopIII stars following models of Ref. *3*. The DCBH model (*4*) predicts very low intensity at 1.1 and 1.6 $\mu$m due to the redshifted Lyman cutoff, as shown by the left-going limit. The color of the measured fluctuations are bluer than any of the modeled components.

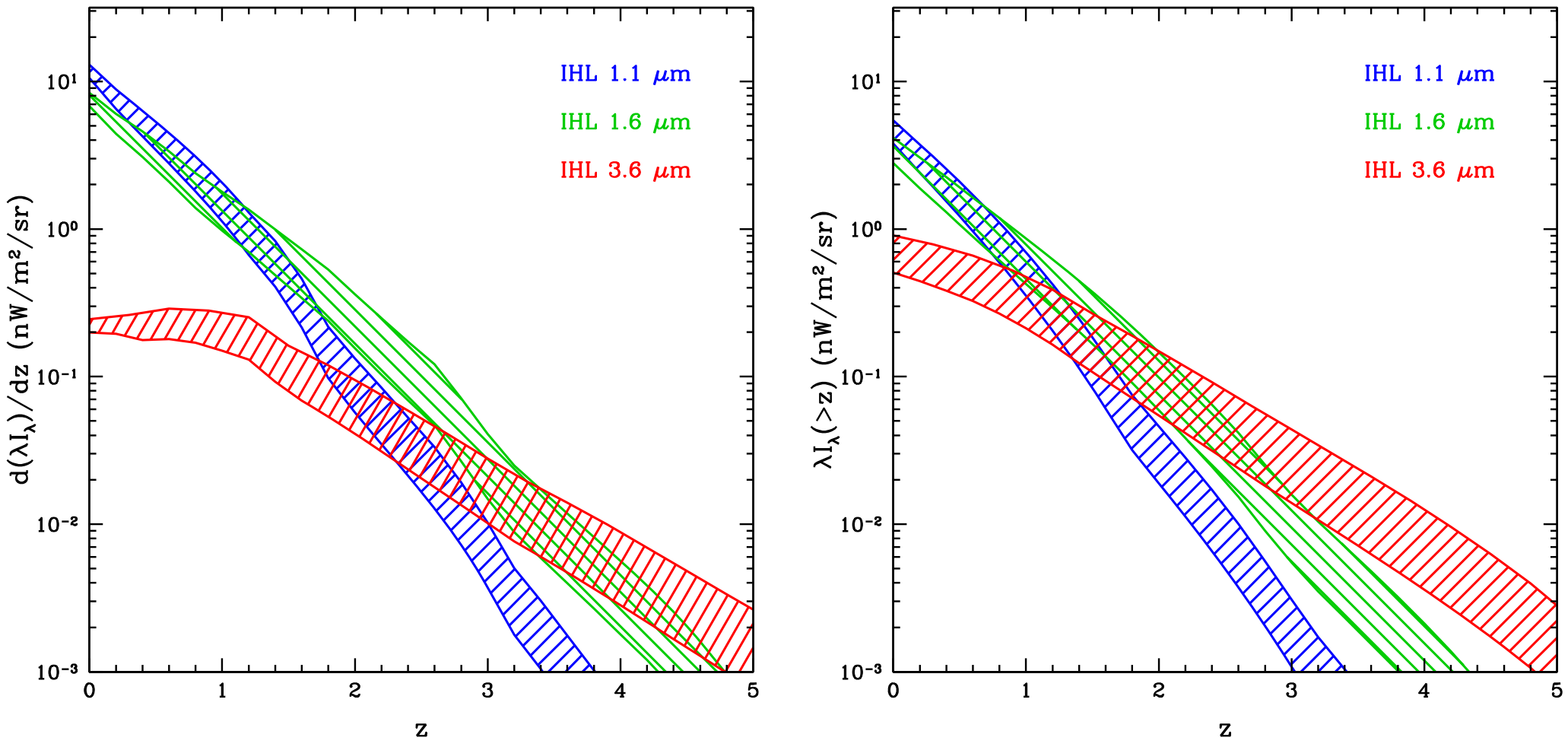

Figure S27: $\mathrm{dF}/\mathrm{dz}$ **and total intensity for the IHL model.** Here we show the IHL intensity as a function of redshift at 1.1, 1.6 and 3.6 $\mu$m (left) and the total integrated intensity (right), based on best-fit models to the CIBER and *Spitzer* fluctuations data. The range of predictions shown here captures the 68% confidence parameter uncertainties in the model fits to the data. The total IHL intensity at 1.1, 1.6 and 3.6 $\mu$m is $4.6 \pm 0.8$, $3.5 \pm 0.7$, and $0.7 \pm 0.2$ nW m$^{-2}$ sr$^{-1}$, respectively.

Given the shallower depth of the CIBER mask, we also find that some of the low-redshift non-linear galaxy clustering is also modeled by the IHL component. To reduce the confusion of the two components requires an exact calculation of the low-redshift galaxy Poisson level, which is not possible with current data. We plan to tackle these issues with future datasets.